\documentclass[10pt, twoside]{article}
\usepackage[paperwidth=170mm, paperheight=238mm]{geometry}
\usepackage{graphicx}
\usepackage{amsmath, amsfonts, amssymb, amsthm, bbm}
\usepackage{bm}
\usepackage{mathtools}
\usepackage{fancyhdr}
\usepackage{subfig}
\usepackage{lettrine}
\usepackage[explicit]{titlesec}
\usepackage{watermark}
\usepackage{color}
\usepackage[final]{pdfpages}
\usepackage{setspace}
\usepackage{chngcntr}
\usepackage{nameref}
\usepackage{enumitem}
\usepackage[perpage]{footmisc}
\usepackage{physics}
\usepackage{float}

\usepackage[font=small]{caption}
\usepackage[colorlinks, citecolor=blue, linkcolor=black, urlcolor=blue]{hyperref}

\usepackage[T1]{fontenc}
\usepackage{lmodern}

\numberwithin{equation}{section}
\numberwithin{figure}{section}
\numberwithin{table}{section}

\definecolor{grey}{rgb}{0.5,0.5,0.5}
\definecolor{l-grey}{rgb}{0.8,0.8,0.8}
\definecolor{white}{rgb}{1,1,1}
\definecolor{black}{rgb}{0,0,0}
\definecolor{myred}{rgb}{0.831, 0.165, 0.184}
\definecolor{myblue}{rgb}{0.149, 0.471, 0.698}
\definecolor{mypurple}{rgb}{0.698, 0.400, 1}

\let\origdoublepage\cleardoublepage
\newcommand{\clearemptydoublepage}{%
  \clearpage
  {\pagestyle{empty}\origdoublepage}%
}

\let\cleardoublepage\clearemptydoublepage

\titleformat{\section}[display]{\vspace*{190pt} \bfseries\sffamily \Huge}
{\begin{picture}(0,0)\put(-60,-30){\textcolor{grey}{\thesection}}\end{picture}}
{0pt}
{#1}
[]
\titlespacing*{\section}{40pt}{10pt}{40pt}[40pt]

\titlespacing*{\subsection}{0pt}{30pt}{20pt}[0pt]
\titleformat{\subsection}[display]{\Large \sffamily}{}{0pt}{\thesubsection \ #1}[]

\begin{document}

\pagestyle{fancy}
\renewcommand{\headrulewidth}{0pt}
\fancyhead{}
\fancyfoot{}




\setcounter{page}{1}
\clearemptydoublepage


\begin{center}

\hrule

\vspace{16pt}
{\huge Digital-analog\\
quantum computing\\
and\\
algorithms\par}
\vspace{16pt}

\hrule

\vspace{35pt}

{\Large {\bf Ana Mart\'in Fern\'andez } }

\vspace{35pt}

\emph{Supervised by} \\

\vspace{15pt}

{\large

Mikel Sanz

}

\vspace*{\fill}

\includegraphics[height=2.5cm]{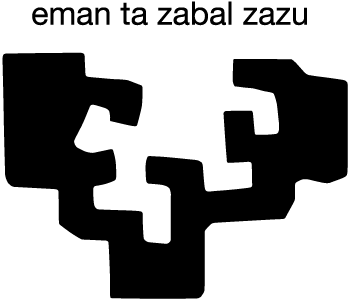}

\vspace{20pt}

Departamento de Qu\'imica F\'isica\\
Facultad de Ciencia y Tecnolog\'ia\\
\vspace{4pt}
Universidad del Pa\'is Vasco

\vspace{15pt}

{\large June 2023}

\end{center}

\pagebreak

This document is a PhD thesis developed during the period from October 2018 to March 2023 at 
the University of the Basque Country.

\vspace*{250pt}

\noindent \textcopyright 2023 by Ana Mart\'in Fern\'andez. All rights reserved.

\bigskip

\noindent An electronic version of this thesis can be found at \href{https://nquirephysics.com}{www.nquirephysics.com}

\bigskip

\noindent Bilbao, April 2023

\vspace*{\fill}

{\setstretch{1.2}


\noindent This document was generated with the 2015  \LaTeX  \ distribution.

\noindent The \LaTeX \  template is adapted from a \href{https://gitlab.com/iagobaapellaniz/PhD-Thesis}{template} by Iagoba Apellaniz.

}

\cleardoublepage

\vspace*{60pt}
\begin{flushright}
\emph{A mi familia, la que toco y la que pienso;\\
a mis amigos, distribuidos en diferentes lugares y momentos;\\
\vspace{1cm} y a Hanse, silenciosa pero siempre presente}
\end{flushright}

\vspace*{45pt}
\begin{flushright}
\rule{.4\textwidth}{0.5pt}
\vspace{45pt}
\end{flushright}
\begin{flushright}
\emph{To my family, tangible and intangible;\\
to my friends, spread over the space and time\\
\vspace{1cm} ... and to my cat, best roomate I can ask for}
\end{flushright}

\cleardoublepage

\vspace*{150pt}
\hspace*{\fill}\begin{minipage}{\textwidth-90pt}
\emph{If we are satisfied with things as they are, nothing will ever be discovered.}
\end{minipage}
\\
\vspace*{10pt}
\begin{flushright}
{\setstretch{2} Seneca \\
Paraphrase of a passage from ``Moral Letters to Lucilius''}
\end{flushright}


\titleformat{\section}[display]
{\vspace*{160pt}
\bf\sffamily \Huge}
{{\textcolor{black}{\thesection}}. #1}
{0pt}
{#1}
[]
\titlespacing*{\section}{75pt}{10pt}{40pt}[40pt]

\textsf{\tableofcontents}


\section*{Abstract}
\pagenumbering{roman}
\fancyfoot[LE,RO]{\thepage}
\phantomsection
\addcontentsline{toc}{section}{Abstract}

The field of quantum computing has experienced significant growth in recent years, driven by breakthroughs in the development of new algorithms, the discovery of novel quantum phenomena, and the construction of increasingly larger and more stable quantum systems. Despite these advances, the implementation of quantum algorithms on real hardware remains a significant challenge, as quantum systems are prone to errors and decoherence that can limit their computational power.

In this context, the digital-analog quantum computational paradigm has emerged as a promising approach to building large-scale quantum computers and implementing quantum algorithms for practical applications. The digital-analog paradigm combines the strengths of both digital and analog quantum computing, allowing for efficient implementation of quantum algorithms with improved accuracy and error mitigation.

The digital-analog paradigm is particularly well-suited for implementing quantum algorithms that require a large number of quantum gates and measurements or that involve complex entanglement structures. By leveraging the robustness of analog quantum simulation and the flexibility of digital quantum computation, the digital-analog paradigm provides a powerful framework for designing and implementing quantum algorithms on real hardware.

In this thesis, we provide a detailed analysis of the digital-analog quantum computational paradigm and its application to the implementation of quantum algorithms. We first present a comprehensive overview of the digital-analog paradigm, including a toolkit for Hamiltonian simulation that enables the implementation of quantum algorithms with improved efficiency and accuracy. In addition, we present a noise model to compare the performance of the digital and digital-analog quantum computation frameworks.

Next, we focus on the implementation of four relevant quantum algorithms using both the digital and digital-analog quantum computational frameworks. These algorithms include the quantum Fourier transform, the quantum phase estimation, the Harrow-Hassidim-Lloyd algorithm for solving linear systems of equations, and the quantum approximate optimization algorithm. By comparing the performance of these algorithms using both frameworks, we demonstrate that the digital-analog paradigm offers significant advantages over the digital approach, particularly in terms of reduced error rates and improved scalability.

Finally, we explore how the cross-resonance effect present in superconducting circuits can be used for digital analog quantum simulations. Our results show that this effect can be leveraged to implement Hamiltonian simulations with high accuracy and efficiency, further demonstrating the potential of the digital-analog paradigm for practical quantum computing applications.

In conclusion, our thesis demonstrates the promise of the digital-analog quantum computational paradigm for implementing quantum algorithms on real hardware. The digital-analog paradigm offers significant advantages over the purely digital approach, including improved scalability, reduced error rates, and greater flexibility in circuit design. We believe that this work will contribute to ongoing efforts to develop practical quantum computing technologies and pave the way for future research in the field. As the field of quantum computing continues to evolve, the exploration of new methods for improving the accuracy and efficiency of quantum computations will be crucial for advancing our understanding of quantum computing and identifying new approaches for implementing quantum algorithms.

\cleardoublepage


\section*{Resumen}
\fancyfoot[LE,RO]{\thepage}
\phantomsection
\addcontentsline{toc}{section}{Resumen}

En las últimas décadas, la computación cuántica ha evolucionado de un concepto abstracto a una tecnología con el potencial de revolucionar la computación tal como la conocemos. Los avances tecnólogicos y el desarrollo de la teoría han permitido el asentamiento este campo en distientas áreas de la ciencia. En esta tesis, exploramos el paradigma computacional cuántico digital-analógico y su potencial para implementar algoritmos cuánticos en los dispositivos actualmente disponibles. Para ello, partimos de la extensa investigación realizada previamente en el campo y contribuimos a ella proponiendo enfoques novedosos para resolver problemas computacionales utilizando la computación cuántica. Nuestro objetivo es contribucir al campo y avanzar en nuestra comprensión del potencial de la computación cuántica.

Fue en la década de los $80$ cuando, Beniof, Manin y Feynman, entre otros, señalaron que existían dificultades esenciales a la hora de simular sistemas cuánticos debido al crecimiento exponencial de sus grados de libertad. Sugirieron entonces la posibilidad de construir ordenadores que estuvieran basados en los principios de la mecánica cuántica para asi solventar esas limitaciones. Más concretamente, propusieron emplear sistemas cuánticos completamente controlables para codificar la dinámica de sistemas quánticos de muchos cuerpos, ya que la complejidad de este tipo de sistemas crece exponencialmente con la dimensión del mismo.

Una década más tarde, en 1996, Lloyd demostró que si los sistemas cuánticos evolucionan a través de interacciones locales, entonces un ordenador cuántico  puede simularlos eficientemente. Argumentó que la aplicación de operaciones que preservan la coherencia --tanto como sea posible--, puede considerarse como una activación y desactivación de Hamiltonianos. Cada operación experimental mueve el sistema a lo largo de cualquier evolución temporal, representada por la operación unitaria $U=e^{iAt}$. Ese resultado implica que existe una variedad de sistemas cuánticos, a los que se les puede aplicar un conjunto de operaciones experimentales simples, pudiendo asi simular otros sistemas cuánticos.

Con estos resultados e inspirados por el desarrollo de la computación clásica, se propuso la idea de la computación digital cuántica basada en puertas. Para ello, resultó fundamental el concepto de {\it conjunto de puertas universales}, que no es más que un set de puertas a las que puede reducirse, con precisión arbitraria, cualquier operación posible en un ordenador cuántico. Esto, junto con el desarrollo de la corrección cuántica de errores, condujo a la computación cuántica basada en puertas tal y como la conocemos hoy en día.

Las puertas utilizadas en los ordenadores cuánticos son pulsos digitales que pueden rotar un solo qubit (puerta de un solo qubit) o activar la interacción entre dos qubits (puerta de dos qubits), según su aplicación. Al alplicar estas puertas de forma secuencial, se pueden ejecutar diferentes algoritmos cuánticos, según la secuendia escogida. La primera demostración concreta de que un ordenador cuántico podría superar a uno clásico fue realizada por Deutsch. A esto le siguió la extensión realizada por Deutsch y Jozsa, que mostró por primera vez una brecha similar para el escalado del tiempo requerido para resolver un problema. Poco después de eso, se propusieron otros algoritmos que ofrecieron una mayor velocidad que sus análogos clásicos, incluido el algoritmo de Shor para factorizar números primos y el algoritmo de Grover para buscar en bases de datos no estructuradas. Desde entonces, se han desarrollado muchos algoritmos en busca de ventajas cuánticas, ampliando enormemente el repertorio de algoritmos cuánticos.

Ha habido varias propuestas de plataformas físicas para realizar computación cuántica. El primer esquema de implementación fue presentado en 1995 por J. I. Cirac y P. Zoller, quienes propusieron puertas lógicas cuánticas utilizando iones atrapados. Posteriormente, se propusieron otros métodos físicos para lograr la computación cuántica, como la resonancia magnética nuclear (RMN), los qubits de espín y los qubits superconductores. Desde entonces, ha habido un progreso experimental significativo en el campo.

La computación cuántica puede llevarse a cabo en estas diferentes plataformas físicas, siempre que éstas cumplan una serie de criterios conocidos como criterios DiVincenzo. Estos criterios establecen que la plataforma utilizada debe: ser un sistema físico escalable con qubits bien caracterizados; tener la capacidad de inicializar el estado de los qubits a un estado fiduciario simple; tener tiempos de decoherencia suficientemente largos; tener un conjunto universal de puertas cuánticas y una capacidad de medición específica de qubit.

El primer resultado experimental de un algoritmo cuántico se logró en un ordenador cuántico de RMN de 2 qubits donde J. A. Jones y M. Mosca resolvieron el problema de Deutsch. Poco después, I. L. Chuang y sus colaboradores realizaron la primera ejecución del algoritmo de Grover también en un ordenador cuántico  de RMN. Un ordenador cuántico de RMN utiliza los estados de espín del núcleo de una molécula como qubits. Estas moléculas podrían estar en una muestra líquida --RMN de estado líquido-- o en una muestra sólida --RMN de estado sólido--, por ejemplo, red de diamante de vacancia de nitrógeno, siendo esta última una mejor aproximación mediante computación cuántica. Desde 1988, los logros experimentales de los ordenadores cuánticos de RMN crecieron rápidamente, con resultados interesantes en la implementación de algoritmos cuánticos y la creación de ordenadores cuánticos RMN híbridos. A pesar de este ritmo de progreso, se reconoció desde el principio que dichos dispositivos RMN se enfrentaban a un pobre escalado de la relación señal/ruido. Además, en 2002, N. C. Menicucci y C. M. Caves demostraron que todos los experimentos en RMN de muchos cuerpos en estado líquido realizados hasta ese momento presentaban poco o ningún entrelazamiento, lo que llevó a afirmar que dichos experimentos podrían haber sido meras simulaciones clásicas de un sistema cuántico. Esta limitación se reduce a la RMN de estado líquido y todavía es posible que la RMN de estado sólido se pueda utilizar para la computación cuántica. Estos resultados llevaron a la investigación en computación cuántica a reenfocarse más en otras tecnologías, como iones atrapados, puntos cuánticos, fotónica y circuitos superconductores, y la idea de computación cuántica de RMN fue relegada progresivamente.

Estos avances abrieron la puerta a las {\it simulaciones analógicas}, que se basan en sistemas cuánticos controlables cuyas dinámicas son similares a las de los sistemas que se simulan. Para lograr este objetivo, se desarrollaron dispositivos cuyo único propósito es estudiar las dinámicas de estos sistemas, conocidos como simuladores cuánticos analógicos. Estos dispositivos han permitido obtener resultados relevantes en una amplia gama de campos, incluido el modelo cuántico de Rabi, la física de Casimir, dinámica de fluidos , química cuántica y física de la materia condensada, entre muchos otros.

Desde entonces y durante la última década, se han logrado avances significativos, llegando incluso a alcanzar uno de los mayores hitos de la computación cuántica: la resolución de un problema que resulta fuera del alcance de los ordenadores clásicos. Este avance, conocido como {\it ventaja cuántica}, fue logrado por primera vez por Google en 2019. En su estudio, F. Arute {\it et al.} utilizó un muestreo de circuito aleatorio en su procesador sycamore de $53$-qubit. Posteriormente, ha habido numerosos experimentos que demuestran la ventaja cuántica en una variedad de sistemas, incluidos los qubits superconductores y los qubits fotónicos. En esta tesis nos hemos centrado en los circuitos superconductores, pero es importante tener en cuenta que lo que presentamos en este sentido puede aplicarse a otras plataformas, como las mencionadas anteriormente.

La idea detrás de la {\it computación cuántica digital-analógica} (DAQC por sus siglas en inglés). es explotar la interacción natural presente en el procesador cuántico y usarla junto con puertas de un solo qubit para controlar el sistema y simular una dinámica deseada. Este nuevo paradigma computacional surgió como una alternativa al conocido como {\it computación cuántica digital} (DQC por su traducción al inglés). Una de las principales diferencias entre los paradigmas DAQC y DQC es que, en este último, la dinámica que se quiere simular se descompone en puertas de uno y dos qubits, en lugar de en términos de la interacción hamiltoniana del sistema. Las puertas de dos qubits requieren que la interacción entre los qubits del sistema que no están involucrados en dicha operación, se atenúe lo más posible, idealmente debe ser casi nula. Esta atenuación parcial de la interacción inherente del sistema es una operación ``antinatural'', que genera una fuente de ruido relevante. No obstante, cabe destacar que el enfoque digital en la computación cuántica ha logrado resultados relevantes en varias áreas, como el aprendizaje automático cuántico, finanzas, sistemas cuánticos abiertos, química cuántica, o teoría cuántica de campos. Sin embargo, dado que estamos en la era de la computación cuántica de escala casi intermedia (NISQ por su siglas en inglés), el paradigma de la computación cuántica digital se ve afectado por las limitaciones experimentales de los dispositivos NISQ. Los dispositivos NISQ son lo más cercano que estamos hoy a los ordenadores cuánticos. Se dice que son ``ruidosos'' porque estos dispositivos no tienen corrección de errores y el ruido limita su poder de cómputo. Aún así, representan una tecnología puntera que nos permite explorar los sistemas cuánticos de muchas partículas en un régimen que nunca antes había sido accesible experimentalmente.

Para superar esta situación e inspirado en el trabajo de Dodd {\it et al.}, en 2020 se propone por parte de Parra {\it et al.} avanzar en este tipo de dispositivos mediante el uso de DAQC. Este enfoque combina la robustez de la simulación cuántica analógica con la flexibilidad de la computación cuántica digital. En el primer artículo, los autores demuestran de manera elegante que es posible generar cualquier Hamiltoniano entrelazado de dos cuerpos usando cualquier otro Hamiltoniano entrelazado arbitrario como generador. En su trabajo, Parra {\it et al.} van un paso más allá y proponen que, a cambio de fijar el hamiltoniano de origen, es posible simular cualquier hamiltoniano entrelazado arbitrario de múltiples cuerpos. Además, proveen instrucciones explícitas sobre cómo usar este paradigma para simular diferentes familias de hamiltonianos. En base a este trabajo, a lo largo de estos cuatro años, hemos desarrollado propuestas para la implementación de algoritmos y subrutinas cuánticas ubicuas utilizando el paradigma DAQC antes mencionado.

La estrategia que seguimos para desarrollar nuestro trabajo fue proponer algoritmos en los que pudiéramos apoyarnos para construir los siguientes, como una matrioska. Por lo tanto, comenzamos con la transformada cuántica de Fourier debido a su relevancia como paso clave en una multitud de algoritmos. El siguiente paso lógico fue trabajar en la estimación cuántica de fase. Finalmente, uniendo estas dos subrutinas e incluyendo una adicional, pudimos completar una propuesta para implementar el algoritmo Harrow-Hassidim-Lloyd para resolver sistemas de ecuaciones lineales. Además, como muchos algoritmos variacionales encajan con el paradigma DAQC, propusimos una implementación para uno de los algoritmos variacionales que han despertado interés más allá de las fronteras de la ciencia, permeando en el campo de la industria debido a sus aplicaciones en logística, estamos hablando del algoritmo de optimización cuántica aproximada (QAOA por sus siglas en inglés). Finalmente, inspirados por las limitaciones físicas de los dispositivos NISQ, consideramos procesadores cuánticos de circuitos superconductores y extendimos el efecto de resonancia cruzada de una interacción de dos qubits a un Hamiltoniano analógico por sí solo. De esa manera, demostramos que es posible generar una amplia gama de hamiltonianos al aumentar la dinámica hamiltoniana analógica con puertas de un solo qubit.

Nuestro trabajo ha tenido un impacto significativo en varios grupos de diferentes universidades, quienes han adoptado algunas de nuestras propuestas para sus investigaciones en el campo de la computación cuántica. En 2020, Babukhin {\it et al.} utilizaron el efecto de la resonancia cruzada en procesadores cuánticos basados en circuitos superconductores para realizar simulaciones DAQC. Sus resultados demuestran que un enfoque DA híbrido permite una simulación eficiente de la dinámica de un modelo Ising de campo transversal sin la necesidad de puertas estándar de dos qubits, que presentan desafíos importantes en la computación cuántica. Al aprovechar el efecto de resonancia cruzada en los procesadores cuánticos superconductores de IBM, simularon con éxito la dinámica Trotterizada de los clústeres de espín y compararon los resultados con los obtenidos mediante la computación digital convencional basada en puertas de dos qubits. Sus hallazgos indican que el enfoque DA supera al enfoque digital estándar, lo que ofrece perspectivas prometedoras para las computadoras NISQ a corto plazo. Este trabajo destaca el potencial de mejora adicional con procesadores más especializados y sugiere la viabilidad de una estrategia de digital a analógico para la computación cuántica.

Otro trabajo destacable es el desarrollado por M. Gong {\it et al.} en el que incluyen una red neuronal cuántica digital-analógica cuya tarea es distinguir la naturaleza de un conjunto de estados cuánticos. Aquí, los autores presentan una técnica novedosa para categorizar estados cuánticos de muchos cuerpos, llamada detección neuronal cuántica. Usando un procesador cuántico superconductor de 61 qubits, demuestran la efectividad de su método para clasificar dos fases distintas de la materia: ergódica y localizada. El corazón de su enfoque de detección neuronal cuántica es una red neuronal cuántica analógica digital (QNN), diseñada para distinguir entre estos estados cuánticos. El QNN toma como entrada un circuito cuántico variacional analógico digital de doble capa, y se muestra que es eficiente en términos de hardware y universalmente capaz de computación cuántica, como lo demostraron previamente Parra {\it et al.}. Este nuevo método representa un paso prometedor hacia la comprensión y caracterización de las propiedades complejas de los sistemas cuánticos de muchos cuerpos.

También ha habido interesantes propuestas a nivel teórico, una de ellas relacionada con el campo de los algoritmos cuánticos variacionales. En su trabajo, A. Michel {\it et al.} proponen usar DAQC para estimar la energía del estado fundamental de varias moléculas usando un algoritmo de solución propia cuántica variacional analógica digital (VQE). El estudio se centra en implementar el algoritmo VQE en un simulador cuántico de Rydberg para moléculas $\text{H}_2$, $\text{Li}\text{H}$ y $\text{Be}\text{H}_2$ . Los autores consideran las limitaciones de la plataforma, como la restricción de la acción local a la preparación y medición del estado inicial y la implementación de la evolución temporal hamiltoniana global. A pesar de esto, destacan que la interacción digital-analógica se puede lograr mediante el uso de una caja de herramientas, como se demostró en un estudio anterior llevado acabo por S. Notarnicola {\it et al.}, que enfatiza el potencial de los dispositivos de átomos neutros de próxima generación.

Adicionalmente, se ha propuesto una arquitectura de circuitos superconductores para realizar DAQC, lo que nos hace vislumbrar un futuro donde DAQC juega un papel importante en el desarrollo del campo de la computación cuántica. En este estudio, los autores se enfocan en el avance de DAQC como un enfoque prometedor hacia el desarrollo de computadoras cuánticas de próxima generación. Los autores proponen un diseño de circuito superconductor para DAQC, que consta de una cadena de qubits de carga acoplados a través de dispositivos superconductores de interferencia cuántica (SQUID por sus siglas en inglés) conectados a tierra. Este diseño permite la manipulación de interacciones individuales entre qubits vecinos más cercanos, produciendo interacciones diferentes e independientes, como un término de intercambio o doble excitación/desexcitación. Mediante el uso de este diseño, los autores pretenden expandir los mapeos algorítmicos de DAQC y la variedad de geometrías y topologías de computadoras cuánticas accesibles. La flexibilidad de este diseño proporciona una gran familia de hamiltonianos multicuerpo analógicos, adecuados para implementaciones eficientes de protocolos DAQC. Este trabajo es un paso significativo hacia el objetivo de alcanzar computadoras cuánticas tolerantes a fallas y sistemas complejos de computación en plataformas cuánticas. Nos acerca al objetivo de lograr una ventaja cuántica con menos recursos algorítmicos y de hardware, contribuyendo a los esfuerzos continuos para avanzar en el paradigma DAQC.

Relacionado con la simulación de sistemas cuánticos de muchos cuerpos, ha habido una interesante propuesta de la mano de L. C. C\'eleri {\it et al.}. En su trabajo, los autores presentan un algoritmo cuántico DA que puede simular una amplia gama de Hamiltonianos fermiónicos, incluido el modelo de Fermi-Hubbard. La simulación de sistemas cuánticos de muchos cuerpos es una tarea desafiante que requiere una cantidad significativa de recursos, que aumentan exponencialmente con el tamaño del sistema. Esto se vuelve aún más difícil con los sistemas fermiónicos debido a las interacciones no locales causadas por la naturaleza antisimétrica de la función de onda fermiónica. Sin embargo, los autores demuestran cómo los métodos DA permiten que los algoritmos cuánticos se ejecuten de manera más eficiente que sus contrapartes digitales al hacer un mejor uso del tiempo de coherencia. Además, en su trabajo, los autores proporcionan una implementación práctica de estas técnicas utilizando una arquitectura de baja conexión para modelos fermiónicos específicos.

Después de cuatro años de intensa investigación, hemos desarrollado un catálogo completo de algoritmos diseñados específicamente para DAQC. Nuestro trabajo ha sentado una base sólida para que futuros investigadores amplíen y exploren más este apasionante campo. Creemos que los algoritmos que hemos desarrollado serán útiles para aquellos que deseen realizar investigaciones en DAQC, y somos optimistas de que esta tesis servirá como un recurso valioso para futuros estudios. Confiamos en haber contribuido al avance del campo y esperamos ver qué progresos adicionales se pueden lograr.

\cleardoublepage


\section*{Acknowledgements}
\phantomsection
\addcontentsline{toc}{section}{Acknowledgements}

\begin{flushright}
\emph{``Cuanto m\'as te acercas a lo esencial, menos puedes nombrarlo''}\\
\begin{flushright} \end{flushright} - Rosa Montero, La rid\'icula idea de no volver a verte\end{flushright}
\vspace{20pt}

Nothing comes from nothing and I know that I never walked alone on the path that has brought me here. The completion of this thesis has been a journey filled with challenges, triumphs, and unforgettable experiences. It is with immense gratitude that I acknowledge the individuals and institutions that have supported me throughout this process. Without their guidance, encouragement, and unwavering belief in my abilities, this accomplishment would not have been possible. Their support has been invaluable and has contributed greatly to my personal and professional growth.

I want to start by thanking those who trusted me when I was just an unknown master's student who had barely been in Bilbao for a few months. Kike Solano invited me to test this research, Mikel Sanz helped me shape it and Iñigo Egusquiza gave me time whenever I needed it. I am enormously grateful to Mikel Sanz for the trust he has always placed in me, as well as for everything he has taught me over the years.

I would like to extend my sincere gratitude to the following individuals and institutions for their support throughout my Ph.D. research. Firstly, I would like to thank Shanghai University for hosting me for the first two months at the beginning of my research journey. I am also grateful to Andreas Wallraff, and Christopher Eichler for hosting me during my visit to their group in ETH Zurich. In particular, I would like to thank them for their hospitality and kindness. During my research, I had the privilege of collaborating with the group of Frank Wilhelm-Mauch at Saarland University. This collaboration also included David Headley and Thorge Müller, and their support and insights were invaluable to the success of my work. I would also like to express my gratitude to Pavel Lougovski, and Eugene F. Dumitrescu, from the Oak Ridge National Laboratory for their work and support and for hosting us at Berkeley Labs. I will always cherish the sundae birthday dessert Eugene arranged for me. Finally, I would like to thank Göran Johansson and Laura Garc\'ia \'Alvarez, from Chalmers University, for their generosity and hospitality in hosting me and my colleagues during our stay in Göteborg. I met many wonderful people during my stay, including Pontus, David, and Tom, who helped me learn a great deal about physics and different ways of working.

Special attention goes to the people that have surrounded me every single day: my colleagues from both QUTIS and NQUIRE groups. To my colleagues in my own research group, thank you for your support and for being a constant source of inspiration and motivation. I have a special mention to G. Gatti, who, unexpectedly, has been the more consistent travel partner I ever had. We have been traveling together since we both started our Ph.D. journey and I am glad that we can share all of those scientific and personal memories. This thing about mentioning one by one the people that have made an impact on my journey can be easily converted to torture, so please, accept my extended gratitude to all of you.

On a personal level, I am deeply grateful for the unwavering support and encouragement of my friends and family. They have been my pillars of strength, cheering me on from the sidelines and always believing in me, even when I doubted myself. I will always be grateful for their love and support, and I hope to make them proud of my achievements. I wish all of my family and friends could read these lines, but I am aware that interdimensional travel is not available for now, so I content myself with being aware of the great impact that they all have had on the researcher that I am today.

\cleardoublepage


\section*{List of Publications}
\phantomsection
\addcontentsline{toc}{section}{List of Publications}

This thesis is based on the following publications and preprints:
\\

\begin{enumerate}
\item {\color{myblue} \underline{A. Martin}}, L. Lamata, E. Solano, and M. Sanz,\\
{\it Digital-analog quantum algorithm for the quantum Fourier transform}, \\
\href{https://doi.org/10.1103/PhysRevResearch.2.013012}{Physical Review Research {\bf 2}, 013012 (2020)}.

\item D. Headley, T. M\"uller,  {\color{myblue} \underline{A. Martin}}, E. Solano, M. Sanz, and F. Wilhelm, \\
{\it Approximating the Quantum Approximate Optimisation Algorithm}, \\
\href{
https://doi.org/10.48550/arXiv.2002.12215}{Physical Review A, {\bf 106}, 042446 (2022)}.

\item T. Gonzalez-Raya, R. Asensio-Perea, {\color{myblue} \underline{A. Martin}}, L. C. C\'eleri, M. Sanz, P. Lougovski, and E. F. Dumitrescu, \\
{\it Digital-Analog Quantum Simulations Using The Cross-Resonance Effect}, \\
\href{https://doi.org/10.1103/PRXQuantum.2.020328}{PRX Quantum {\bf 2}, 020328 (2021)}.

\item P. Garc\'ia-Molina, {\color{myblue} \underline{A. Martin}}, M. Garcia de Andoin, and M. Sanz, \\
{\it Noise in Digital and Digital-Analog Quantum Computation}, \\
\href{https://doi.org/10.48550/arXiv.2107.12969}{arxiv: 2107.12969 (2021)}.

\item {\color{myblue} \underline{A. Martin}}, R. Ibarrondo, and M. Sanz \\
{\it Digital-analog co-design of the Harrow-Hassidim-Lloyd algorithm}, \\
\href{https://doi.org/10.1103/PhysRevApplied.19.064056}{Physical Review Applied {\bf 19}, 064056 (2023)}.
\end{enumerate}

\newpage

Other publications not included in this thesis:

\begin{enumerate}[resume]

\item {\color{myblue} \underline{A. Martin}}, B. Candelas, \'A. Rodr\'iguez-Rozas, J. D. Mart\'in-Guerrero, X. Chen, L. Lamata, R. Orús, E. Solano, and M. Sanz, \\
{\it Towards Pricing Financial Derivatives with an IBM Quantum Computer}, \\
\href{https://doi.org/10.1103/PhysRevResearch.3.013167}{Physical Review Research {\bf 3}, 013167 (2021)}.

\end{enumerate}

\cleardoublepage

%
%


\section*{Abbreviations and \\ conventions}
\phantomsection
\addcontentsline{toc}{section}{Abbreviations and conventions}

We use the following abbreviations throughout the thesis
\begin{enumerate}[leftmargin=5cm]
	\item [\bf ATA]{All-to-all}
	\item [\bf bDAQC]{Banged digital-analog quantum computing}
	\item [\bf CR] {Cross-resonance}
	\item [\bf DA]{Digital-analog}
	\item [\bf DAQC]{Digital-analog quantum computing}
	\item [\bf DFT]{Discrete Fourier transform}
	\item [\bf DQC]{Digital quantum computing}
	\item [\bf HHL]{Harrow--Hassidim--Lloyd}
	\item [\bf RWA]{Rotating-wave approximation}
	\item [\bf NISQ]{Near intermediate-scale quantum computing}
	\item [\bf NMR]{Nuclear Magnetic Resonance}
	\item [\bf NN]{Nearest-neighbor}
	\item [\bf QAOA]{Quantum approximate optimization algorithm}
	\item [\bf QFT]{Quantum Fourier transform}
	\item [\bf QPE]{Quantum phase estimation}
	\item [\bf sDAQC]{Step-wise digital-analog quantum computing}
	\item [\bf SQG] {Single-qubit gate}
	\item [\bf SQR]{Single-qubit rotation}
	\item [\bf TQG] {Two-qubit gate}  
 \end{enumerate}

We set the reduced Planck constant $\hbar = 1$ throughout the thesis.

\cleardoublepage


\renewcommand{\headrulewidth}{0.5pt}
\fancyfoot[LE,RO]{\thepage}
\fancyhead[LE]{\rightmark}
\fancyhead[RO]{\leftmark}


\titleformat{\section}[display]
{\vspace*{190pt}
\bfseries\sffamily \LARGE}
{\begin{picture}(0,0)\put(-50,-23){\textcolor{grey}{\thesection}}\end{picture}}
{0pt}
{\textcolor{white}{#1}}
[]
\titlespacing*{\section}{80pt}{10pt}{40pt}[20pt]
\vfill


\pagenumbering{arabic}

\section[Introduction]{Introduction}


\thiswatermark{\put(1,-280){\color{l-grey}\rule{70pt}{42pt}}
\put(70,-280){\color{grey}\rule{297pt}{42pt}}}

%
%

\lettrine[lines=2, findent=3pt,nindent=0pt]{O}{ver} the past few decades, quantum computing has evolved from an abstract concept to a technology with the potential to revolutionize computing as we know it. In this thesis, we explore the digital-analog quantum computational paradigm and its potential to implement quantum algorithms on currently available devices. We build upon the extensive research conducted in this field and contribute to it by proposing novel approaches to solving computational problems using quantum computing. By doing so, we aim to make valuable contributions to the field and advance our understanding of the potential of quantum computing.

In the 1980s, Benioff \cite{Benioff1980}, Manin \cite{Manin1980}, and Feynman \cite{Feyn1982} recognized the need to use a quantum platform for simulating quantum systems, since that task was unattainable for classical computers due to the exponential growth of degrees of freedom. This was triggered by the aim of finding a computation model that could efficiently simulate any other model of computation. To address this challenge, they proposed to use fully controllable quantum systems to encode the dynamics of many-body quantum systems. The complexity of these systems also grows exponentially with the dimension of the system, making them difficult to simulate classically. By employing a quantum platform, it is possible to achieve a more efficient and accurate simulation of quantum systems, which could lead to a better understanding of their behavior and potential applications.

A decade later, Lloyd showed that if quantum systems evolve through local interactions, then a quantum computer can efficiently simulate them \cite{Lloyd_1996}. He argued that the application of operations that --approximately-- preserve coherence, can be thought of as turning on and off Hamiltonians. Each experimental operation drives the system along any time evolution, represented by the unitary operation $U=e^{iAt}$. That result implied that a variety of quantum systems, to which a set of simple experimental operations can be applied, can simulate other quantum systems. With these results and inspired by the development of classical computing, the idea of gate-based quantum digital computing was proposed. Fundamental to this was the concept of the universal gate set, a set of gates to which any possible operation in a quantum computer can be reduced with arbitrary precision. This, along with the development of quantum error correction led to gate-based quantum computing.

The gates used in quantum computers are digital pulses that can rotate a single qubit (single-qubit gate) or activate the interaction between two qubits (two-qubit gate), depending on their application. By sequencing these gates in different ways, various quantum algorithms can be executed. The first concrete demonstration that a quantum computer could outperform a classical one was shown by Deutsch \cite{Deutsch85}. This was followed by the extension made by Deutsch and Jozsa \cite{DeutschJotzsa1992}, which showed for the first time a similar gap for the scaling of the time required to solve a problem. Shortly after that, other algorithms were proposed that offered speed-up over their classical counterparts, including Shor's algorithm for prime number factorization \cite{Shor1996} and Grover's algorithm for searching unstructured databases \cite{Grover1996}. Since then, many algorithms have been developed in the search for quantum advantage, greatly expanding the repertoire of quantum algorithms \cite{Alba2022}.

There have been several proposals for physical platforms to perform quantum computing. The first implementation scheme was introduced in 1995 by J. I. Cirac and P. Zoller, who proposed quantum logic gates using trapped ions \cite{Cirac1995}. Subsequently, other physical methods for achieving quantum computing, such as nuclear magnetic resonance (NMR) \cite{Cory1997, Gershenfeld1997, Cory1998}, spin qubits \cite{Kane1998, Loss1998}, and superconducting qubits \cite{Nakamura1999}, were proposed. Since then, there has been significant experimental progress in the field.

Quantum computing can be carried out on these different physical platforms, as long as they meet a series of criteria known as the DiVincenzo criteria \cite{DiVincenzo1997}. These criteria establish that the platform used must be a scalable physical system with well-characterized qubits; have the ability to initialize the state of the qubits to a simple fiducial state; have sufficiently long decoherence times; have a universal set of quantum gates and a qubit-specific measurement capability. 

The first experimental result of a quantum algorithm was achieved in a 2-qubit NMR quantum computer where J. A. Jones and M. Mosca solved Deutsch’s problem \cite{JonesMoscaNMR1998}. Shortly after, I. L. Chuang and collaborators performed the first execution of Grover's algorithm also on an NMR quantum computer \cite{ChuangNMR1998}. An NMR quantum computer uses the spin states of a molecule nucleus as qubits. These molecules could be in a liquid sample -- liquid state NMR -- or in a solid sample -- solid-state NMR --, being the latter a better approach by means of quantum computing. Since 1988, the experimental achievements of NMR quantum computers grew rapidly, with interesting results in implementing quantum algorithms and creating hybrid NMR quantum computers \cite{Kane1998}. Despite this pace of progress, it was recognized from the beginning that NMR quantum computers face a poorly scaled signal-to-noise ratio. Moreover, in 2002, N. C. Menicucci and C. M. Caves \cite{MC2002} showed that all experiments in liquid state bulk ensemble NMR carried out until then, presented little or no entanglement, which led to the affirmation that liquid-state NMR experiments could have been mere classical simulations of a quantum system. This limitation reduces to liquid-state NMR and it is still possible that solid-state NMR could be used for quantum computing. These results led research in quantum computing to refocus more on other technologies, such as trapped ions, quantum dots, photonics, and superconducting circuits, and the idea of NMR quantum computing was progressively relegated.

The advances in quantum computing have paved the way for {\it analog quantum simulations} \cite{Buluta2009, Georgescu2014}, which rely on controllable quantum systems whose dynamics are similar to those of the systems being simulated. Single-purpose devices known as analog quantum simulators have been developed to achieve this goal. These simulators have been used to obtain numerous results in a wide range of fields, including the quantum Rabi Model \cite{BRGDS12, MLHPDSL, PLFRLS2015, BMSSRWU17, LALZPLSK2018}, Casimir physics \cite{FSLRJDS2014, RFERSS2016, SWGS2018}, fluid dynamics \cite{MSLESS2015}, quantum chemistry \cite{Zoller2019}, and condensed matter physics \cite{Hofstetter2018}, among many others.

Since then, significant progress has been made in the last decade, culminating in one of the biggest milestones for quantum computing: solving a problem that is beyond the reach of classical computers. This breakthrough, known as {\it quantum advantage}, was first achieved by Google in 2019 \cite{Arute2019}. In their study, F. Arute {\it et al.} used a randomized circuit sampling on their $53$-qubit sycamore processor. Subsequently, there have been numerous experiments demonstrating quantum advantage in a variety of systems, including superconducting qubits \cite{Wu2021, Zhu2022} and photonic qubits \cite{Zhong2020, Zhong2021, Madsen2022}. In this thesis, we have focused on superconducting circuits, but it is important to keep in mind that what we present along these lines can be applied to other platforms, such as the ones previously mentioned.

The idea behind {\it digital-analog quantum computation} (DAQC) is to exploit the natural interaction present in the quantum processor and use it together with single-qubit gates to control the system and simulate a desired dynamic. This computational paradigm emerged as an alternative of the widespread digital quantum computational (DQC) one. One of the main differences between DAQC and DQC is that, in the latter, the dynamic that one wants to simulate is decomposed into single- and two-qubit gates, rather than in terms of the interaction Hamiltonian of the system. The two-qubit gates require that the interaction among the qubits of the system that are not involved in such operation, has to be attenuated as much as possible, ideally,  it has to be almost null. This partial attenuation of the system's inherent interaction is a rather ``anti-natural'' operation, resulting in a relevant noise source. Nevertheless, the digital approach in quantum computing has achieved relevant results in several areas such as quantum machine learning \cite{ARSL2018, OSCSL2018}, finance \cite{MCRMCLOSS2019, DLMGLMS2019, Javi2021}, open quantum systems \cite{SSSPS2016}, quantum chemistry \cite{GALMSSL2016}, or quantum field theories \cite{LSKDBLD2017}. However, since we are in the near intermediate scale quantum computing (NISQ) era \cite{Preskill2018}, the DQC paradigm is affected by the experimental limitations of NISQ devices. NISQ devices are the closest we are today to quantum computers. They are said to be ``noisy'' because these devices are not error corrected, and the noise limits their computational power. Still, they represent an exciting technology that allows us to explore the complex many-particle quantum systems in a regime that has never been experimentally accessible before \cite{Preskill2021}.

In order to overcome this situation and inspired by the work of Dodd {\it et al.} \cite{Dodd_2002_UniversalQCandS}, it has been proposed by Parra {\it et al.} in 2020, to get advantage of such devices by using DAQC \cite{Parra2020}. This approach merges the robustness of analog quantum simulation with the flexibility of digital quantum computation. In \cite{Dodd_2002_UniversalQCandS}, the authors elegantly prove that it is possible to generate any two-body entangling Hamiltonian using any arbitrary entangling Hamiltonian as the generator. In their work, Parra {\it et al.} go a step further and propose that, in exchange for fixing the origin Hamiltonian, it is possible to simulate any arbitrary multi-body entangling Hamiltonian. Furthermore, they give explicit instructions on how to use this paradigm to simulate different families of Hamiltonians. Based on this work, throughout these four years, we have developed proposals for the implementation of algorithms and ubiquitous quantum subroutines using the aforementioned DAQC paradigm.

The strategy that we followed to develop our work was to propose algorithms on which we could rely to build the following ones, like a matryoshka doll. Thus, we start with the quantum Fourier transform \cite{Martin2020} due to its relevance as a key step in a multitude of algorithms. The next logical step was to work on the quantum phase estimation \cite{GMS2021}. Finally, joining these two subroutines and including an additional one, we were able to complete a proposal to implement the Harrow-Hassidim-Lloyd (HHL) algorithm to solve linear systems of equations \cite{Martin2022}. In addition, as many variational algorithms fit with the DAQC paradigm, we proposed an implementation for the quantum approximate optimization algorithm (QAOA) \cite{Headley2020}. Finally, inspired by the physical limitations of NISQ devices, we considered superconducting circuits quantum processors and extend the cross-resonance effect from a two-qubit interaction to an analog Hamiltonian on its own. That way, we showed that it is possible to generate a wide range of Hamiltonians by augmenting the analog Hamiltonian dynamics with single-qubit gates \cite{GonzalezRaya2021}.

Our work has had a significant impact on various groups from different universities, who have adopted some of our proposals for their research in the field of quantum computing. In 2020, Babukhin {\it et al.} \cite{Babukhin2020} showcase the potential of using superconducting circuit quantum processors for DAQC by utilizing the cross-resonance effect in transmon qubits. Their results demonstrate that a hybrid digital-analog (DA) approach allows for efficient simulation of the dynamics of a transverse-field Ising model without the need for standard two-qubit gates, which present significant challenges in quantum computing. By leveraging qubit-qubit crosstalks in IBM superconducting quantum processors, they successfully simulate the Trotterized dynamics of spin clusters and compare the results with those obtained from conventional digital computation based on two-qubit gates. Their findings indicate that the DA approach outperforms the standard digital approach despite the relatively small crosstalks in IBM quantum processors, offering promising prospects for near-term NISQ computers. This work highlights the potential for further improvement with more specialized processors and suggests the feasibility of a digital-to-analog strategy for quantum computing.

Another remarkable work is the one developed by M. Gong {\it et al.} \cite{JWPan2022} in which they include a DA quantum neural network whose task was to distinguish the nature of a set of quantum states. Here, the authors introduce a novel technique for categorizing many-body quantum states, called quantum neuronal sensing. Using a 61-qubit superconducting quantum processor, they demonstrate their method's effectiveness in classifying two distinct phases of matter: ergodic and localized. The heart of their quantum neuronal sensing approach is a DA quantum neural network (QNN), designed to distinguish between these quantum states. The QNN takes as input a double-layered DA variational quantum circuit, and is shown to be hardware efficient and universally capable of quantum computing, as previously demonstrated by Parra et al. in Ref. \cite{Parra2020}. This new method represents a promising step towards understanding and characterizing the complex properties of many-body quantum systems.

There have been also interesting proposals at the theoretical level, one of them related to the field of variational quantum algorithms  \cite{Michel2023}. In their work,  A. Michel {\it et al.} propose to use DAQC to estimate the ground-state energy of several molecules using a DA variational quantum eigensolver (VQE) algorithm. The study focuses on implementing the VQE algorithm on a Rydberg quantum simulator for $\text{H}_2$, $\text{Li}\text{H}$ and $\text{Be}\text{H}_2$ molecules. The authors consider the platform's constraints, such as the restriction of the local action to initial state preparation and measurement and the implementation of global Hamiltonian time evolution. Despite this, they highlight that the DA interplay is attainable by using a toolbox, as demonstrated in a prior study \cite{Notarnicola2021}, emphasizing the potential for next-generation neutral atom devices.

Additionally, it has been proposed a superconducting circuit architecture for performing DAQC \cite{Albarran2022}, which makes us foresee a future where DAQC plays an important role in the development of the quantum computing field. In this study, the authors focus on the advancement of DAQC as a promising approach toward the development of next-generation quantum computers. The authors propose a superconducting circuit design for DAQC, consisting of a chain of charge qubits coupled through grounded superconducting quantum interference devices (SQUIDs). This design allows for the manipulation of individual interactions between nearest-neighbor (NN) qubits, producing different and independent interactions such as an exchange or double excitation/de-excitation term. By using this design, the authors aim to expand the DAQC algorithmic mappings and the variety of accessible quantum computer geometries and topologies. The flexibility of this design provides a large family of analog multibody Hamiltonians, suitable for efficient implementations of DAQC protocols. This work is a significant step towards the goal of reaching fault-tolerant quantum computers and computing complex systems on quantum platforms. It brings us closer to the goal of achieving a quantum advantage with fewer algorithmic and hardware resources, contributing to the ongoing efforts to advance the DAQC paradigm.

Related to the simulation of quantum many-body systems, there has been an interesting proposal by L. C. C\'eleri {\it et al.}. In \cite{Celeri2021}, the authors present a DA quantum algorithm that can simulate a wide range of fermionic Hamiltonians, including the Fermi-Hubbard model. Simulating quantum many-body systems is a challenging task that requires a significant amount of resources, which increase exponentially with the system's size. This becomes even more difficult with fermionic systems due to the nonlocal interactions caused by the fermionic wave function's antisymmetric nature. However, the authors demonstrate how DA methods enable quantum algorithms to run more efficiently than their digital counterparts by making better use of coherence time. Additionally, they provide a practical implementation of their techniques using a low-connected architecture for specific fermionic models.

After four years of intensive research, we have developed a comprehensive catalog of algorithms specifically designed for DAQC. Our work has laid a solid foundation for future researchers to expand upon and explore this exciting field further. We believe that the algorithms we have developed will be useful for those who wish to pursue research in DAQC, and we are optimistic that this thesis will serve as a valuable resource for future studies. We are confident to have contributed to the advancement of the field and look forward to seeing what further progress can be made.

\subsection{What you will find in this thesis}

\sloppy
In this thesis, we make an in-depth description of the DAQC paradigm and present several implementations of quantum algorithms, tailored for this computational paradigm. The performance of these proposals is compared with the one of the usual digital approach. We discuss its resilience against noise sources, especially those coming from the experimental constraints of NISQ devices. This thesis is structured into seven chapters, including this introduction and a concluding chapter, and it is devoted to the description of several ubiquitous quantum algorithms, using the DAQC approach. 

As we mention, in Chapter \ref{chapter2}  we present the fundamentals and main techniques for DAQC. Although these techniques may be extended to any resource, we propose to use propagators generated by the ubiquitous Ising Hamiltonian for the analog entangling blocks and provide proof of its universal character. We construct explicit DAQC protocols for efficient simulations of arbitrary inhomogeneous Ising, two-body, and M-body spin Hamiltonian dynamics by means of single-qubit gates and a fixed homogeneous Ising Hamiltonian. Additionally, we compare a sequential approach where the interactions are switched on and off (step-wise DAQC) with an always-on analog interaction interspersed by fast single-qubit pulses (banged DAQC). We also present a DAQC algorithm  to enhance the physical connectivity among qubits coupled by an arbitrary inhomogeneous NN Ising Hamiltonian and generate an arbitrary all-to-all (ATA) Ising Hamiltonian by employing single-qubit rotations. Finally, we introduce the theoretical description of some of the most relevant noise sources present in current superconducting quantum computers, in order to be able to compare the resilience under those noises of both the DQC and the DAQC methods in the following chapters.

In Chapter \ref{chapter3} we introduce an efficient DAQC algorithm to compute the quantum Fourier transform, a subroutine widely employed in several relevant quantum algorithms, such as the quantum phase estimation, which is also presented in this chapter. We show that, under reasonable assumptions about noise models, the fidelity of the quantum Fourier transformation improves considerably using this approach when the number of qubits involved grows. This suggests that, in the NISQ era, hybrid protocols combining digital and analog quantum computing could be a sensible approach to reaching useful quantum supremacy.

In Chapter \ref{chapter4} we use the quantum subroutines that we have described in the previous chapter to propose a DA implementation of the Harrow-Hassidim-Lloyd algorithm to solve linear systems of equations. Additionally, a co-designed architecture is proposed, which allows for a reduction of the number of SWAP gates required for its implementation. Merging a co-design quantum processor architecture with a DA implementation contributes to the reduction of noise sources during the experimental realization of the algorithm.

In Chapter \ref{chapter5} it is shown that the DAQC paradigm is suited to the QAOA due to the algorithm's variational resilience against the coherent errors introduced by the scheme. By performing large-scale simulations and providing analytic bounds for its performance in devices with finite single-qubit operation time, we observe regimes of single-qubit operation speed in which the considered variational algorithm provides a significant improvement over nonvariational counterparts in the DAQC scheme.

In Chapter \ref{chapter6}, we examine the feasibility of using superconducting circuit quantum processors, specifically transmon qubits, for DAQC by exploiting the cross-resonance effect. Our aim is to assess the potential of this approach as a solution for implementing DAQC in superconducting circuit quantum processors. To achieve it, we consider superconducting architectures and extend the cross-resonance effect, up to first order in perturbation theory, from a two-qubit interaction to an analog Hamiltonian acting on 1D chains and 2D square lattices which,  in an appropriate reference frame, results in a purely two-local Hamiltonian. By augmenting the analog Hamiltonian dynamics with single-qubit gates we show how one may generate a larger variety of distinct analog Hamiltonians. We then synthesize unitary sequences, in which we toggle between the various analog Hamiltonians as needed, simulating the dynamics of Ising, $XY$, and Heisenberg spin models. Our dynamics simulations are Trotter error-free for the Ising and $XY$ models in 1D. We also show that the Trotter errors for 2D $XY$ and 1D Heisenberg chains are reduced, with respect to a digital decomposition, by a constant factor. In order to realize these important near-term speedups, we discuss the practical considerations needed to accurately characterize and calibrate our analog Hamiltonians for use in quantum simulations. We conclude with a discussion of how the Hamiltonian toggling techniques could be extended to derive new analog Hamiltonians which may be of use in more complex DA quantum simulations for various models of interacting spins.

Finally, we dedicate a chapter to discuss the overall conclusions of this thesis, and we provide an appendix section with additional material that complements the discussions held during the text. A complete bibliography can be found at the end of this document.

The main focus of this thesis is the DAQC, which is why the most relevant section is the second chapter. There, the protocol bases are established and useful tools are provided so the following chapters are easy to follow. Chapter \ref{chapter2} closely follows sections of the dedicated work carried out by Parra {\it et al.} in 2020 \cite{Parra2020} since this constitutes a foundational piece of the DAQC.


\section[Digital--analog quantum computation]{Digital--analog \\ quantum computation}
\label{chapter2}

%
%

\thiswatermark{\put(1,-302){\color{l-grey}\rule{70pt}{60pt}}
\put(70,-302){\color{grey}\rule{297pt}{60pt}}}

%
%

\newtheorem*{definition}{Definition}

\vfill
\lettrine[lines=2, findent=3pt,nindent=0pt]{I}{n} this chapter, we present an in-depth description of the DAQC paradigm to simulate any arbitrary Hamiltonian. In the following chapters, these techniques will be extended to implement quantum algorithms. The DAQC paradigm requires the implementation of entangling multipartite evolutions and fast single-qubit gates (SQGs), which universality has already been proven \cite{Dodd_2002_UniversalQCandS, Masanes_2002, Bennett_2002, Jane_2003, Martin2020}. In this approach, we make use of a sequence of entangling time slices, called analog blocks, and fast single-qubit rotations (SQRs), which comprise our class of digital steps. In this chapter, we also introduce protocols to simulate arbitrary inhomogeneous two-body and $M$-body spin Hamiltonians. Then, we present the {\it bang DAQC} (bDAQC) approach, to mitigate the errors that derive from the {\it stepwise DAQC} (sDAQC) version. Afterwards, we show that it is possible to enhance the connectivity of quantum hardware on the software level with digital-analog control. Finally, we present a theoretical description and models of some of the most relevant noise sources present in current superconducting quantum computers, since they will be employed to compare the performance of the DQC and DAQC paradigms when implementing quantum algorithms.

As we mention in the introduction, we want to follow significantly the work carried out by Parra et al. in 2020 \cite{Parra2020}. In their work, the authors established formal definitions for the main pieces of the DAQC paradigms, as well as smart systematic algorithms to simulate relevant families of Hamiltonians. The rest of the chapter include part of the work that we developed in \cite{Martin2020} and \cite{GMS2021}.

\subsection{Analog, digital, and digital-analog quantum computing. Formal definitions.}\label{sec21}

The terms ``digital'' and ``analog'' are commonly utilized in the context of quantum computing, however, it is important to provide precise definitions for these concepts in a formal manner. In the following, we will present clear definitions for these terms as they relate to quantum gates, digital, and analog blocks in $N$-qubit systems. These definitions are extracted from Ref. \cite{Parra2020}.

\begin{definition}[Quantum gate]
A quantum gate is a fixed unitary evolution $U_n$,  $U_n\in\mathcal{B}((\mathbbm{C}^2)^{\otimes n})$. 
\end{definition}

\begin{definition}[Digital block]
A $k$-parametric continuous family of unitary operators $U_n(\vec{\phi})$, with $\phi_l\in \mathcal{I}_l(\mathbbm{R})$ and $1\leq l\leq k$,  comprises a {\it digital block} if it is equivalent to a fixed unitary evolution $U_n$ up to a set of local rotations $W_i(\vec{\phi})$, i.e.,  $U_n (\vec{\phi}) = \left (\bigotimes_{i}^n W_{i}(\vec{\phi})\right ) U_n$.
\end{definition}

It is important to note that digital blocks, which include both parameter-fixed entangling quantum gates and single qubit rotations with arbitrary angles, generate an entanglement that remains the same for all values of the parameters. On the other hand, analog blocks generate entanglement that varies with the values of the parameters.

\begin{definition}[Analog block]
We call {\it analog block} a $k$-parameter-dependent entangling unitary evolutions $V_n(\vec{\phi})$ with a semigroup structure $V_n(\vec{\phi})=V_n(\vec{\phi}_1)V_n(\vec{\phi}_2)$; $\vec{\phi}=\vec{\phi}_1+\vec{\phi}_2$. For $k=0$, it obviously reduces to a quantum gate.
\end{definition}

Under these definitions, for instance, $U_n=e^{i\frac{\pi}{4}\sigma_z^1 \sigma_z^2}$ is a quantum gate, both $U_n(\phi)=(e^{i\phi_1 \sigma_z^1}\otimes e^{i\phi_2 \sigma_z^2})e^{i\frac{\pi}{4}\sigma_z^i \sigma_z^j}$ and $W_i(\phi)=e^{i\phi \sigma_z^1}$ are digital blocks, while $V_n(\phi)=e^{i\phi \sigma_z^i \sigma_z^j}$ is an analog block.

When a quantum protocol only employs digital blocks, it is referred to as a digital quantum algorithm. However, if the protocol consists of a single analog block for different values of the parameters, it is referred to as an analog quantum algorithm. Additionally, a protocol that contains both digital and analog blocks is referred to as a digital-analog protocol.

In this paper, we have limited the digital blocks to arbitrary single qubit rotations, which allows us to write the total evolution of the system as $\prod_{j} \left (U_j(\vec{\phi}_j)\left [\bigotimes_i W_{i}^{(j)}(\vec{\alpha}_{ji})\right ]\right)$.

With these concepts formally defined, we can go deeper into the description of the DAQC paradigm.

\subsection{Digital-analog quantum computation} \label{sec22}
In this section, our focus will be on two quintessential models found in quantum computing systems, serving as examples to highlight the DAQC paradigm, though it can be easily extended to other scenarios. Our focus will be on utilizing either a homogeneous nearest-neighbor (NN) or a homogeneous all-to-all (ATA) two-body Ising Hamiltonian as analog blocks. By combining these evolutions with SQRs, we will demonstrate their universality, meaning that any unitary can be simulated with arbitrary accuracy using these resources. We show the protocols, sometimes optimal, to generate increasingly complex families of Hamiltonians.

First, we will detail how to generate arbitrary inhomogeneous two-body Ising Hamiltonians. The subsequent step will be to create an inhomogeneous two-body Hamiltonian, where the terms do not commute, requiring the application of Trotterization to the algorithm. Finally, we will show how to generate an arbitrary $M$-body Hamiltonian with a polynomial number of resources in terms of the number of spins, for a fixed value of $M$.

The foundation of the DAQC algorithms presented in this section lies in three key components:
\begin{enumerate}
    \item The ability to accurately evolve for a specific time $t$ with a Hamiltonian $H$, while incorporating local rotations to create an evolved Hamiltonian $H' = RHR^\dagger$. The impact of limited precision during the physical implementation of engineered times will also be considered.
    \item The utilization of a Trotter decomposition, $e^{-itH} = (e^{-iH_1t/n_T} e^{-iH_2t/n_T})^{n_T} + O(t^2/n_T)$, which is necessary for simulating Hamiltonians with non-commuting terms, such as $H=H_1 + H_2$, where $[H_1, H_2]\neq 0$. The error introduced through this approximation is of second order in the time error $\Delta t$. Advanced decompositions, like the symmetrized Trotter decomposition, are also considered and can reduce the error to $O(\Delta t)^3$.
    \item The ability to evolve with Hamiltonian $\lambda H$, where $\lambda > 0$ is a continuous positive parameter. This provides the flexibility to adjust the strength of the Hamiltonian evolution to meet the needs of the simulation.
\end{enumerate}

\subsubsection*{Ising model}

Let us illustrate the problem of implementing the inhomogeneous all-to-all (ATA) two-body Ising model, described by the Hamiltonian $H_{ZZ}=\sum_{j<k}^{N}g_{ij}\sigma_z^{(j)}\sigma_z^{(k)}$. The matrices $\sigma_{x, y, z}^{(n)}$ denote the application of the Pauli matrices $x$, $y$, or $z$ on the $n$-th qubit. The Hamiltonian describes the interaction between qubits through the coupling parameters $g_{ij}$. The goal is to simulate the unitary evolution $U_{ZZ}(t)=e^{i H_{ZZ} t}$ of this Hamiltonian using a combination of digital and analog blocks.

For this example, the analog block that will be used is the unitary evolution, $U_{zz}(t)$, of the homogeneous ATA two-body Ising model whose Hamiltonian reads
\begin{equation}\label{eq:Ham_sz_sz_base}
H_{zz}=g\sum_{j<k}^{N}\sigma_z^{(j)}\sigma_z^{(k)},
\end{equation}
where $g$ is a fixed coupling strength and the independent time parameter $t$ is used in the calculation of the unitary evolution. We set $\hbar=1$ in the whole chapter (and thesis).

The digital blocks used will be single qubit rotations around the $x$-axis, which can be performed with a continuous range of phases $\theta\in[0,2\pi]$ and are defined as
\begin{equation}
R_x\left(\theta\right) = e^{-i\theta\sigma_x/2}.
\end{equation}
 only digital blocks employed are single qubit rotations around the $x$-axis with the continuous range of phases $\theta\in[0,2\pi]$. The protocol can be easily modified to incorporate a fixed inhomogeneous evolution as an analog block, with Hamiltonian
\begin{equation}
    H_{ZZ}=\sum_{j<k}^{N}g_{jk}\sigma_z^{(j)}\sigma_z^{(k)}.
\end{equation}
Such Hamiltonians, which feature scaling of coupling parameters of the form $\bar{g}_{jk}\propto 1/|j-k|^{\alpha}$ with $0<\alpha<3$, commonly arise in various quantum platforms, such as ion-trap setups \cite{Porras2004}. In effective Hamiltonian models, where qubits are coupled through linear multi-mode systems \cite{LehmbergI_1970,LehmbergII_1970}, even more complex coupling distributions may arise and can be tailored or designed \cite{GonzalezTudela_2015,Solgun_2017}. A different decomposition into most entangling Hamiltonians and single qubit gates has been shown to be a universal quantum machine by Dodd et al. in Ref. \cite{Dodd_2002_UniversalQCandS}. Under the digital-analog paradigm, the Ising Hamiltonian serves as an example of a universal Hamiltonian, as it enables the construction of a universal ZZ gate \cite{vandenNest_2008} between any two arbitrary qubits. This is detailed in Appendix \ref{Appsec:Universality}.

\begin{figure}[t]
\centering
	\includegraphics[width=.85\linewidth]{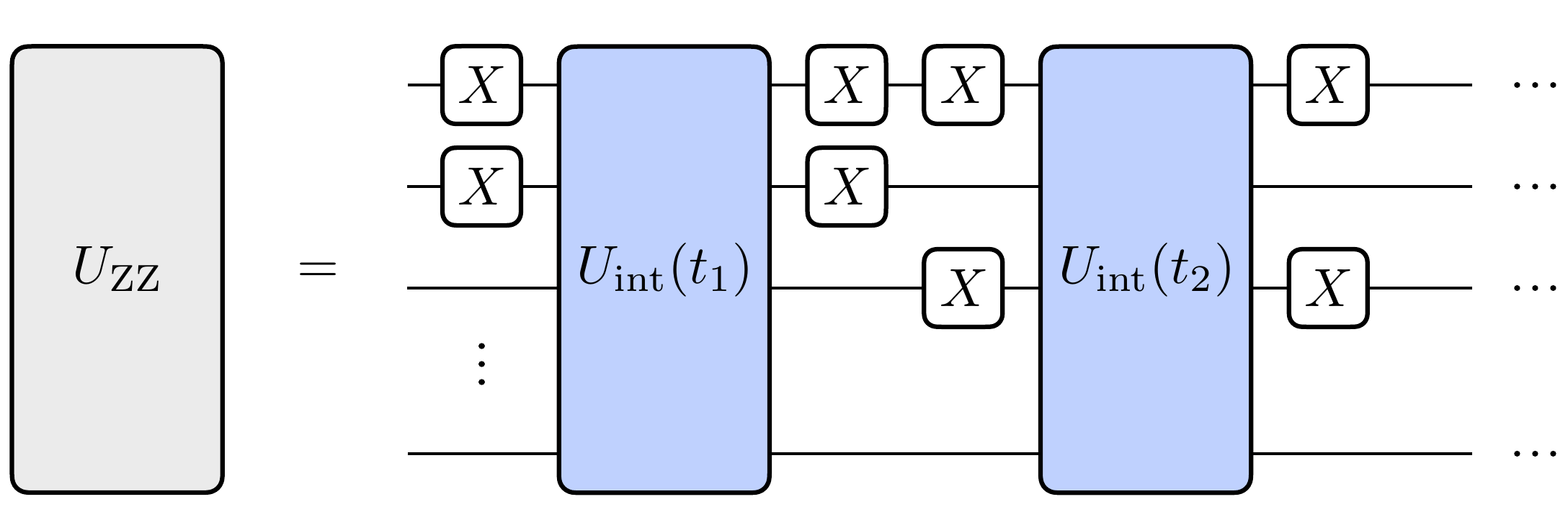}
	\caption{\label{Fig1chap2} \textbf{Algorithm to simulate the general inhomogeneous Ising model from a fixed one}. Each analog block consisting on a time step evolution of time $t_\alpha$ is sandwiched by a pair of single qubit gates, $X$, which refers to $\pi$ rotation arround $x$-axis, $R_x(\pi)$) applied to qubits $(n,m)$, with $\alpha(n,m)$. It is possible to use optimal sequences of SQRs to simplify the number of pulses.}
\end{figure}

The target Hamiltonian evolves according to the unitary 
\begin{equation}
    U_{ZZ}=e^{i t_F H_{ZZ}}\text{, where }H_{ZZ}=\sum_{j<k}g_{jk}\sigma_z^j\sigma_z^k, 
\end{equation} 
where $t_F$ is the final time. The objective is to determine a suitable mapping between the expressions $g_{jk}t_F$ and $g t_{nm}$ through the decomposition of the homogeneous time evolution into a maximum of $N(N-1)/2$ analog blocks with varying durations $t_{nm}$, which are bracketed by local rotations $\sigma_x^n\sigma_x^m$,
\begin{eqnarray}
H_{ZZ}&=& \sum_{j<k}^N g_{jk}\sigma_z^j\sigma_z^k=\frac{g}{t_F}\sum_{j<k}^N \sum_{n<m}^N t_{nm}\sigma_x^{(n)} \sigma_x^{(m)} \sigma_z^{(j)} \sigma_z^{(k)} \sigma_x^{(n)} \sigma_x^{(m)}\label{eq:H_I_ZZ}\nonumber\\
&=& \frac{g}{t_F} \sum_{j<k}^N \sum_{n<m}^N t_{nm}(-1)^{\delta_{nj}+\delta_{nk}+\delta_{mj}+\delta_{mk}}\sigma_z^j \sigma_z^k,
\end{eqnarray}
as depicted in Fig. \ref{Fig1chap2}. As $\sigma_x^{(n)}\sigma_z^{(k)}\sigma_x^{(n)}$ is equal to $-\sigma_z^{(n)}$ if $n=k$, and it is equal to $\sigma_z^{(k)}$ otherwise, then,
\begin{equation}
H_{ZZ}=\frac{g}{t_F}\sum_{j<k}^N\sum_{n<m}^N t_{nm}(-1)^{\delta_{nj}+\delta_{nk}+\delta_{mj}+\delta_{mk}}\sigma_z^{(j)} \sigma_z^{(k)}.
\end{equation}

Thus, the problem of finding the value of each time $t_{nm}$ is a matrix-inversion problem,
\begin{equation}
g_\beta=t_\alpha M_{\alpha \beta}\frac{g}{t_F}\quad\rightarrow\quad t_\alpha=M_{\alpha\beta}^{-1}g_\beta \frac{t_F}{g},
\end{equation}
where $\alpha$ and $\beta$ are introduced to vectorize each pair of indices $(n,m)$ and $(j,k)$ as
\begin{equation}
\alpha=N(n-1)-\frac{n(n+1)}{2}+m,\qquad\beta= N(j-1)-\frac{j(j+1)}{2}+k ,
\end{equation}
and $M$ is a sign matrix built up by the elements,
\begin{equation}
M_{\alpha\beta}=(-1)^{\delta_{nj}+\delta_{nk}+\delta_{mj}+\delta_{mk}},\label{eq2.9}
\end{equation}
where the inverted relations for the indices $n$ and $j$ are 
\begin{eqnarray}
    j&=&1+\left[\frac{\beta}{N}\right],\\
    k&=&\beta+\frac{1}{2}\left(1+\left[\frac{\beta}{N}\right]\right)\left(2+\left[\frac{\beta}{N}\right]\right)-N\left[\frac{\beta}{N}\right],\\
    n&=&1+\left[\frac{\alpha}{N}\right],\\
    m&=&\alpha+\frac{1}{2}\left(1+\left[\frac{\alpha}{N}\right]\right)\left(2+\left[\frac{\alpha}{N}\right]\right)-N\left[\frac{\alpha}{N}\right].
\end{eqnarray}

The matrix $\mathsf{M}_{\alpha\beta}$ is a non-singular matrix $\forall N\in \mathbb{Z}-\{4\}$ since it has three degenerate eigenvalues, namely, $\lambda_1=N(N-9)/2+8$, $\lambda_2=2(4-N)$ and $\lambda_3=4$ with degeneracies $1$, $N-1$, and $N(N-1)/2-N$, respectively. The corner case $N=4$ requires the use of a slightly modified set of SQRs, e.g. single $\sigma_x$-rotations per site is sufficient for a NN Hamiltonian. The total unitary evolution is $U_{ZZ}(t_F) = e^{i t_F \sum_{\beta}H_{ZZ}^\beta }$, where 
\begin{equation}
H_{ZZ}^\beta = g_\beta \sigma_z^{j(\beta)}\sigma_z^{k(\beta)},\label{eq:H_beta}
\end{equation}
and $g_\beta= t_\alpha \mathsf{M}_{\alpha\beta}(g/t_F)$. 

By solving the linear problem, we determine the durations $t_\alpha = \mathsf{M}{\alpha\beta}^{-1}g\beta(t_F/g)$ for each analog block, interspersed with pairs of SQRs, as depicted in Figure \ref{Fig1chap2}. The uniqueness of the solution is guaranteed by the Rouché-Frobenius theorem, as the matrix $\mathsf{M}_{\alpha\beta}$ is invertible. It is noteworthy that some of the $t_\alpha$ may be negative, indicating the need for inverted coupling signs during the corresponding analog evolutions. However, this issue can be easily resolved by evolving with modified durations $\tilde{t}_\alpha=t_\alpha+|t_{\mathrm{min}}|$. In Appendix \ref{AppSec:Negative_times} we provide further information.

In conclusion, this section presents an optimized DAQC protocol for the construction of an arbitrary inhomogeneous ATA two-body Ising Hamiltonian utilizing a homogeneous ATA two-body Ising Hamiltonian and $x$-rotations as resources. The protocol, which is quadratic in the total number of qubits, is optimal for a generic Hamiltonian as it utilizes the same number of resources as degrees of freedom are introduced in the Hamiltonian. In the case of NN Hamiltonian, the protocol is even simpler and requires only $N-1$ SQRs, with one SQR per site. These protocols demonstrate the universality of the Ising model, as it can be utilized to generate a ZZ gate between any two arbitrary qubits.

\subsubsection*{$XZ$ model}
\begin{figure}[t]
\centering
	\includegraphics[width=0.98\linewidth]{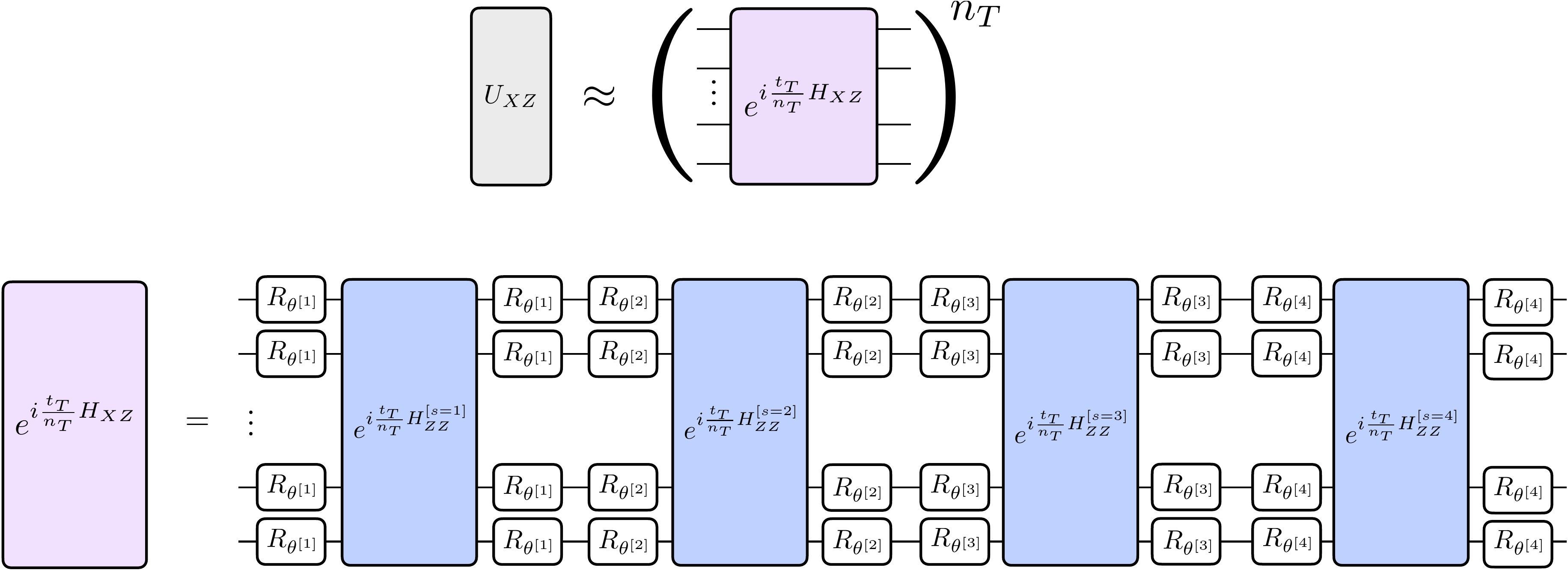}
	\caption{\label{Fig2chap2} \textbf{Algorithm to simulate the inhomogeneous XZ model from an Ising model}. The combination of four different general Ising evolutions and rotations handles the minimum degrees of freedom for each term $g_{jk}^{\mu\nu}$. $R_{\theta^{}}$ rotations can be combined with inner $\sigma_x^n \sigma_x^m$ rotations  required to implement the inhomogeneous ZZ Ising model into unique blocks of SQRs.}
\end{figure}

As the Ising model is a universal Hamiltonian within the DAQC paradigm, it can be used to simulate the evolution of any other Hamiltonian. To accomplish this in a systematic manner, we will utilize a Trotter decomposition in conjunction with the previous algorithm. As an example, we will demonstrate the construction of an inhomogeneous general two-body $XZ$ Hamiltonian, as depicted in Figure \ref{Fig2chap2}.

We want to simulate the unitary evolution
\begin{equation}
    U_{XZ} = e^{it_FH_{XZ}}, \quad \text{with } H_{XZ}= \sum_{j<k}^N \sum_{\mu,\nu=\{x,z\}}g_{jk}^{\mu\nu}\sigma_\mu^{(j)} \sigma_\nu^{(k)}.
\end{equation}
To do so, we start performing a Trotter decomposition $U_{XZ} \approx (e^{i \frac{t_F}{n_T} H_{XZ}})^{n_T}$, with $n_T$ the number of Trotter steps,
\begin{eqnarray}
H_{XZ}&=&  \sum_{j<k}^N \sum_{\mu,\nu}g_{jk}^{\mu\nu}\sigma_\mu^{(j)} \sigma_\nu^{(k)}= \sum_{j<k}^N \sum_{\mu,\nu,s}g_{jk}^{[s]}\alpha_j^{(\mu,s)}\alpha_k^{(\nu,s)}\sigma_\mu^{(j)} \sigma_\nu^{(k)}\label{eq:H_I_XZ}\nonumber\\
&=&\sum_{j<k}^N \sum_{s=1}^4 g_{jk}^{[s]}\left(\cos(\theta_j^{[s]})\sigma_z^{(j)}+\sin(\theta_j^{[s]})\sigma_x^{(j)}\right)\times\nonumber\\
&&\left(\cos(\theta_k^{[s]})\sigma_z^{(k)}+\sin(\theta_k^{[s]})\sigma_x^{(k)}\right),
\end{eqnarray}
with $s$ running from $\{1,\dots, 4\}$, which are the number of combinations of the two types of couplings (for instance, for the XYZ Hamiltonian, it would run from $1$ to $9$). The parameters of the above formula are defined as 
\begin{equation}
\alpha_j^{(x,s)}=\sin(\theta_j^{[s]}) \qquad \text{and }\qquad \alpha_j^{(z,s)}=\cos(\theta_j^{[s]}).
\end{equation}
The pairs of operators are decomposed with their coefficients in homogeneous $\sigma_z^{(j)}\sigma_z^{(k)}$ operators with local rotations
\begin{equation}
R_{\theta_j^{[s]}}=\left(\cos(\theta_j^{[s]}/2)\sigma_z^{(j)}+\sin(\theta_j^{[s]}/2)\sigma_x^{(j)}\right),
\end{equation}
thus
\begin{equation}
R_{\theta_j^{[s]}} \sigma_z^{(j)} R_{\theta_j^{[s]}}=\left(\cos(\theta_j^{[s]})\sigma_z^{(j)}+\sin(\theta_j^{[s]})\sigma_x^{(j)}\right),
\end{equation}
for all pairs of qubits. This rotation is produced by the Hamiltonian 
\begin{equation}
H_{\theta_j^{[s]}}=(\pi/2)\left(-\mathbbm{1}+\cos(\theta_j^{[s]})\sigma_z^{(j)}+\sin(\theta_j^{[s]})\sigma_x^{(j)}\right).
\end{equation}
The SQRs have to be performed in all qubits $R_{\theta^{[s]}}=\otimes_w^N R_{\theta_w^{[s]}}$. The total Hamiltonian is reconstructed as 
\begin{equation}
H_{XZ}=\sum_{s=1}^4 R_{\theta^{[s]}}\left(g_{jk}^{[s]}\sigma_z^{(j)} \sigma_z^{(k)}\right) R_{\theta^{[s]}}^\dagger.
\end{equation}
This equality is only valid in the Hamiltonian and not in the total unitary evolution. This means that, in each Trotter time step $t_T=\left(t_F/n_T\right)$, we must evolve according to $H_{ZZ}^{[s]}=g_{jk}^{[s]}\sigma_z^{(j)} \sigma_z^{(k)}$ between a pair of sets of SQRs $R_{\theta^{[s]}}$
\begin{eqnarray}
U_{XZ}\approx\left(\prod_{s=1}^4 R_{\theta^{[s]}}e^{i\frac{t_T}{n_T}(H_{ZZ})^{[s]}} R_{\theta^{[s]}}\right)^{n_T}.
\end{eqnarray}
The objective then, is to find a set of phases $\theta_w^{[s]}$ that satisfy the equation $g_{jk}^{\mu\nu}=g_{jk}^{[s]}\alpha_j^{(\mu,s)}\alpha_k^{(\nu,s)}$. A suitable set of phases is given by $\theta_w^{[s]}=\frac{s\pi w}{2(w+1)}$, with a polynomially-scaled distance between nearest-neighbor qubit phases of $d(s,w)=\frac{s\pi}{2(w^2+3w+2)}$. Further details can be found in Appendix \ref{Appsec:XZmodel_couplings}.

Thus, we have presented a protocol for constructing an arbitrary two-body Hamiltonian using the freedom in the phases of SQRs and Trotterization. This protocol requires $4N(N-1)$ ($4(N-1)$) analog blocks per Trotter step for the ATA (NN) XZ model, making it optimal in terms of the number of free coefficients. Additionally, it can be shown that extending the protocol to the general two-body $XYZ$ model would require even more angles per rotation, summing up to $9N(N-1)$ ($9(N-1)$) blocks per Trotter step for the ATA (NN) model.

\subsubsection*{$M$-body Hamiltonian}

\begin{figure}[t]
\centering
	\includegraphics[width=0.8\linewidth]{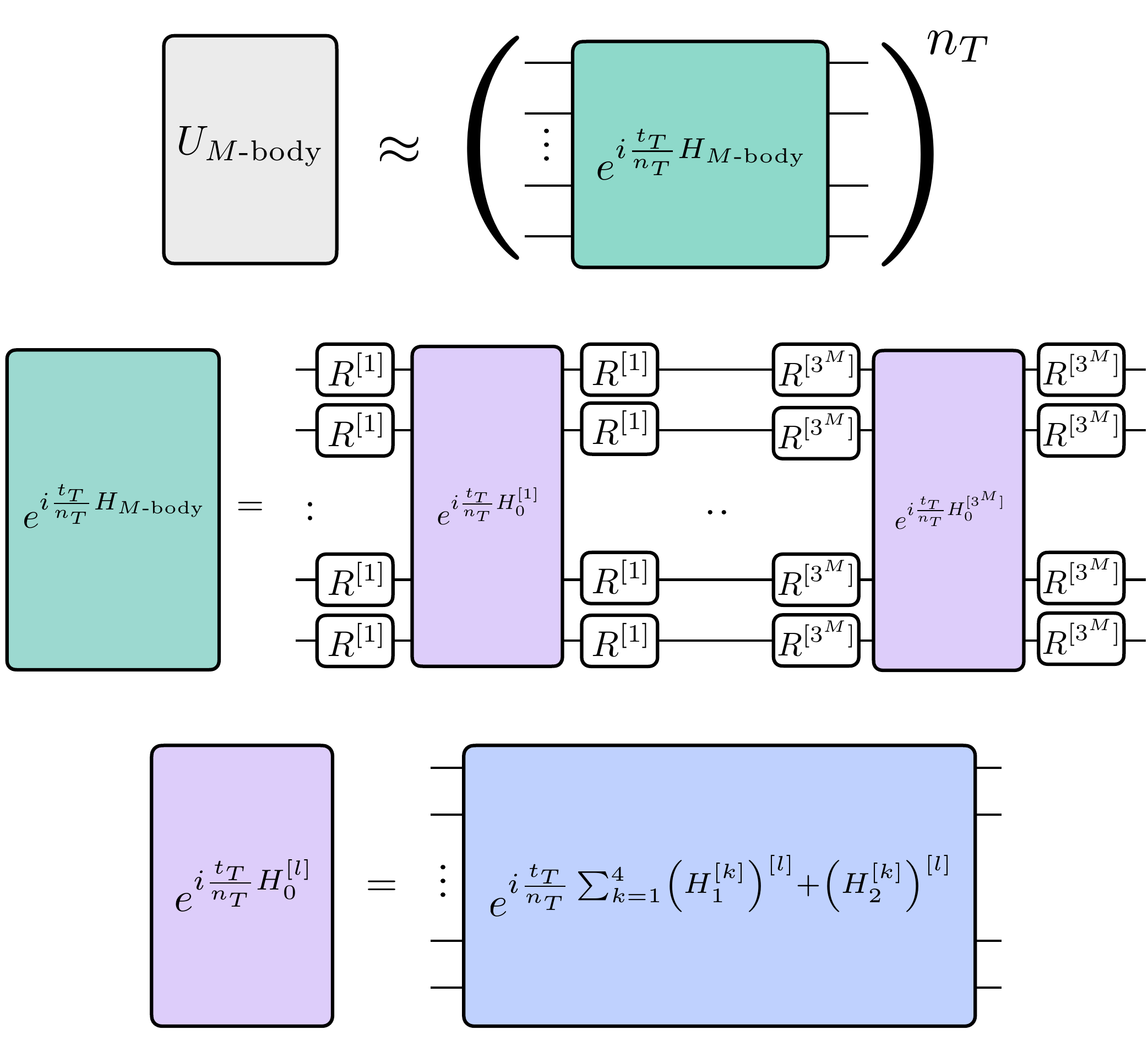}
	\caption{\label{Fig3chap2} \textbf{Algorithm to implement the $M$-body evolution.} The time-evolution of $H_0^{[l]}=\sum_{k=1}^4 (H_1^{[k]})^{[l]}+(H_2^{[k]})^{[l]}$ is sandwiched by the generalized rotations $R^{[l]}=\otimes_j^N r_j^{[l]}\sigma_x^{(j)}+s_j^{[l]}\sigma_y^{(j)}+t_j^{[l]}\sigma_z^{(j)}$. The Hamiltonians $H_1^{[k]}$ and $H_2^{[k]}$ are defined in Eqs. (\ref{eq:Mbody_H1k}) and (\ref{eq:Mbody_H2k}). For an example with $M=4$, we refer the reader to Appendix \ref{Appsec:Mbody}.}
\end{figure}

By utilizing similar methods, one can systematically build the evolution of any completely general Hamiltonian with interactions of up to $M$ bodies. The procedure can be visualized in the diagram shown in Fig. \ref{Fig3chap2}. For the sake of clarity, we present a clear outline of the steps necessary to simulate an arbitrary 4-body nearest-neighbor Hamiltonian and provide a detailed explanation on how to extend the protocol to Hamiltonians with up to $M$ bodies in the appendix (see Appendix \ref{Appsec:Mbody} for further details). The $4$-body Hamiltonan would be defined as
\begin{eqnarray}
	H_{4b}&=&\sum_{j \chi\eta} g_{(2,j)}^{\chi\eta}\sigma_{\chi}^{(j)}\sigma_{\eta}^{(j+1)}+\sum_{j\chi\eta\gamma} g_{(3,j)}^{\chi\eta\gamma}\sigma_{\chi}^{(j})\sigma_{\eta}^{(j+1)}\sigma_{\gamma}^{(j+2)}\nonumber\\
	&&+\sum_{j \chi\eta\gamma\rho}g_{(4,j)}^{\chi\eta\gamma\rho}\sigma_{\chi}^{(j)}\sigma_{\eta}^{(j+1)}\sigma_{\gamma}^{(j+2)}\sigma_{\rho}^{(j+3)},
\end{eqnarray}
where the indices run over $\{\chi,\eta,\gamma,\rho\}\in\{x,y,z\}$ and $j \in\{1,...,N\}$. First, an inhomogeneous ZZ-Ising Hamiltonian is sandwitched between rotated XX-Ising evolutions and their conjugate transposes, 
\begin{eqnarray}
	H_1^{[k]}=e^{-iO_{XX}^{1[k]}}H_{ZZ}^{1[k]}e^{iO_{XX}^{1[k]}},\label{eq:Mbody_H1k}\\
	H_2^{[k]}=e^{-iO_{XX}^{2[k]}}H_{ZZ}^{2[k]}e^{iO_{XX}^{2[k]}},\label{eq:Mbody_H2k}
\end{eqnarray}
with 
\begin{eqnarray}
O_{XX}^{1[k]} &=& \Phi^{[k]}_1\sigma_x^{(1)}\sigma_x^{(2)}+\Phi^{[k]}_3\sigma_x^{(3)}\sigma_x^{(4)}+\Phi^{[k]}_5\sigma_x^{(5)}\sigma_x^{(6)}+...\\
O_{XX}^{2[k]} &=& \Phi^{[k]}_2\sigma_x^{(2)}\sigma_x^{(3)}+\Phi^{[k]}_4\sigma_x^{(4)}\sigma_x^{(5)}+\Phi^{[k]}_6\sigma_x^{(6)}\sigma_x^{(7)}+...
\end{eqnarray}
built from evolutions of ZZ-Ising models rotated with the single qubit gates $R$ in all qubits, defined as
\begin{equation}
R=\otimes_j\left(\cos(\pi/4)\sigma_z^j+\sin(\pi/4)\sigma_x^j\right).
\end{equation}
It is important to note that the operators $O_{XX}^{1[k]}$ and $O_{XX}^{2[k]}$ contain interactions that are separated by an interaction length $L=M/2=2$. For example, $O_{XX}^{1[k]}$ includes a term $\sigma_x^{(1)}\sigma_x^{(2)}$, but not $\sigma_x^{(2)}\sigma_x^{(3)}$. To simulate a three-body (five-body) Hamiltonian, a different decomposition with 3 (5) translationally invariant sets of blocks would be required.

In other words, $H_1^{[k]}$ and $H_2^{[k]}$ together form a complete representation of the arbitrary Hamiltonian with up to four-body interactions. To further extend this idea to simulate Hamiltonians with $M$-body interactions, it is necessary to decompose the Hamiltonian into $M/2$ translationally invariant sets of blocks, similar to the procedure described for the 4-body Hamiltonian. The detailed explanation on how to achieve this generalization can be found in the appendix section \ref{Appsec:Mbody}.

The coefficients of the $h_{ij...}$ operators are interrelated. To simulate an arbitrary four-body nearest-neighbor Hamiltonian, it is sufficient to sum four copies each of the blocks $H_0=\sum_{k=1}^4 H_1^{[k]} + H_2^{[k]}$, in order to separate the parameters and generate at least one term that operates on each support. To obtain all XYZ operators, it is necessary to concatenate $3^M=81$ $H_0$ blocks, interleaved by SQRs described by 
\begin{equation}
R^{[l]} = \otimes_j^N (r_j^{[l]} \sigma_x^{(j)} + s_j^{[l]} \sigma_y^{(j)} + t_j^{[l]} \sigma_z^{(j)}).
\end{equation}
In other words, 
\begin{equation}
	H_{4{b}}=\sum_l R^{[l]} H_0^{[l]} R^{[l]},
\end{equation}
with $(r_j^{[l]},s_j^{[l]},t_j^{[l]})$ fulfilling the constraint $|r_j^{[l]}|^2+|s_j^{[l]}|^2+|t_j^{[l]}|^2=1$. It is important to note that the above construction is applied to a single Trotter step, meaning that it must be repeated $n_T$ times to approximate the evolution of the Hamiltonian, which can be expressed as $U_{4{b}}=e^{-iH_{4{b}}t}\approx (e^{-iH_{4{b}}t/n_T})^{n_T}$. The total number of analog blocks, or engineered time slices, required for the simulation of an arbitrary Hamiltonian with up to $M$-body interactions and $N$ qubits is given by $a(M)N+b(M)$ with, 
\begin{eqnarray}
	a(M)&=\frac{9}{4}\left(3^{M-1}-3\right),\\
	b(M)&=\frac{3^{M-1}}{2}\left(\frac{3}{2}-M\right).
\end{eqnarray}
Then, it can be stated that the computational cost of simulating NN Hamiltonians grows linearly with the number of qubits and exponentially with the number of bodies. For a specific example of a four-body system, the total number of required analog blocks is $117N-306$. For a more in-depth understanding of the simulation algorithm for general $M$-body interactions, readers can refer to the appendix section (Appendix \ref{Appsec:Mbody}).

\vfill
\subsubsection{Enhancing the connectivity of quantum hardware using DAQC}

\begin{figure}
  \centering \includegraphics[width=0.9\textwidth]{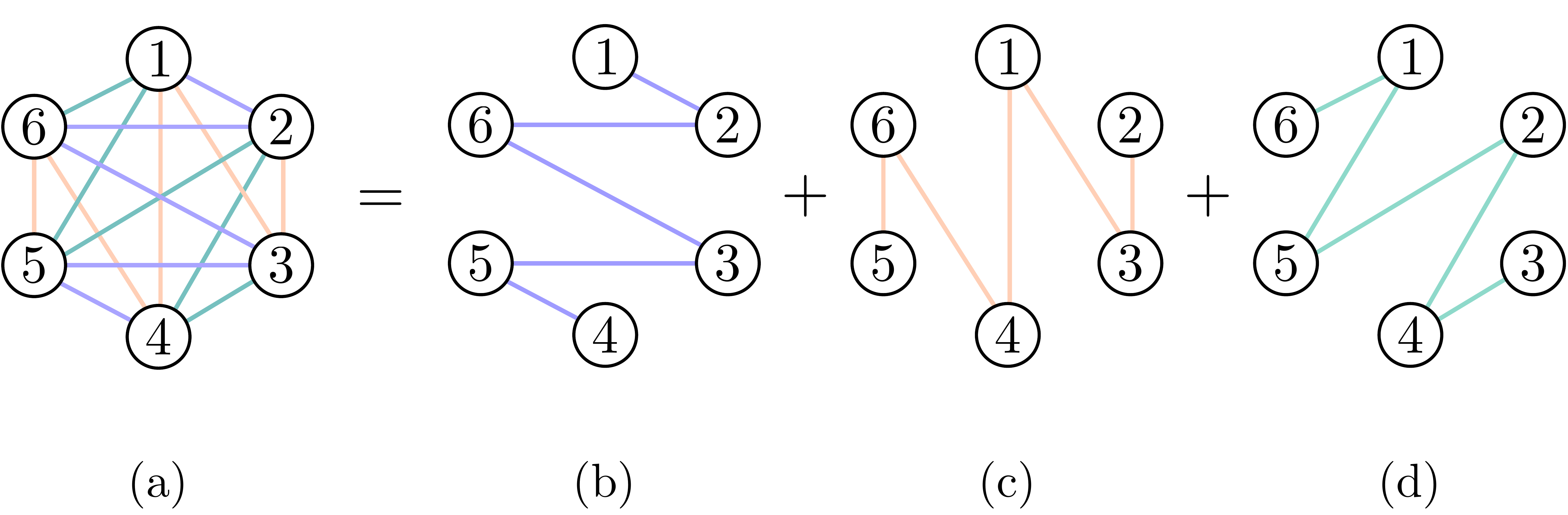}
  \caption{{\bf Decomposition of a $6$-qubit system with ATA connectivity into $3$ NN $6$-qubit system. (a) }Complete graph that one wants to simulate. {\bf (b - d)} Hamiltonian paths that are used as the resource. Figure (b) is built by starting in the first node, going forward to the next node, then 2 nodes backward, 3 forward... The other two Hamiltonian paths are obtained rotating the first one.}
  \label{Fig4chap2}
\end{figure}

Throughout the entire section, we are assuming that the quantum hardware that we are employing presents an ATA connectivity. But it does not necessarily has to be like this, in fact, a realistic quantum processor is expected to present lower connectivity. This is given by the fact that ATA connections are extremely demanding in the amount of wiring among qubits.

In Ref. \cite{galicia2019enhanced}, the authors design an algorithm that simulated an arbitrary ATA Ising Hamiltonian employing as a resource a given inhomogeneous NN Ising model and SQRs, in an optimal manner. They do so using $\mathcal{O}(5N^2)$ analog block, where $N$ stands for the number of qubits of the system. The algorithm is based in two main ideas:
\begin{itemize}
\item[1.] The Ising Hamiltonian for $N$ qubits can be represented as a weighted graph with $N$ vertices, where the weight of the edge connecting vertex $i$ to vertex $j$ is the coupling constant between qubits $i$ and $j$, denoted as $g_{ij}$. If two qubits are not connected, the vertices $i$ and $j$ representing them are also not connected, resulting in a null coupling constant ($g_{ij}=0$).

\item[2.] In this representation, an ATA Ising Hamiltonian for $N$ qubits can be expressed as a complete graph $K_N$, with edges between every possible vertex without repetition. On the other hand, the NN Ising Hamiltonian is represented as a Hamiltonian path that visits all possible vertices only once.
\end{itemize}
Figure \ref{Fig4chap2} shows an example of an ATA $6$-qubit system represented as a complete $K_6$ graph and its path decomposition into $3$ NN $6$-qubit systems.

The algorithm first splits the complete graph, representing the ATA Hamiltonian, into a set of Hamiltonian paths that can be efficiently decomposed into NN resources. Next, the algorithm obtains each of the Hamiltonian path connections using a SWAP-like gate $U$, which changes the connections between the qubits of the system. The SWAP gate is defined as
\begin{equation}
U_\text{SWAP} = e^{i\frac{\pi}{4}( \sigma_x^{i}\sigma_x^{j} + \sigma_y^{i}\sigma_y^{j} + \sigma_z^{i}\sigma_z^{j}) },
\end{equation}
where the superscripts refer to the qubit on which the gate acts.

Partitioning a complete graph into a set of paths, allow to simulate an ATA Hamiltonian by using only NN-connected resources. This result extends the DAQC paradigm to a wide range of quantum platforms that  require low connectivity. For a more detailed explanation of the algorithm, we refer the reader to the main article.  

\subsection{Stepwise and banged DAQC}\label{sec:sDAQC_and_bDAQC_SC}

The method we have described to simulate different families of Hamiltonians using the DAQC paradigm is the sDAQC. Under ideal circumstances, i. e. without taking into account noise sources, would lead to the same result as the DQC method. But this noiseless situation is just an illusion. The reality is that turning on and off multiqubit quantum gates introduce an important source of errors in realistic quantum algorithms. In the interest of reducing its effect, we introduce the concept of bDAQC as a different way to perform quantum algorithms.

The term ``bang bang'' is a well-established concept in classical control theory, as seen in references \cite{ControlTheory_Bangbang, ControlTheory_Bangbang2}. It was later introduced into the field of quantum physics to serve as a means of dynamically decoupling controlled quantum systems from their environment \cite{Viola_1998_DynamicalDec, Morton_2006_BangBang}. In this work, we use the term ``banged'' to describe the situation in which the analog evolution is not interrupted while fast SQRs pulses are being applied. Compared to sDAQC, the bDAQC introduces an additional digital error, which could be of the third order in the duration of the pulse, at best. However, we argue that this error is in competition with the experimental error produced by the switch of multi-qubit gates and may, in certain cases, prove to be less harmful to the overall algorithm.

\subsubsection*{Stepwise DAQC}

\begin{figure}[t]
\centering
	\includegraphics[width=0.98\linewidth]{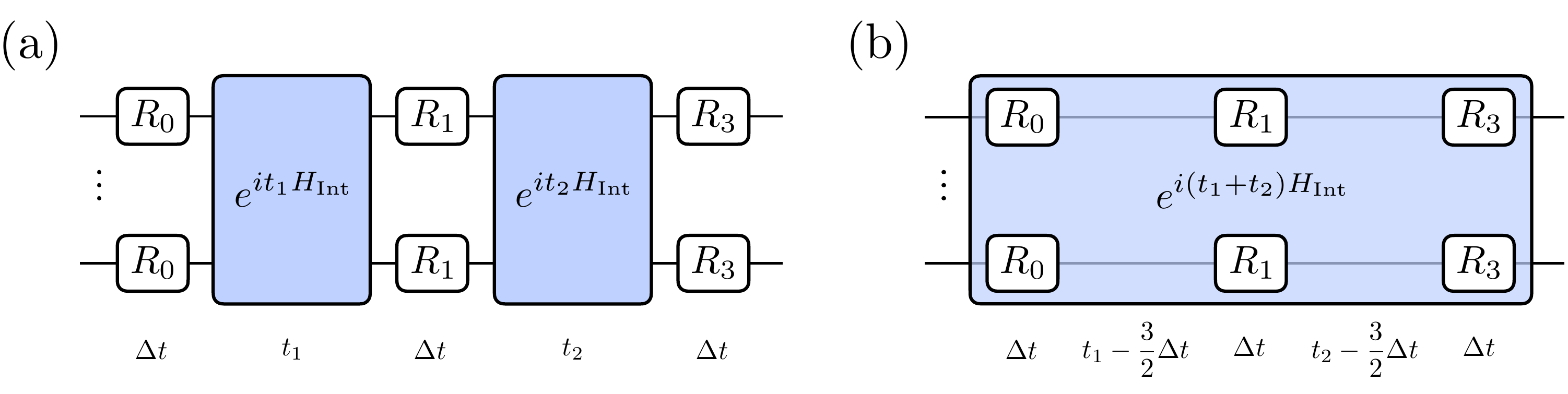}
	\caption{\label{Fig5chap2} \textbf{Schematic representation of the sDAQC and the bDAQC protocol. (a)} sDAQC: digital evolutions with Hamiltonian $H_{R_k}$, and analog blocks ones evolving with Hamiltonians $H_I$ are well separated in time under the sudden approximation. {\bf (b)} bDAQC: the adiabatic evolution $H_I$ is on during the whole time and fast rotations are added ($H_{R_k}+H_I$). The sudden approximation is employed.}
\end{figure}

In a sDAQC scenario, a total unitary evolution is constructed by alternating an evolution under a fixed entangling Hamiltonian $H_\text{Int}$ with sequences of SQRs
\begin{equation}\label{eq:sDAQC_UT}
    U_T = \dots U_{R_2} e^{-iH_\text{Int}t_2} U_{R_1} e^{-iH_\text{Int}t_1} U_{R_0}, 
    \end{equation}
with $U_{R_n}$ a general digital rotation operator acting on any subset of the Hilbert space --in this case, two qubits. In Fig. \ref{Fig4chap2}(a), we show a schematic representation of the procedure. According to the sudden approximation, we can implement such evolution $U(t)\approx e^{-it_FH_(t)}$ with a time-dependent Hamiltonian,
\begin{equation}
    H(t)= \sum_{n=0}H_{R_n} \Pi_{\Delta t}(t-[T_n+n\Delta t]) + H_I \Pi_{T_{n+1}}(t-[T_{n}+(n+1)\Delta t]),\label{eq:H_time_PDAQC}
\end{equation}
where $H_{R_n}$ gives rise to the digital unitary evolution $U_{R_n}=e^{-iH_{R_n}\Delta t}$. We have defined $T_n=\sum_{r=0}^{n}t_{r}$, assuming $t_0=0$, and the rectangular window function 
\begin{equation}
\Pi_{T}(t-t_s)= \theta(t-t_s)-\theta(t-(t_s+T)),
\end{equation}
with $\theta(t)=1$ for $t\ge0$ and 0 otherwise. 

\subsubsection*{Banged DAQC}
The bDAQC approach involves continuously evolving the system with the entangling Hamiltonian $H_I$, while applying short, intense pulses to implement the SQRs $H_{R_n}$. In contrast, the stepwise protocol involves switching the Hamiltonians $H_I$ and $H_{R_n}$ on and off, an unrealistic assumption due to the finite speed of operation in physical systems.

By utilizing the symmetrized exponential decomposition for all the SQRs blocks, the ideal evolution of the system described in Eq. (\ref{eq:sDAQC_UT}) can be accomplished without having to turn off the entangling Hamiltonian
\begin{eqnarray}
H(t)&= H_I+\sum_n^{N-1} H_{R_n}\Pi_{\Delta t}(t-[T_n-\Delta t/2])\nonumber\\
&+ H_{R_N}\Pi_{\Delta t}(t-[T_N-\Delta t]),\label{eq:H_time_bDAQC}
\end{eqnarray}
where $T_n=\sum_{r=0}^{n}t_{r}$. In Fig. \ref{Fig4chap2}(b), we show an schematic representation of the bDAQC approach.

\subsubsection{Error estimation}
The error-per-step introduced can be estimated using the Schatten norm by calculating the difference between the evolution of the system using the sDAQC protocol and the bDAQC. The error would then be defined as
\begin{eqnarray}
e_n &=& ||1-e^{-i H_{I} \Delta t/2}e^{-i H_{R_n} \Delta t}e^{-i H_{I} \Delta t/2}e^{i (H_I+H_{R_n}) \Delta t} ||\nonumber\\
&=&\frac{(\Delta t)^3}{4}||[[H_I,H_{R_n}],H_I+2H_{R_n}]||+O((\Delta t)^4),
\end{eqnarray}
where we have made use of the Zassenhaus formula. The first and last blocks at the boundaries result in second-order errors, $e_{0,N}$, which are proportional to $(\Delta t)^2$ due to the lack of symmetry in their evolution. The overall error in the banged protocol is calculated as the sum of these errors, $E=\sum_n e_n = A e_n$, where the proportionality constant $A$ is proportional to either $O(N)$ or $O(N^2)$, depending on whether the coupling configuration is NN or ATA, respectively. This is because the total number of analog blocks increases either linearly or quadratically with the number of qubits, $N$.

Naturally, the bDAQC does not generate the same result as the sDAQC or the DQC method, since there is an intrinsic error on the bDAQC which does not depend on either the experimental conditions or the noise sources. This error is due to the superposition between the Hamiltonians of the SQRs and the analog Hamiltonian. However, if SQRs are performed in a time $\Delta t$ much smaller than the intrinsic time scale of the analog block, the error will be smaller than the one coming from switching on and off the analog Hamiltonian. Indeed, we have just shown that the additional error per SQR introduced by not turning off the evolution of the Hamiltonian is on the order $\mathcal{O}[(\Delta t)^3]$. The reason why we aim at using the bDAQC protocol despite its intrinsic error is that, it accumulates less experimental error. Experimentally, switching on and off the Hamiltonian is not an exact step function, it takes some time to stabilize. Quantum control tries to suppress these errors, but it turns cumbersome when the system scales up and cannot be solved in a classical computer. If we keep the analog block on during the evolution, we will avoid these errors. This will be of great importance when we explore a more realistic implementation of the DAQC protocol in the following chapters.

\vspace{1cm}

In this section, we presented systematic techniques to simulate different families of Hamiltonians using the DAQC paradigm. At some point, these Hamiltonians would represent certain quantum algorithms, such as the quantum Fourier transform or the Harrow-Hassidim-Lloyd. What we want to do next, is to compare the performance of these quantum algorithms when we implement them in a real quantum computer using either the DAQC or the DQC paradigm. For that, we first need to model the noisy environment that nowadays quantum computers represent. Afterward, we will be able to compare the results of each method in a realistic simulation.

\subsection{Performance of the DAQC compared to other paradigms}\label{sec:NOISE_SC}

In this section, we describe the noise model that we use to compare the performance of a quantum algorithm when it is implemented in a noisy quantum computer using either the DQC or the DAQC model. We begin by briefly describing the formalism that we will use to model the effect of the main noise sources of a NISQ device. Afterward, we will show how, using the aforementioned formalism, it is possible to simulate the implementation of a given quantum algorithm with experimental errors.

\subsubsection{Theoretical description of the effect of noise sources in quantum systems.}\label{sec241}

{\bf Quantum operation formalism and Operator-sum representation}

Quantum computers are open quantum systems, meaning that they cannot be isolated from the environment, and thus, they are susceptible to the noise caused by the control and the interaction with the environment. We can say that, in general, the noise sources depend on the experimental setup used to run the quantum experiments. As the field develops, these sources tend to be less significant, but in the context of the NISQ era, we cannot obviate the experimental constraints.

We will employ the {\it quantum operation formalism} as a general tool for describing the evolution of a quantum system in different scenarios. A deeper description of this formalism can be found in chapter $8$ of Ref. \cite{Nielsen2000}, but here we will describe the main aspect of it. 

The quantum states are well described by the {\it density operator} $\rho$, thus, they transform as 
\begin{equation}
    \rho' = \mathcal{E}\left(\rho\right),
\end{equation}
where the map $\mathcal{E}$ is a quantum operation. By using this formalism, we can describe any unitary transformation $U$ as $\mathcal{E}\left(\rho\right) = U\rho U^\dagger$, and, similarly, the measurement process as $\varepsilon_m\left(\rho\right) = M_m\rho M_m^\dagger$. Thus, the quantum operation tells us about the dynamic change of a state as a result of a physical process, where the initial state is $\rho$ and the final state is $\mathcal{E}\left(\rho\right)$. 

\begin{figure}[t]
\centering
	\includegraphics[width=0.95\linewidth]{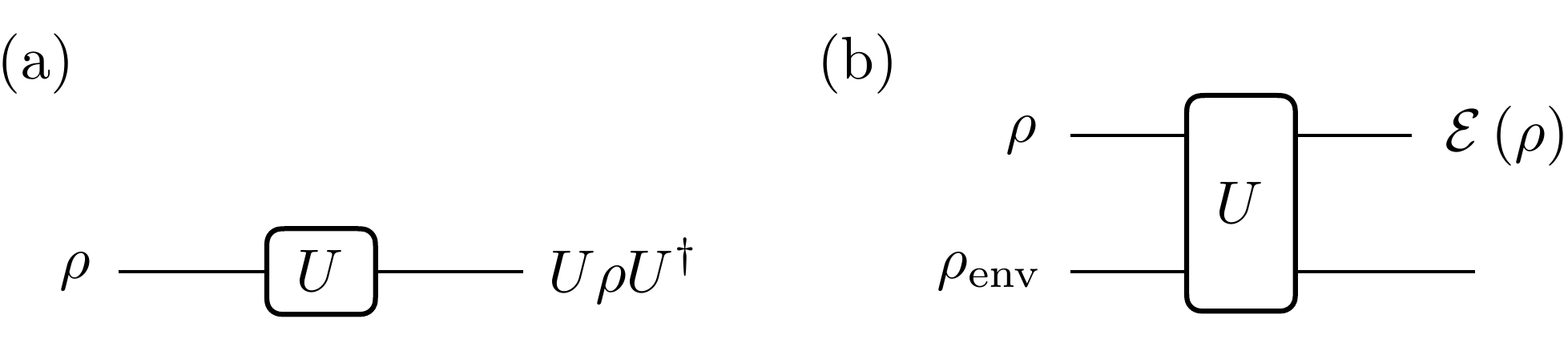}
	\caption{\label{Fig6chap2} \textbf{Schematic representation of a unitary operation $U$ in two different quantum systems. (a)} Closed quantum system. {\bf (b)} Open quantum system, composed of the principal system and the environment.}
\end{figure}

Unitary transformations can describe the dynamics of a closed quantum system, where we have an input state $\rho$, and an output one $U\rho U$. We can see the unitary transform $U$ as a black box whose internal details do not concern us. But quantum computers do not behave as closed quantum systems but rather like open quantum systems, whose dynamics can be described as an interaction between a principal system, which is the one that we want to study, and an environment. Together, they form a closed quantum system. This way, when we send the initial state $\rho$ into a box that is coupled to the environment, the final state $\mathcal{E}(\rho)$ may not be related to the initial state by a unitary transformation. After the unitary transformation $U$, the system no longer interacts with the environment and we can perform the partial trace over the environment top obtain the reduced state of the system, i.e.
\begin{equation}
    \mathcal{E}\left(\rho\right) = \Tr_\text{env}\left[ U\left( \rho\otimes\rho_\text{env}\right)U^\dagger\right],
\end{equation}
assuming that the system-environment input state can be described as a product state, $\rho\otimes\rho_\text{env}$. Fig. \ref{Fig6chap2} shows a schematic picture of a unitary transformation taking place on a closed (Fig. \ref{Fig6chap2}(a)) and an open (Fig. \ref{Fig6chap2}(b)) quantum system. A useful manner to represent quantum operations is the {\it operator-sum} representation, which we will introduce below. 

Let us suppose that we have an open quantum system that includes our system of interest $\mathcal{S}$ and the environment $E$, with Hilbert spaces $\mathcal{H_S}$ and $\mathcal{H}_E$ respectively, and being $|e_k\rangle$ an orthonormal basis for the finite-dimensional Hilbert space corresponding to the environment. Let us assume that the initial state of the environment is $\rho_\text{env} = |e_0\rangle\langle e_0|$. Then, for a unitary transformation $U$, the quantum operation can be written as
\begin{eqnarray}
    \mathcal{E}(\rho)  & =&\ \sum_k|e_k\rangle U\left(\rho\otimes|e_0\rangle\langle e_0|\right)U^{\dagger}|e_k\rangle \\
      & =& \sum_k E_k\rho E_k^{\dagger},\label{eq_operatorSum}
\end{eqnarray}
where $E_k\equiv \langle e_k| U |e_0\rangle$ is an operator which acts on the state space of the principal system. The operators ${E_k}$ are the {\it operation elements} of the quantum operation $\mathcal{E}$, also known as {\it Kraus operators} \cite{Kraus1983} and they play a key role in the description of the effect of noise sources in quantum systems. We will use these operators to characterize the quantum channel for each noise source. In the case where the initial state of the environment is not a pure state, we can always introduce an extra system to purify the environment. This extra system can be used as an intermediate step in calculations since it does not interfere with the dynamics of the principal system, and thus there is no loss of generality in assuming that the environment starts in a pure state.

The Kraus operators $E_k$ satisfy the completeness relation
\begin{equation}
    \sum_k E_k^\dagger E_k = \mathbbm{1}.
\end{equation}
This arises from the condition that the trace of $\mathcal{E}$ must be preserved and hence equal to one, i.e.
\begin{eqnarray}
    1 &=& \Tr\left(\mathcal{E}\left(\rho\right)\right)\nonumber \\
    &=& \Tr\left(\sum_k E_k\rho E_k^{\dagger}\right)\nonumber \\
    &=& \Tr\left(\sum_k E_k E_k^{\dagger}\rho\right)\nonumber,
\end{eqnarray}
and, since this condition has to be fulfilled for all $\rho$, then we must have $\sum_k E_k^\dagger E_k = \mathbbm{1}$.

The power of the operator-sum representation is that it allows us to characterize the principal system without having to consider the properties of the environment. We just need the Kraus operators $E_k$ that act on the principal system. In the following, we will specify the form of these operators for the different incoherent noise sources that can arise during the implementation of a quantum algorithm into a NISQ device.

\subsubsection*{Incoherent noise sources: quantum channels}
Noise channels are non-unitary quantum channels that reproduce the effect of noise in the state of the system of interest, in our case the qubits comprising a quantum computer. Let us first focus on single-qubit quantum channels, which are those that only act on one qubit. More concretely, we study the noise channels that account for the three most common noise sources in superconducting quantum processors. These are bit-flip, decoherence, and measurement error.

\begin{enumerate}
    \item {\bf Bit-flip and measurement channel.} It is a model for errors in which the state of the qubit is randomly switched with a probability $p$ and it is also present in classical computers. The energy to cause the flip of the qubit can be accidentally provided by pulses used to control the qubits or by thermal fluctuations. The Kraus operators of the bit-flip channel are
    \begin{eqnarray}
        & E_0=\sqrt{1-p}\begin{pmatrix}
    1 & 0\\
    0 & 1
    \end{pmatrix}=\sqrt{1-p}\text{ }\mathbbm{I},\\
    & \quad E_1=\sqrt{p} \begin{pmatrix}
    0 & 1\\
    1 & 0
    \end{pmatrix}=\sqrt{p}\text{ }X,
    \end{eqnarray}
    where $p$ is the bit-flip probability, $\mathbb{I}$ is the $2\times 2$ identity matrix and $X$ is the $\sigma_x$ Pauli matrix. When the measurement of the system is performed, the qubits might be affected by a bit-flip error. Based on this idea, we will introduce the measurement error as a bit-flip error preceding a perfect measurement operation.
    
    \item {\bf Amplitude-damping channel.} Decoherence is represented by means of the generalized amplitude-damping channel. This quantum channel accounts for the de-excitation of a two-level system, i.e., the loss of amplitude of the excited state $\ket{1}$ that will decay into the ground state $\ket{0}$ with probability $p$. It can describe different physical phenomena related to energy dissipation such as the spontaneous emission of a photon and the scattering and attenuation of the state of a photon in a cavity, among others \cite{Nielsen2000}. The Kraus operators of the generalized amplitude-damping channel are
    \begin{eqnarray}
       & E_0=\sqrt{p}\begin{pmatrix}
    1 & 0\\
    0 & \sqrt{1-\gamma}
    \end{pmatrix}, \quad \\
    & E_1=\sqrt{p}\begin{pmatrix}
    0 & \sqrt{\gamma}\\
    0 & 0
    \end{pmatrix}, \\
       & E_2=\sqrt{1-p}\begin{pmatrix}
    \sqrt{1-\gamma} & 0\\
    0 & 1
    \end{pmatrix}, \quad \\
    & E_3=\sqrt{1-p}\begin{pmatrix}
    0 & 0\\
    \sqrt{\gamma} & 0
    \end{pmatrix},
    \end{eqnarray}
    where the stationary state of the environment is
    
    \begin{equation}\label{environment}
        \rho_{\text{env}}=\begin{pmatrix}
        p & 0 \\
        0 & 1-p
        \end{pmatrix},
    \end{equation}
    with $p$ the thermal population of the ground state.
    
    The damping process can be understood as a time-accumulative problem, in which the probability grows with time \cite{Preskill2018_2019}. It can be shown that $\gamma$ can be written as $\gamma=1-e^{-t/T_1}$, where $t$ is the time and $T_1$ is the thermal relaxation time, and it describes processes due to the coupling of the qubit to its neighbors, which is a large system in thermal equilibrium at a temperature much higher than the one of the qubit \cite{Nielsen2000}.
  
\end{enumerate}

\subsubsection*{Coherent noise sources: quantum trajectories}
For the noise model, we also took into account coherent control-related experimental errors. In this case, instead of the quantum-channel approach, we employed quantum trajectories. These errors include those that might appear when we apply a SQG in both the DQC and the DAQC paradigm, the two-qubit gate (TQG) exclusive for the DQC, and the analog block noise exclusive for the DAQC. 

\begin{enumerate}
    \item {\bf Single-qubit gate error.} For this, we have included a magnetic-field noise $\Delta B$ added to the phases of the SQGs. This parameter $\Delta B$ is taken from a uniform probability distribution $\mathcal{U}(-r_D \Delta t/2,r_D \Delta t/2)$, where $\mathcal{U}(a,b)$ stands for a uniform noise distribution with range boundaries $(a,b)$. The deviation ratio is given by $r_D$. This can be modeled as
    \begin{equation}
    e^{iH_\text{SQG} }\ \rightarrow \ e^{i \Delta B H_\text{SQG}},
    \end{equation}
    where $H_\text{SQG}$ is the corresponding Hamiltonian for the applied SQG.

    \item {\bf Two-qubit gate control error.} As the CZ gate can be written in terms of SQG's and $\sigma_z\sigma_z$ interactions, the effects of noise in any TQG can be modeled using the magnetic field noise for SQG's and a Gaussian phase noise for the two-qubit $\sigma_z\sigma_z$ interaction. This Gaussian phase noise is represented by a random parameter $\varepsilon\in\mathcal{N}(0,\sigma_D)$ affecting the phase of the interaction, where $\mathcal{N}(\mu,\sigma)$ stands for a Gaussian distribution with mean $\mu$ and standard deviation $\sigma$. Its effect on a fixed $\pi/4$ phase interaction, which is usually the interaction of a digital quantum computer, is described by the change
    \begin{equation}
    e^{i\frac{\pi}{4}\sigma_z\sigma_z} \ \rightarrow \ e^{i\frac{\pi}{4}(1+\varepsilon)\sigma_z\sigma_z}.
    \end{equation}

    \item {\bf Analog block control error.} To model this error, we include a Gaussian coherent noise to the time that the analog blocks are applied. It produces a switch in the time from $t_{\alpha} \ \rightarrow \ t_{\alpha}+\delta$, where $\delta \in \mathcal{N}(0,r_b\Delta t)$, with $r_b$ the deviation ratio of the time $\Delta t$ required for a SQR. This can be implemented as
    \begin{equation}
            e^{it_{\alpha}H_\text{int}} \ \rightarrow \ e^{i(t_{\alpha}+\delta)H_\text{int}},
    \end{equation}
    where $H_\text{int}$ is the natural interacting Hamiltonian of the quantum processor. The Gaussian coherent phase noise has a greater value for sDAQC than for bDAQC, as a result of the effect of switching on and off the interaction in sDAQC. The value of these parameters changes each time a gate is applied to the circuit, as the effects of the noise described can vary in each application.
\end{enumerate}

\subsubsection{Error mitigation in DAQC}
As the NISQ era progresses, new methods for mitigating errors in quantum computing have emerged  \cite{Endo2018,Kandala2019,Endo2021}. One approach is the use of quasi-probability methods on analog, DAQC, or DQC platforms \cite{Endo2019,Sun2021}. These techniques have been shown to effectively eliminate or suppress errors, such as Trotter error \cite{Hakoshima2021}. In order to demonstrate the true performance of DAQC, it is important to apply these error mitigation techniques. In chapter 3, we demonstrate this by incorporating these techniques in the implementation of the quantum Fourier transform on a superconductive circuit quantum computer. For more detailed information on the results, we refer the reader to that chapter.

\vspace{1cm}

In this chapter, we have described the DAQC paradigm and presented the main tools that will allow us to work within this framework. In the following, we will show the different quantum algorithms we have proposed that emanate from the DAQC method.


\section[Digital-analog quantum Fourier transform and phase estimation]{Digital-analog quantum Fourier\\ transform and phase estimation}
\label{chapter3}

\fancyhead[RO]{3\quad DA QUANTUM FOURIER TRANSFORM AND PHASE ESTIMATION}

%
%

\thiswatermark{\put(1,-302){\color{l-grey}\rule{70pt}{60pt}}
\put(70,-302){\color{grey}\rule{297pt}{60pt}}}

%
%

\vfill
\lettrine[lines=2, findent=3pt,nindent=0pt]{T}{he} algorithms that we presented in chapter \ref{chapter2} allow us to describe relevant quantum algorithms in terms of the DAQC framework. In this chapter, we present a DA implementation of both the quantum Fourier transform (QFT) and the quantum phase estimation (QPE). We chose the QFT as the first quantum subroutine since it is a key ingredient for both more elaborated quantum subroutines, such as the QPE, and relevant quantum algorithms, such as Shor's algorithm for factorization \cite{Shor1996}. The content of this chapter include the work developed in \cite{Martin2020} and \cite{GMS2021}.

In section \ref{sec31}, we introduce the QFT, describing it in both the DQC and the DAQC paradigms. We compare the performance of both paradigms in a NISQ device, for which we have included the noise sources listed in section \ref{sec:NOISE_SC} in our simulations. We also included an analysis of how error mitigation techniques would improve the DAQC implementation. Afterward, in section \ref{sec32}, we study the QPE algorithm in a similar manner as we did for the QFT. We first find its description using DAQC and then show its performance in a realistic noisy scenario. The DAQC description of these two quantum algorithms will allow us to face more complex algorithms, such as the HHL, in future chapters.

\subsection{Quantum Fourier transform}\label{sec31}

\subsubsection{Description of the QFT algorithm and its DAQC implementation}

The discrete Fourier transform (DFT) is a mathematical tool that is widely used in signal processing, image processing, and physics. In classical computing, the DFT is used to transform a discrete signal from the time domain to the frequency domain, which can be useful for analyzing the frequency components of a signal. It is a computationally expensive task, and it can take a significant amount of time to perform on a classical computer.

In quantum computing, the DFT is performed by the QFT algorithm. The QFT is a quantum algorithm that performs the Fourier transform on a quantum state. It is a key building block in many quantum algorithms, such as Shor's algorithm for factoring integers and the QPE. The QFT can be implemented using a series of quantum gates and can be applied to a system of qubits. The main advantage of the QFT over classical algorithms is that it can be performed exponentially faster on a quantum computer.

The classical DFT takes a complex vector of length $N$, $\left(x_0, x_1, ..., x_{N-1}\right)$ and transforms it into another complex vector of the same length $\left(y_0, y_1,...y_{n-1}\right)$ whose $k$-th element is defined as
\begin{equation}
y_k\equiv \frac{1}{\sqrt{N}}\sum_{j=0}^{N-1}x_j e^{2\pi i j}.
\end{equation}

\begin{figure}
\centering
\includegraphics[width=0.98\textwidth]{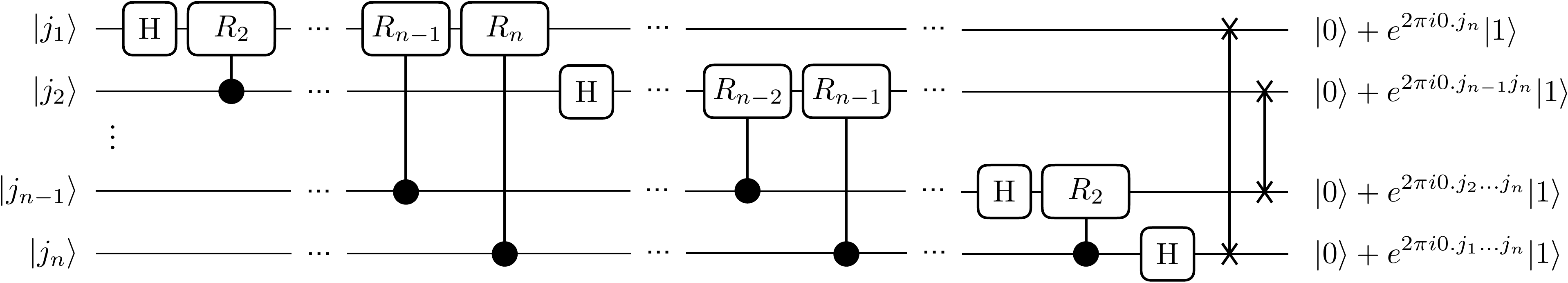} 
\caption{{\bf Digital implementation of the QFT for an $n-$qubit system}. Each set of controlled operations $cR_k$ is preceded by a Hadamard gate, $\text{H}$. The family of controlled rotations $cR_k$ is defined as $cR_k=\dyad{0}\otimes\mathbbm{1}+\dyad{1}\otimes R_k$, with $R_k=\dyad{0}+e^{2\pi i/2^k}\dyad{1}$. The final state is written using the binary representation for $|j\rangle$, where $j = j_1 j_2 ... j_n = \sum_{l=1}^n j_l 2^{n-l}$, and  $0.j_l j_{l+1} ... j_m$, refers to the binary fraction $j_l/2+j_{l+1}/4 + ... + j_m / 2^{m-l+1}$.}\label{Fig1chap3}
\end{figure}

QFT, its quantum counterpart, is a linear operator $\mathcal{F}$ with the following action on the basis states:
\begin{equation}
\mathcal{F}|j\rangle \equiv \frac{1}{\sqrt{N}}\sum_{k=0}^{N-1} e^{2\pi i j k/N}|k\rangle,
\end{equation}
where $N=2^n$ and $n$ is the number of qubits of the system. In Fig.~\ref{Fig1chap3} we show the implementation of the QFT using single- and two-qubit gates. The only single-qubit gates applied are Hadamard gates $H$, whose unitary matrix and Hamiltonian expressions are
\begin{equation}\label{eq: H}
\text{H}=e^{iH_\text{H}}=\frac{1}{\sqrt{2}}\begin{pmatrix}
1&1\\
1&-1
\end{pmatrix},~~H_\text{H}=\frac{\pi}{2}\left[\mathbbm{1}-\frac{1}{\sqrt{2}}(Z+X)\right],
\end{equation}
respectively. The entangling two-qubit gates of the circuit implementation are the controlled-$R_k$ ($cR_k$) rotations with
\begin{equation}
R_k=\begin{pmatrix}
1&0\\
0&e^{2\pi i/2^k}
\end{pmatrix},
\end{equation}
\begin{equation}\label{eq: cR_k}
cR_k=\dyad{0}\otimes\mathbbm{1}+\dyad{1}\otimes R_k={\small\begin{pmatrix}
1&0&0&0\\
0&1&0&0\\
0&0&1&0\\
0&0&0&e^{2\pi i/2^k}
\end{pmatrix}}.
\end{equation}
They appear in $(n-1)$ different blocks of controlled rotations, all of them preceded by a Hadamard gate as shown in Fig.~\ref{Fig1chap3}.

In order to apply the DAQC protocol to implement the QFT, we first have to describe the complete QFT as one unitary operation whose Hamiltonian we can simulate using the DAQC algorithms presented in section \ref{sec21}. To do that, we express the unitary matrices involved in the algorithm in terms of the resource Hamiltonian. Without loss of generality, we will assume that the interaction in the quantum processor can be described by an ATA two-body Ising Hamiltonian. Thus, s complete quantum subroutine can be defined as
\begin{equation} \label{eq: U_QFT}
U_{\text{QFT}}=\left[\prod_{m=1}^{n-1}U_{\text{SQG},m}U_{\text{TQG},m}\right] U_{H,m},
\end{equation}
where
\begin{eqnarray}
U_{\text{SQG},m}&=&\exp\left[i\sum_{k=2}^{n-(m-1)} \theta_k\left(\mathbbm{1}_{N\times N}-\sigma_z^{(k+m-1)}-\sigma_z^{(m)}\right)\right]\nonumber\\
&&\times \exp\left[ \frac{i\pi}{2}\left(\mathbbm{1}-\frac{\sigma_z^{(m)}+\sigma_x^{(m)}}{\sqrt{2}}\right)\right], \label{eq:U_SQG}\\
U_{\text{TQG}}&=&\exp\left(i\sum_{c<k}^n\alpha_{c,k,m}\sigma_z^{(c)}\otimes \sigma_z^{(k)}\right),\label{eq:H_ZZ}\\
U_{H,m}&=&\exp\left(\frac{i \pi}{2}\left[\mathbbm{1}^{(m)}-\frac{(\sigma_z^{(m)}+\sigma_x^{(m)})}{\sqrt{2}}\right]\right),\label{eq:U_Hm}\\
\theta_k&=&\frac{\pi}{2^{k+1}},\qquad\text{and}\qquad\alpha_{c,k,m}=\delta_{c,m}\frac{\pi}{2^{k-m+2}}.\label{eq:theta_alpha}
\end{eqnarray}
The superindices in parenthesis specify the qubit in which the unitary operation is performed and the operators $\sigma_j$ refers to the Pauli matrices. 

\begin{figure*}[t]
\centering
\includegraphics[width=1\textwidth]{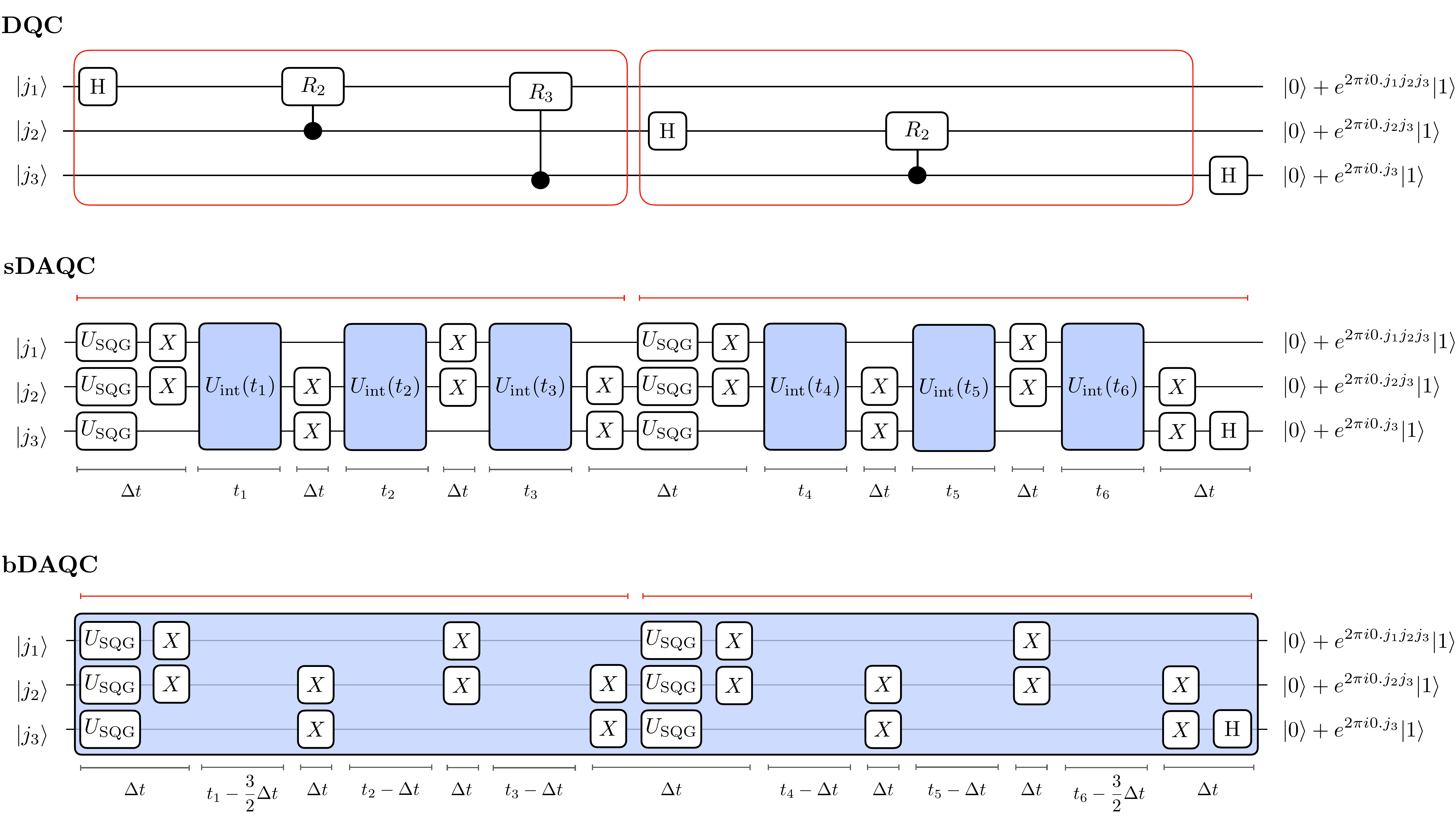} 
\caption{{\bf Implementation of the QFT for a $3-$~qubit system using three different protocols: DQC, sDAQC and bDAQC.} The blue blocks $U_\text{int}(t_k)$ represent the analog blocks, and each of them is applied during different times $t_k$. The single-qubit rotations $R_x(\pi)$, which correspond to an $X$ gate, act for a time $\Delta t$. We apply the DAQC protocol for each block of controlled rotations of the DQC implementation, which is detailed by the red line over each of those blocks. The sDAQC switches on and off the analog evolution before applying the single-qubit rotations $X$. In contrast, in the bDAQC protocol, the single-qubit rotations are performed on top of the analog evolution. The total analog interaction time in the bDAQC must equal the sum of each individual analog block in the sDAQC.} \label{Fig2chap3}
\end{figure*}

In Fig.~\ref{Fig2chap3}, we depict the DQC implementation of the QFT using Eqs.~(\ref{eq:U_SQG})-(\ref{eq:U_Hm}). As one can see, each controlled-rotation block can be implemented by applying first a set of single-qubit gates, and then a set of two-qubit gates. This is why we decompose the complete unitary transformation into three different operations. The subindices SQG and TQG stand for single-qubit gates and two-qubit gates, respectively. The inhomogeneous ATA two-body Ising Hamiltonian we want to write in the DAQC framework [see Eq.~\eqref{eq:H_ZZ}] represents a complete block of controlled rotations, and it is different for each block. This means that we need to apply the DAQC protocol $(n-1)$ times, one time per controlled-rotation block.

Every two-qubit gate is applied following the ATA DQC protocol and using a fixed $\pi/4$ phase
\begin{equation}
e^{i\varphi_{jk}^{\mu\nu}\sigma_\mu^{(j)}\sigma_\nu^{(k)}}=e^{i(\pi/4)\sigma_y^{(j)}}e^{i(\pi/4)\sigma_\mu^{(j)}\sigma_\nu^{(k)}}e^{i\varphi_{jk}^{\mu\nu}\sigma_y^{(j)}}e^{-i(\pi/4)\sigma_\mu^{(j)}\sigma_\nu^{(k)}} e^{-i(\pi/4)\sigma_y^{(j)}}.
\end{equation}
In our case, $\mu=\nu=Z$ and the phase $\varphi_{jk}^{\mu\nu}$ correspond to the coefficient $\alpha_{c,k,m}$, given in Eq.~\eqref{eq:theta_alpha},
\begin{eqnarray} \label{eq:ZZ gates}
e^{i\alpha_{c,k,m}\sigma_z^{(c)} \sigma_z ^{(k)}}&=&e^{i(\pi/4)\sigma_y^{(c)}}e^{i(\pi/4)\sigma_z^{(c)} \sigma_z^{(k)}}e^{i\alpha_{c,k,m} \sigma_y^{(c)}} R_x(\pi)^{(k)} \nonumber\\
&& e^{i(\pi/4) \sigma_z^{(c)} \sigma_z^{(k)}} R_x(\pi)^{(k)} e^{-i(\pi/4)\sigma_y^{(c)}},
\end{eqnarray}
where the rotation $R_x(\theta)^{(k)}$ is applied to the $k-$th qubit and its defined as $R_x(\theta) = e^{-i\theta \sigma_x/2}$. We could also have decomposed the single- and two-qubit gates involved in the circuit into a base gate-set that forms a universal set of gates, such as $\{Rz(\theta), Rx(\theta), \text{CNOT}\}$. In this case, we decided to go this way so we could be as close as possible to the work of Parra et. al \cite{Parra2020}. In the following works, we decided to go the other way, so we could include in our simulations the same kind of experimental errors that companies like IBM specify about their devices. These specifications were not available when we first presented our work on the QFT, so we took as reference the state-of-art resonators that were known at that time.

\subsubsection{Realistic implementation with experimental errors}

Now we will compare the performance of the QFT using two different paradigms, the DQC and the DAQC. To do so, we will compute the Fourier transform of a family of initial states $|\psi_0\rangle = \sin \beta |\text{W}_N\rangle + cos \beta |\text{GHZ}_N\rangle$, with $\beta \in [0,\pi]$. We apply the QFT for a 3-, 5-, and 6-qubit system to gasp the effect of the noise sources as the number of qubits increases. As a figure of merit we have used the quantum fidelity, \begin{equation}
    F(\rho_{\text{ideal}},\rho_{\text{noisy}})=\left[\text{tr}\left(\sqrt{\sqrt{\rho_{\text{ideal}}}\rho_{\text{noisy}}\sqrt{\rho_{\text{ideal}}}}\right)\right]^2,
\end{equation}
where $\rho_{\text{ideal}}$ is the ideal state after the application of the QFT, while $\rho_{\text{noisy}}$ is the resulting state when the noise model is implemented. The fidelity verifies $0\leq F(\rho_{\text{ideal}},\rho_{\text{noisy}})\leq 1$, thus the larger the values the more similar are $\rho_\text{ideal}$ and $\rho_\text{noisy}$, meaning that the noise sources has less effect on the performance of the experiment.

The noise model that we consider encloses the most relevant noise sources in superconducting quantum computers: bit-flip, decoherence, and measurement error, as well as control-related errors such as magnetic field fluctuations. However, the difference in the implementation of the entangling blocks in DQC and DAQC leads to a different experimental error associated with them, which is modeled as a Gaussian phase noise on the application time of the TQG's in DQC and the analog blocks in DAQC, as we discussed in section \ref{sec241}, chapter \ref{chapter2}.

The values of the parameters used to simulate the different noise sources are chosen to model realistic NISQ superconducting devices. In single-qubit gates, we have introduced a magnetic-field noise $\Delta B_\gamma$ by adding to the Hamiltonian of the single-qubit gate a random variable taken from a uniform probability distribution centered in $1$, i.e., $\mathcal{U}(1-\text{SQGN}, 1+\text{SQGN})$. We have chosen SQGN~$=0.0005$. For the two-qubit gates, we add a Gaussian phase noise $\epsilon\in\mathcal{N}(0,\text{TQGN})$, with variance TQGN~$=0.2000$, to the $\pi/4$ phases in the DQC protocol. Finally, to model the experimental control error on the analog blocks, we include a Gaussian coherent noise to the time those blocks are applied, this is $t\rightarrow t+\delta$, where $\delta\in\mathcal{N}(0,\text{ABN})$. The value of the variance ABN depends on which DAQC protocol we are using. The value used on the sDAQC is double the value used for the bDAQC case. The values we have considered are $\text{ABN}_\text{s}=0.0200$ for the sDAQC case and $\text{ABN}_\text{b}=0.0100$ for the bDAQC case. Thus, each ideal gate transforms as
\begin{eqnarray}
e^{i\theta_k Z}&\rightarrow& e^{i\theta_k \Delta B\, Z},\\
e^{i(\pi/4)Z Z} &\rightarrow&e^{i(\pi/4)(1+\epsilon)Z Z},\\
e^{it_\alpha H_{int}}&\rightarrow&e^{i(t_\alpha +\delta) H_{\text{int}}}.
\end{eqnarray}

On the other hand, for the incoherent noise sources introduced in section \ref{sec241}, we have a bit-flip error with probability $p_\text{b-f}=0.005$ and a measurement error with probability $p_\text{meas}=0.01$. Notice that measurement error is currently the technological limiting factor for quantum algorithms run in superconducting quantum processors. Finally, for the decoherence channel, the relaxation time $T_1$ is $50\ \mu \text{s}$, with the thermal population of the ground state $p=0.35$. The length of the SQGs is $\Delta t_\text{SQG}= 1/(100g_0)$, while for the two-qubit interaction, it is $\Delta t_\text{TQG}= (1+\varepsilon)100\Delta t_\text{SQG}\pi/4$ for TQG's, and for the analog blocks, it is the one obtained from the decomposition of the interaction. 

\begin{figure}[t]
\centering
\includegraphics[width=0.95\textwidth]{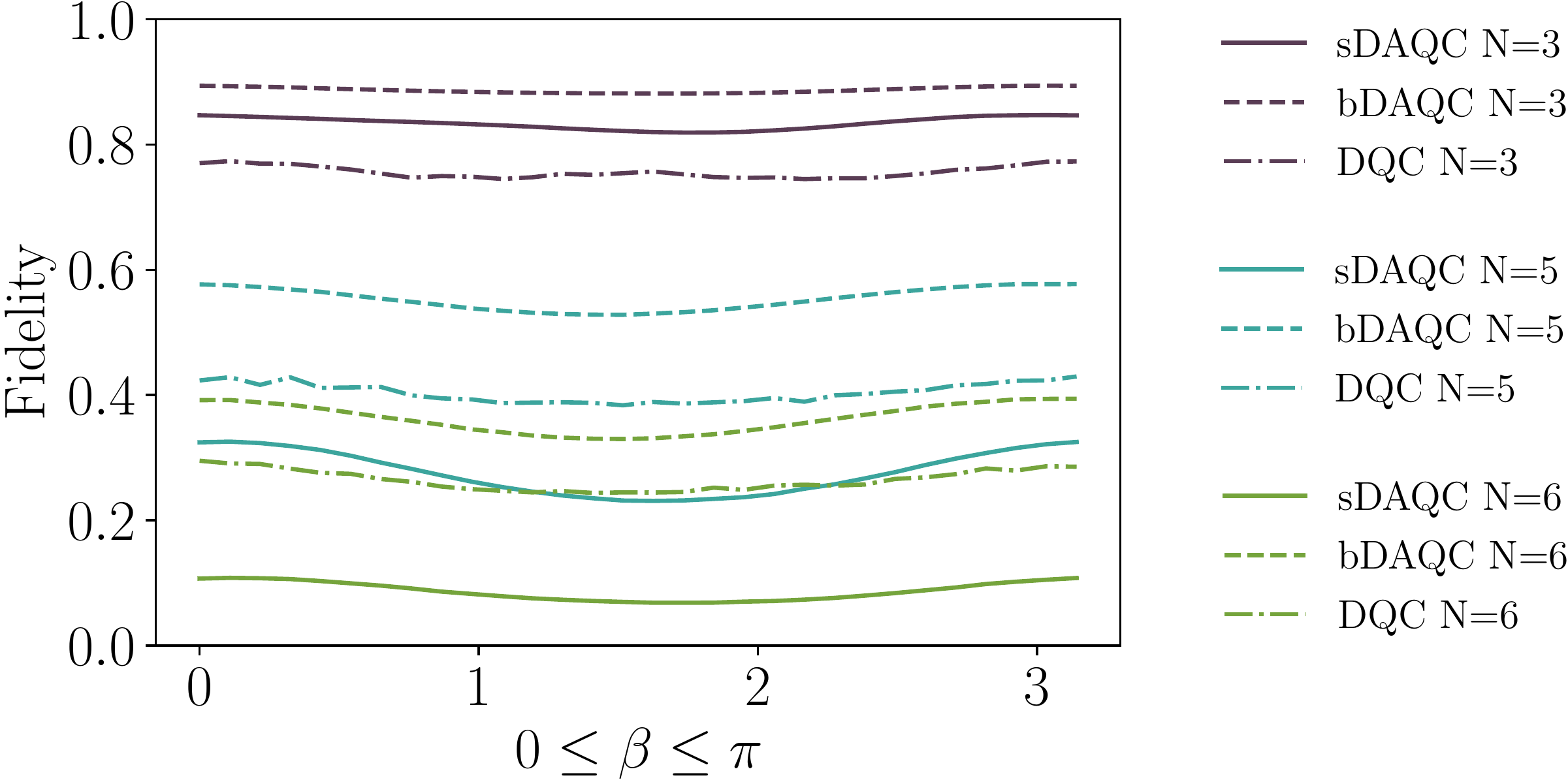} 
\caption{{\bf Comparison of the fidelity of the QFT algorithm for 3-, 5-, and 6-qubit systems using the DQC, sDAQC, and bDAQC protocols.} The highest fidelity is obtained using the bDAQC protocol, which demonstrates better tolerance to decoherence and reduced bit-flip errors compared to the other protocols. The study was conducted using the state $\ket{\psi_0}=\sin\beta\ket{W_N}+\cos\beta\ket{GHZ_N}$.}
\label{Fig3chap3}
\end{figure}

In Fig. \ref{Fig3chap3}, we present the results of our realistic implementation of the QFT, which were obtained by averaging over 1000 repetitions to ensure a reliable statistical sample. The results indicate that the bDAQC protocol outperforms the others when noise sources are taken into account. This is because the error caused by turning the interaction on and off is less significant than the inherent error of having the interaction always on. Additionally, the fact that the interaction is never switched off also means that bDAQC is only affected by bit-flip errors for the SQGs, as opposed to DQC and sDAQC, which are affected by bit-flip errors due to the entangling blocks. For the parameters used in our simulations, the total time of application of the QFT algorithm using the DAQC paradigm is shorter than the total time using DQC, thus making DAQC less susceptible to decoherence. The measurement error is common for all three paradigms, and its effect on fidelity is reflected as a shift in all values. However, it should be noted that as the number of qubits increases, the performance of the QFT algorithm decreases rapidly due to the increase in depth of the algorithm, which scales as $\mathcal{O}(N^2)$, where $N$ is the number of qubits.

It is important to note that the results shown in Fig. \ref{Fig3chap3} should not be taken as definitive, as the actual fidelity obtained in an experiment will depend on factors such as fabrication, architecture, and materials. However, the simulation results suggest that under fair comparison and considering the same noise sources, bDAQC obtains the highest value of fidelity. The distance between the fidelities of bDAQC and DQC is always remarkable in favor of bDAQC. Even though we have used a sensible choice for the noise model of the DAQC implementation, this model will not be accurate until it is compared against experimental data.

Our simulations indicate that bDAQC is a viable alternative to NISQ processors. With 3 qubits, bDAQC has a fidelity of approximately $90\%$, compared to $85\%$ for sDAQC, and $80\%$ for DQC. As the number of qubits increases, we have found that bDAQC's fidelity values continue to surpass those of DQC, although sDAQC's performance deteriorates due to the increasing number of analog blocks affected by bit-flip errors. With a larger number of qubits, the impact of noise sources results in fidelity values that are too low to yield meaningful results from the experiments.

We would like to point out that, while most quantum systems do not have ATA connectivity, with the notable exceptions of trapped ions and NMR, we have chosen to use ATA for the purpose of simplicity. Using the DAQC paradigm, it can be demonstrated that it is possible to simulate an $N$-qubit ATA Ising Hamiltonian with no more than $\frac{1}{2}N(N-1)$ NN Hamiltonians. This quadratic overhead is the worst-case scenario, but it can be significantly reduced if a pattern exists in our couplings \cite{galicia2019enhanced}. However, this overhead (or worse) also applies to the DQC paradigm. Therefore, the fidelity values may differ when considering NN Hamiltonians, but the comparison between DQC and DAQC is fair, and the conclusion that bDAQC outperforms DQC remains valid.

\subsubsection{Error mitigation in DA-QFT}

To show the actual performance of DAQC, it is necessary to show how to adapt quantum error mitigation techniques to this paradigm. Specifically, we will prove that these techniques can effectively eliminate the inherent error that arises from bDAQC.

We utilized the zero noise extrapolation technique \cite{Li2017,Temme2017} to demonstrate the impact of decoherence on the bDAQC QFT circuit. Specifically, we applied the circuit to an initial state of $\ket{\psi_0}=\sin\pi/4\ket{\text{W}8}+\cos\pi/4\ket{\text{GHZ}8}$ for varying coupling constants $g_j, \ j=0,\dots,n_g$. By modifying the total circuit time as $t{j{\alpha}} \propto g_j^{-1}$, we could vary the effect of decoherence. However, changes in $g_j$ also affected the intrinsic error of the bDAQC paradigm. To address this issue, we selected $\Delta t_{\text{SQG}_{i,j}} = b_i/g_j$, such that the digital block time was inversely proportional to the coupling constant for a given $g_j$. By extrapolating the fidelity results for several values of $g_j$ to the limit of zero decoherence and total time, we could eliminate the error caused by decoherence.

\begin{figure}[t]
\centering
\includegraphics[width=0.96\textwidth]{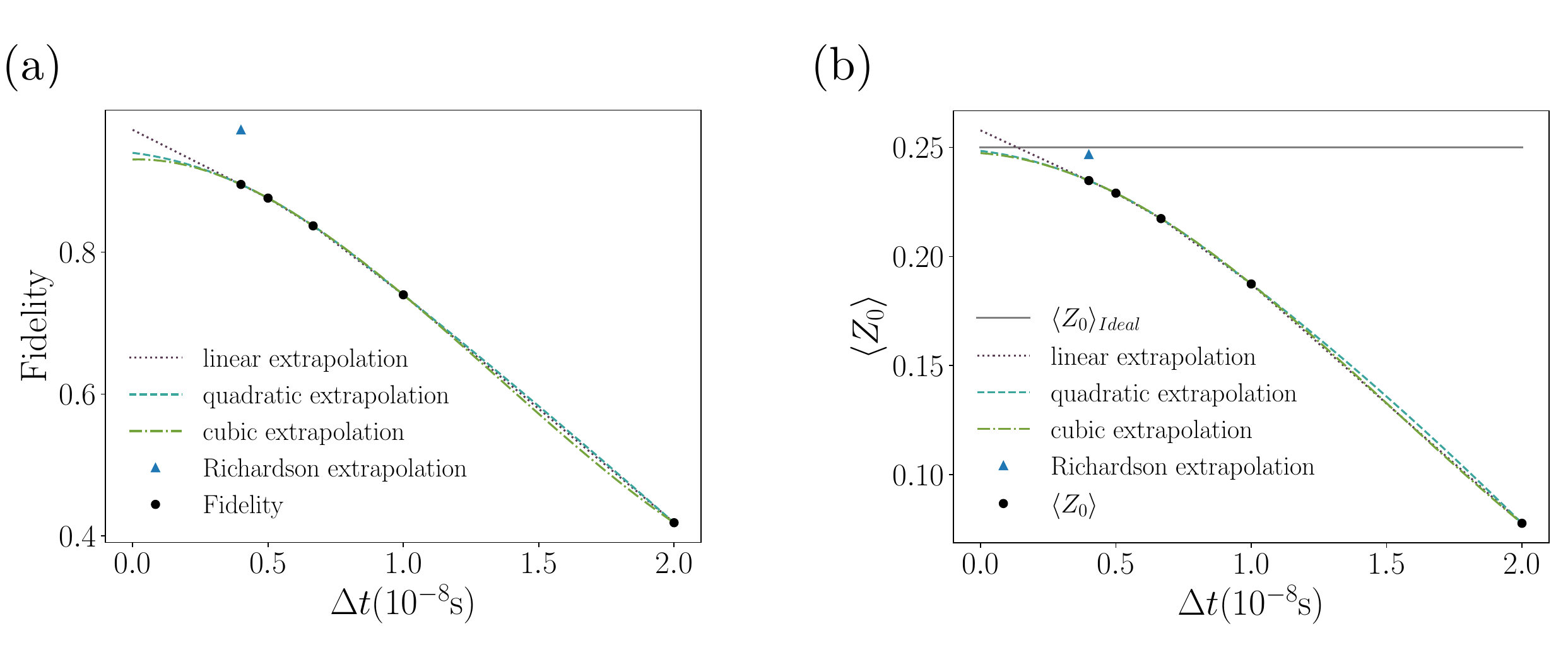} 
\caption{{\bf Results of the error mitigation. (a)} Fidelity. {\bf (b)} $\boldsymbol{\langle Z_0 \rangle}$. Both fidelity and $\langle Z_0 \rangle$ attain accurate approximations of the ideal values. The simulations were made for the $8$-qubit QFT circuit with $T_1 = 50 \ \mu$s, $p=0.35$ and $g_0 = 1$ MHz.}
\label{Fig: error mitigation}
\end{figure}

Furthermore, it is feasible to expand this method to eradicate the inherent error brought about by bDAQC. To achieve this, we iterate the prior approach for $i = 0, ..., n_t$ values of $\Delta t_{\text{SQG}{i,j}} = b_i/g_j$, utilizing the zero decoherence values to carry out the extrapolation to the zero $\Delta t\text{SQG}$ limit. We select an 8-qubit QFT circuit to implement this method as the number of qubits and quantum gates pose a significant challenge for cutting-edge quantum devices. We determine the fidelity of the ideal circuit, both without the error due to decoherence and bDAQC. The optimal fidelity is given by the linear extrapolation, which is approximately $0.97$, indicating that the error-mitigated state is an excellent approximation of the ideal one. These results are depicted in Fig. \ref{Fig: error mitigation}(a). The same result is obtained for Richardson extrapolation, as in this case, the best outcome is obtained for the first-order approximation.

However, obtaining the quantum state after applying a quantum circuit is not always feasible as it can be a costly operation due to the use of tomography. Therefore, to assess the effectiveness of error mitigation for bDAQC, we need to select a second figure of merit that can be easily computed from the measurements of a quantum circuit. In this regard, we choose the expectation value of the observable $Z_0 = Z\otimes \mathbb{I}\otimes \dots \otimes \mathbb{I}$. The results of the expectation value approximation are presented in Fig. \ref{Fig: error mitigation}b. It can be observed that the best approximation of $\langle Z_0 \rangle$ is approximately $0.25$, which is computed using Richardson extrapolation. The error $\varepsilon = | \langle Z_0 \rangle-\langle Z_0 \rangle_\text{approx} | = 0.0016$ indicates that we can obtain an extremely accurate approximation of the ideal expectation value from the noisy one.

\begin{figure}[t]
\centering
\includegraphics[width=0.96\textwidth]{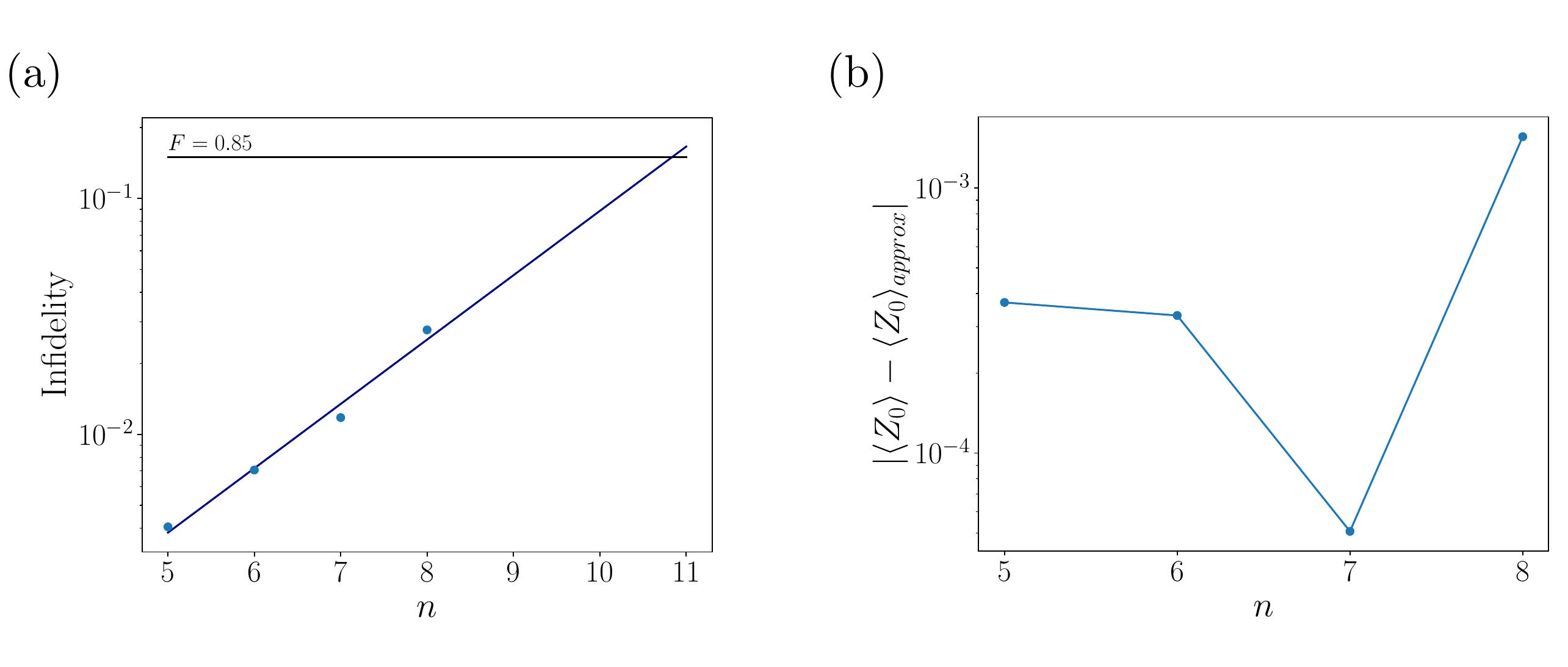} 
\caption{{\bf Scaling of the error mitigation results with the number of qubits $n$ for $n=5,6,7,8$. (a)} Infidelity. {\bf (b)} $|\boldsymbol{\langle Z_0 \rangle}-\boldsymbol{\langle Z_0 \rangle}_\text{approx}|$. The simulations were made with $T_1 = 50 \ \mu$s, $p=0.35$ and $g_0 = 1$ MHz.}
\label{Fig: error mitigation 2}
\end{figure}

The scaling behavior of the infidelity and the error in the estimation of $\langle Z_0 \rangle$ can also be studied as a function of the number of qubits. As depicted in Fig. \ref{Fig: error mitigation 2}(a), we can observe a decrease in fidelity as the number of qubits increases. However, by applying linear regression, we can still achieve fidelities of approximately $0.85$ for up to 11 qubits. On the other hand, for the computation of $\langle Z_0 \rangle$ (Fig. \ref{Fig: error mitigation 2}(b)), the value to be approximated changes with each number of qubits, since the quantum circuit is modified accordingly. Therefore, we cannot observe a similar scaling as for the infidelity. Nonetheless, we obtain errors of a similar order of magnitude for the different number of qubits.

The figures presented in Fig. \ref{Fig: error mitigation} demonstrate the effectiveness of using an adapted zero noise extrapolation to recover the ideal results from the bDAQC paradigm. This approach can be extended to sDAQC by solely studying the zero decoherence limit. Although other error mitigation techniques, such as an adaptation of Clifford Data Regression \cite{Czarnik2021}, have been considered, they result in a suboptimal outcome as they do not allow for independent error correction of the bDAQC extra interaction.


\vfill
\subsection{Quantum phase estimation}\label{sec32}

We studied the implementation of the QFT using the DAQC framework. Our findings prompted us to propose further investigation into the QPE algorithm. The QFT transforms the system state to reveal the phase information of eigenvalues, which can be estimated with high precision. In our previous research, we found that the DAQC paradigm outperforms the digital approach for larger numbers of qubits. As the inverse QFT is a part of the QPE algorithm, we expect similar performance. Our goal is to test the limits of these paradigms by obtaining concrete eigenvalues through the QPE algorithm. This poses a greater challenge for DQC, as it involves more two-qubit operations which are a major source of noise in current superconducting devices. We believe that using the analog evolution of DAQC will result in higher fidelity compared to a digital approach.

\subsubsection{Description of the QPE algorithm and its DAQC implementation}

The QPE algorithm estimates the phase $\varphi$ that corresponds to the eigenvalue $e^{2\pi i \varphi}$ of the eigenvector $|u\rangle$ of a unitary operator $U$, where $\varphi$ is unknown. This algorithm utilizes two quantum registers. The first register consists of $t$ qubits, which determine the precision of the $\varphi$ approximation. The second register encodes the eigenvector $|u\rangle$. The first register is initialized by applying a Hadamard gate to each qubit, followed by a series of controlled rotations on the second register using increasingly powers of two exponentiated by $U$. Finally, the inverse Fourier transform is applied to the first register, yielding an approximation of $\varphi$ and the desired eigenvalue.

\begin{figure}[t]
\centering
\includegraphics[width = 0.90\textwidth]{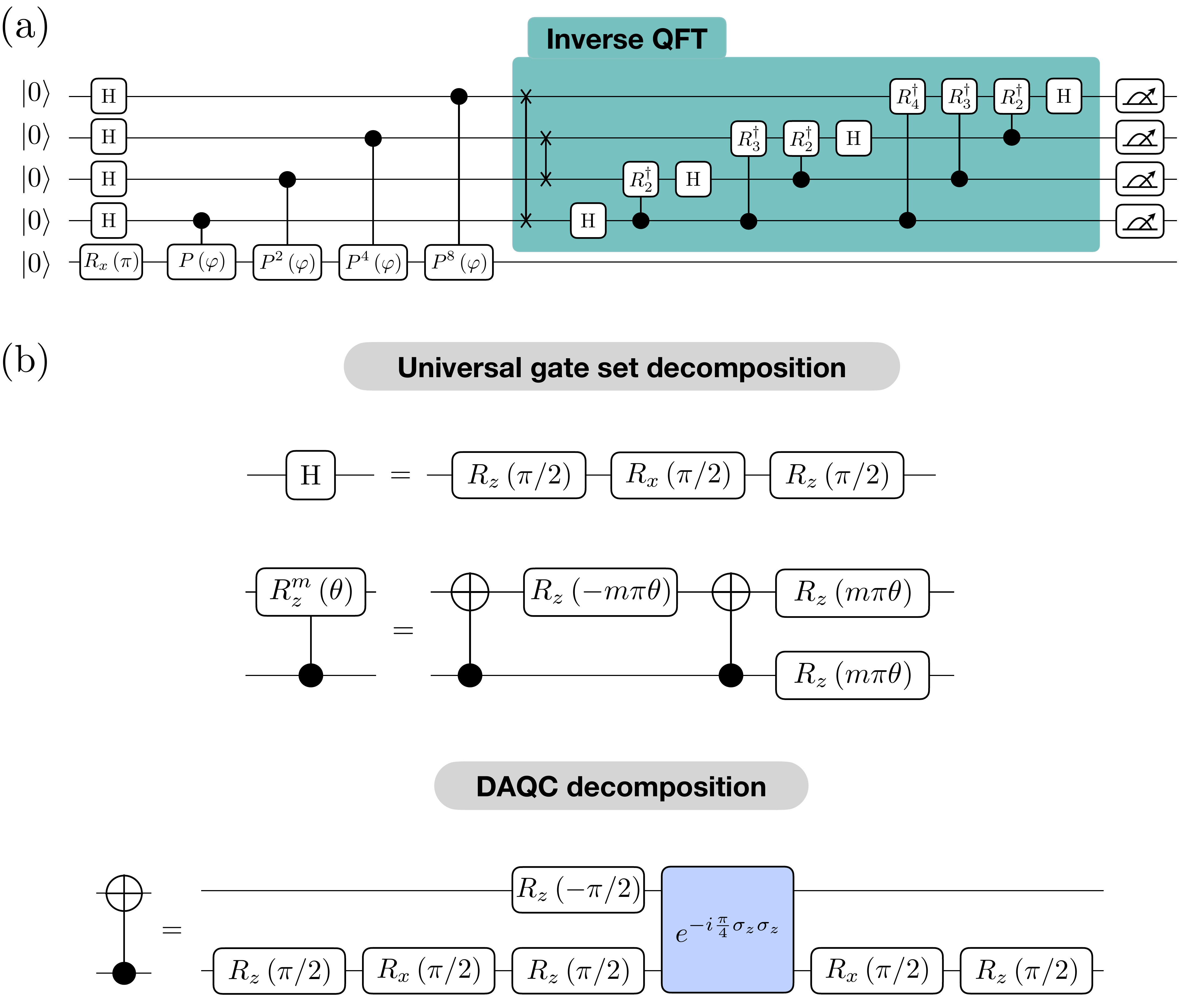}
\caption{{\bf QPE algorithm and its decomposition into a universal set of gates. (a)} QPE algorithm for $P\left(\varphi\right)$ \eqref{eq: QPE_matrix}, such that $P(\varphi)|u\rangle=e^{i2\pi\varphi}|u\rangle$. Hence, the final value of $\varphi$ should be $\varphi = 1/3$. {\bf (b)} Decomposition of the single- and two-qubit rotations in terms of the  universal set of gates $\lbrace R_x(\theta), R_z(\theta), \text{CNOT}\rbrace$, as well as the decomposition of the CNOT gate using the DAQC method. Using this universal set of gate, the controlled rotations involved in the QPE algorithm are defined as $P^m(\varphi)\equiv R_z^m(\theta = \varphi)$ and $R_k^\dagger\equiv R_z^{m=1}(\theta=1/2^k)$}
  \label{Fig4chap3}
\end{figure}

Given the unitary operator $P\left(\varphi\right)$,
\begin{equation} \label{eq: QPE_matrix}
    P\left(\varphi\right) = \begin{pmatrix} 1 & 0 \\
                        0 & e^{i2\pi \varphi}
        \end{pmatrix},
\end{equation}  
we aim at obtaining $\varphi$ such that $P(\varphi)|u\rangle=e^{i2\pi\varphi}|u\rangle$. We choose a 4-qubit register to estimate the phase $\varphi$ that corresponds to the eigenvector $|u\rangle = |1\rangle =(0 \quad 1)^T$. The complete QPE circuit is depicted in Fig. \ref{Fig4chap3}(a).

For the implementation of the QPE circuit, we took a different approach compared to the one used for the QFT. Rather than directly simulating the complete QPE algorithm Hamiltonian using the DAQC toolkit described in Chapter \ref{chapter2}, we first decomposed the QPE quantum circuit into a universal set of gates, $\lbrace R_x(\theta), R_z(\theta), \text{CNOT}\rbrace$, which are commonly used in real superconducting quantum computers. Then, we translated each two-qubit gate (which, after the previous decomposition, would only be CNOT gates) into the DAQC paradigm by representing it as an analog block surrounded by single-qubit rotations. This approach allows us to explore different and more systematic alternatives to build DAQC quantum algorithms. In Fig. \ref{Fig4chap3}(b) we show both the decomposition of the QPE single- and two-qubit gates in terms of the aforementioned set of gates, and the DAQC decomposition of the entangling CNOT gate.

\subsubsection{Realistic implementation with experimental errors}

\begin{figure}
\centering
\includegraphics[width=0.8\textwidth]{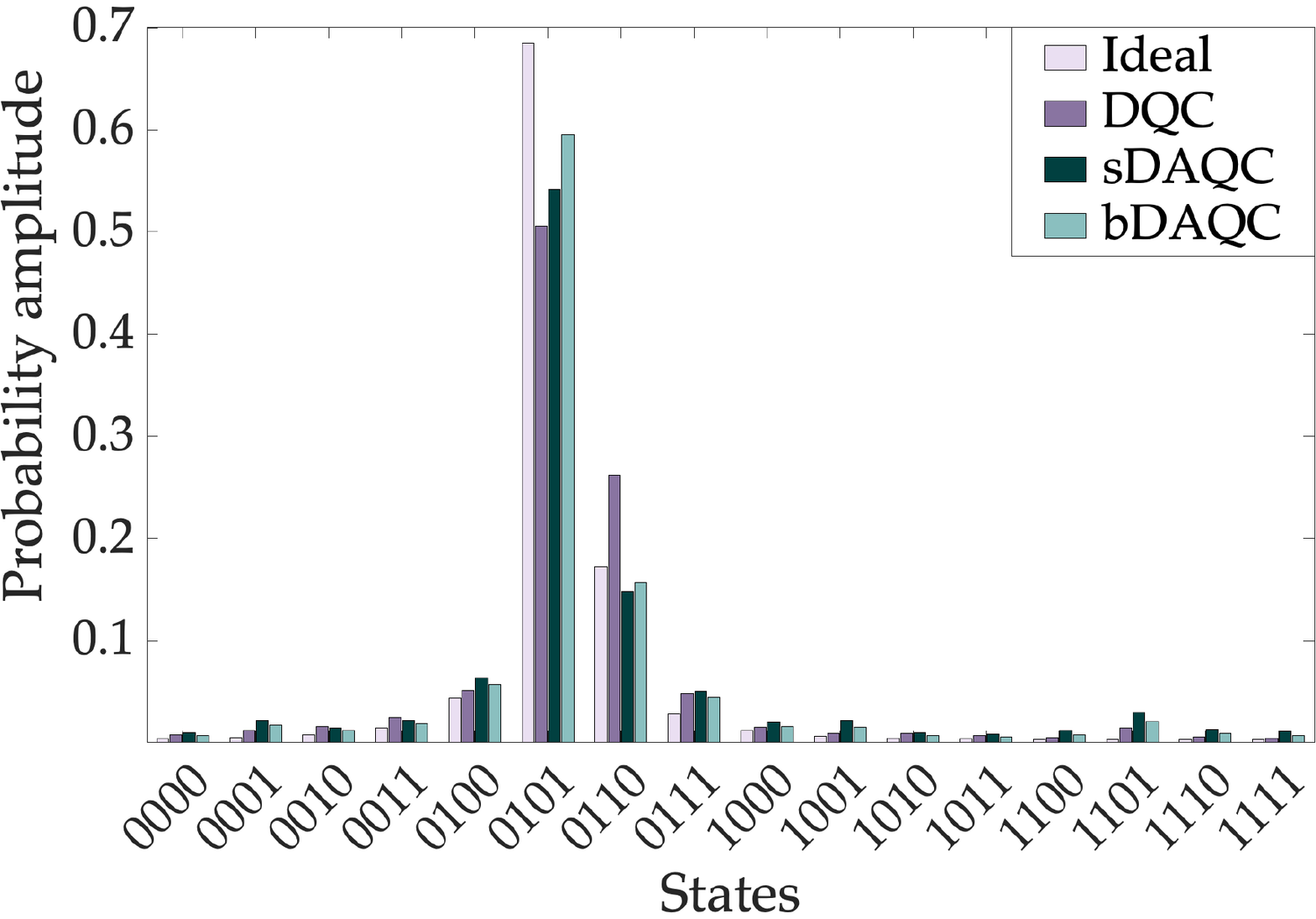} 
\caption{{\bf QPE results for DQC, sDAQC and bDAQC.} Probability amplitudes of the final states generated by the QPE algorithm for various paradigms and sources of noise, with $\varphi = 1/3$ and the unitary operator specified in equation \eqref{eq: QPE_matrix}}
\label{Fig5chap3}
\end{figure}

Our developed tool simulates noise sources in quantum computers similarly to the QFT simulation in Sec.\ \ref{sec31}, but with a more versatile approach for implementing different algorithms. It takes into account the residual signals in neighboring qubits during SQG pulses (cross-talk), leading to a bit-flip error on all qubits after each digital block.

Simulations were conducted using noise parameters based on current state-of-the-art quantum computers. Approximate values of execution times for IBM quantum computer's basis gates were used, with $t_{R_x}=10\text{ns}$, $t_{R_z}=1\text{ns}$, and $t_{\text{CNOT}}=300\text{ns}$. These values indicate that TQG execution times are one to two orders of magnitude larger than SQGs, consistent with increased noise from decoherence. The approximated error for these gates is $\text{TQGN}=0.08$, with $\text{CNOT}=e^{-isH_\text{CNOT}}$ and $s \in \mathcal{N}(\pi/2,\text{TQGN})$. The error was also estimated from real IBM devices. A 0.0001 bit-flip probability was chosen for SQGs and $p_\text{meas}=0.01$ for error measurement. Decoherence was considered with a $T_1$ time of $50\mu$s and thermal ground state population of $p=0.35$. Experimental errors defined in Ref. \cite{Martin2020} were simulated using $\text{SQGN}=0.0005$, $\text{sABN}=0.002$, and $\text{bABN}=0.001$. The coupling strength was set to $g_0=1$ MHz.

Using these parameters, we estimate the phase $\varphi$ such that $P(\varphi)|u\rangle=e^{i2\pi\varphi}|u\rangle$, choosing $|u\rangle = |1\rangle =(0 \quad 1)^T$. In ideal circumstances, Fig. \ref{Fig5chap3} shows the highest probabilities belong to the $0101$ and $0110$ states, leading to the approximations $\varphi_{0101}=5/16=0.3125$ and $\varphi_{0110}=6/16=0.3750$ with relative errors of $6.25 \% $ and $12.50 \%$ respectively, compared to the exact result $\varphi=1/3$. Results show that under noisy conditions, both DQC and DAQC can still recover the maximum probability states. Fidelity was calculated with $F(\rho_\text{ideal},\rho_\text{DQC})=0.88$, $F(\rho_\text{ideal},\rho_\text{bDAQC})=0.87$, and $F(\rho_\text{ideal},\rho_\text{sDAQC})=0.79$. DQC and bDAQC lead to similar fidelity values, but bDAQC shows better approximation of the solution, as maximum probability peaks are closer to the ideal.

If we calculate the weighted mean of the estimated phase, we obtain an estimate of phase of $\varphi_\text{bDAQC}=0.348\pm0.119$ for bDAQC, $\varphi_\text{sDAQC}=0.360\pm0.107$ for sDAQC and $\varphi_\text{DQC}=0.347\pm0.111$ for DQC. Using this figure of merit we conclude that the results for bDAQC and DQC are practically the same. However, if we use the majority voting as the estimation strategy, we can see that bDAQC outperforms the rest of circuits as we reach the closest approximation ($\varphi=5/16$) faster than with the other approaches. This aligns with the results obtained when implementing the QFT in the previous section.

\vspace{1cm}

The results of our simulations of DAQC quantum algorithms, which included realistic noise sources in quantum computers and residual signals during SQG pulses, are highly promising. These findings demonstrate the significant potential of the DAQC paradigm for quantum computation. In the following chapter, we outline the implementation of a quantum algorithm to solve linear system of equations, that leverages the QPE subroutine and the QFT.


\section[Digital-analog algorithm for solving linear systems of equations]{Digital-analog algorithm for solving \\ linear systems of equations}
\label{sec4}
\label{chapter4}

\fancyhead[RO]{4\quad DA QUANTUM ALGORITHM FOR SOLVING LINEAR SYSTEMS}

%
%

\thiswatermark{\put(1,-302){\color{l-grey}\rule{70pt}{60pt}}
\put(70,-302){\color{grey}\rule{297pt}{60pt}}}

%
%

\vfill
\lettrine[lines=2, findent=3pt,nindent=0pt]{I}n this chapter, we will discuss the Harrow-Hassidim-Lloyd (HHL) quantum algorithm for solving linear systems of equations using the QPE as a subroutine. Similar to our approach for the QFT, we will first describe the algorithm as a Hamiltonian, which can then be simulated using the DAQC paradigm.  We will mainly focus on the proposal for the DQC implementation of this algorithm since the DAQC simulations become trivial by using the toolkit presented in chapter \ref{chapter2}. The content of this chapter is based on the work developed in \cite{Martin2022}.

In Section \ref{sec41}, we will provide a theoretical overview of the HHL algorithm, discussing how it works and its theoretical underpinnings. In Section \ref{sec42}, we will focus on the implementation of the HHL algorithm using the DQC paradigm, describing the specific steps involved and discussing potential issues that may arise during implementation. In Section \ref{sec43}, we will present an extension of the HHL algorithm to the DAQC framework, which allows for even more efficient simulations of the algorithm. Finally, in Section \ref{sec44}, we will use the HHL algorithm as an example to demonstrate how experimental noise can be mitigated through the co-design of quantum processor architecture.

\subsection{Quantum algorithm for solving linear systems of equations}\label{sec41}

The Harrow-Hassidim-Lloyd (HHL) quantum algorithm was proposed in 2009 ~\cite{HHL2009} to solve linear systems of equations: given a matrix $A$ and a vector $\vec{b}$, find a vector $\vec{x}$ such that $A\vec{x} = \vec{b}$. Due to the significance of linear systems of equations in several fields of science and engineering, this algorithm has attracted significant attention. For a $s$-sparse system matrix of size $N\times N$ with condition number $\kappa$, which is given by the ratio of the maximal and minimal singular values of $A$, the HHL algorithm achieves a desired accuracy $\epsilon$ in a running time of $\mathcal{O}(\log (N) s^2\kappa^2 \epsilon)$, compared to the best known classical algorithm with a running time of $\mathcal{O}(N_s \kappa/\log \epsilon)$ ~\cite{A2015}.

The HHL algorithm requires three sets of qubits: an ancilla qubit to store the inverse of the eigenvalues of the matrix problem in its amplitude, a register of $n_R$ qubits to encode the binary representation of the eigenvalues of the problem matrix, and $n_M = \log_2(\dim |b\rangle)$ memory qubits to load $|b\rangle$ and store the output $|x\rangle$. The number of qubits needed depends on the size of the matrix $A$ and the desired precision.

\begin{figure}[t]
    \centering
    \includegraphics[width=0.98\textwidth]{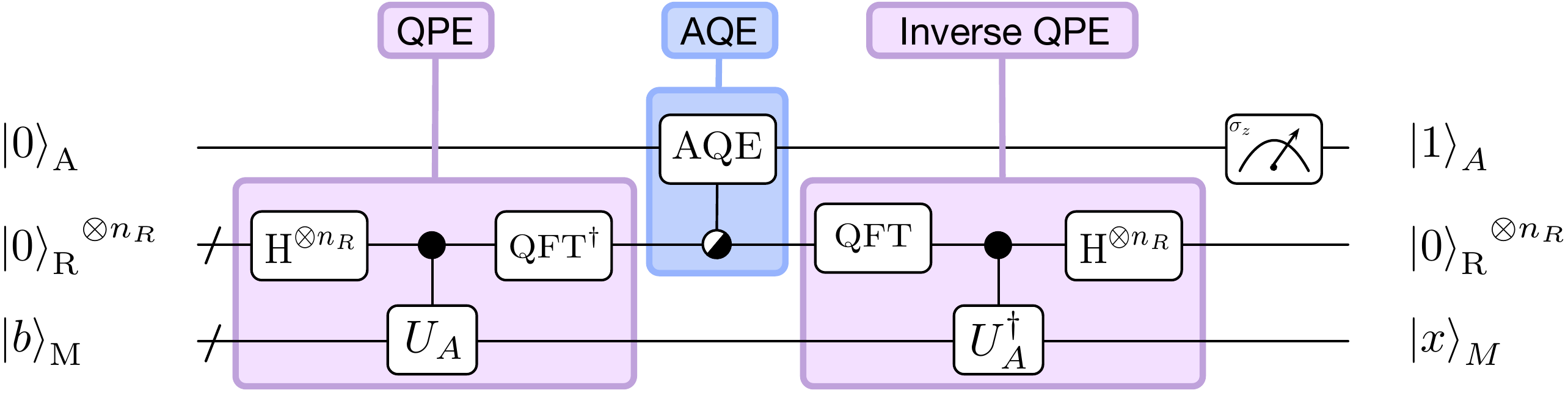}
    \caption{{\bf Quantum circuit for the digital implementation of the HHL.} The implementation of the HHL algorithm can be divided into three subroutines: QPE to compute the eigenvalues of the problem matrix $A$, AQE to store the inverse of the eigenvalues in the amplitude of the ancilla qubit, and inverse QPE to reset the registers back to the ground state $|0\rangle^{\otimes n_R}_\text{R}$.}
    \label{Fig1chap4}
\end{figure}

Additionally, the algorithm consists of three steps: QPE to compute the eigenvalues of $A$, ancilla quantum encoding (AQE) to map the inverse of the eigenvalues to the amplitude of the ancilla qubit, and inverse QPE to reset the registers back to the ground state $|0\rangle^{\otimes n_R}_R$. Fig.~\ref{Fig1chap4} shows a schematic representation of the algorithm, with the qubits divided into three sets and the steps delimited in colored boxes.

Since the HHL can be employed as a subroutine of a larger algorithm, the memory qubits might be initialized in the state $|b\rangle_\text{M} = \sum_{i=1}^N b_i|i\rangle$ as an outcome of the previous operations in the algorithm. If this is not the case, it is possible to prepare $|b\rangle$ using the procedure in Ref.~\cite{GR2002} if $b_i$ and $\sum_i |b_i|^2$ are efficiently computable. In any case, the system is initially in the state $|0\rangle_\text{A} \otimes |0\rangle_\text{R}^{\otimes n_R} \otimes |b\rangle_\text{M}$.

The first step of the algorithm employs the QPE to calculate the eigenvalues $\lambda_j$ of $A$ and encode them in binary form into the state of the register qubits. Given $|b\rangle=\sum_{j=1}^N\alpha_j|u_j\rangle$, where $|u_j\rangle$ is the eigenvector basis of matrix $A$, the state of the system after QPE is 
\begin{equation}
    |0\rangle_A \otimes \sum_{j=1}^N\sum_{k=0}^{2^{n_R}-1} \alpha_j\beta_{k|j}|k\rangle_R \otimes |u_j\rangle_M,
\end{equation}
with the coefficient $\beta_{k|j}$ defined as
\begin{equation}
    \beta_{k|j}=\frac{1}{2^{n_R}}\sum_{y=0}^{2^{n_R}-1}e^{2\pi i y(\lambda_j-k/2^{n_R})}.
\end{equation}
If the eigenvalues $\lambda_j$ are perfectly $n_R$-estimated, they can be relabeled as $\widetilde{\lambda_k} \equiv k/2^{n_R}$ such that $\beta_{k|j}=\delta_{\lambda_k,\lambda_j}$ and the final state of the system is 
\begin{equation}
    |0\rangle_\text{A}\otimes \sum_{j=1}^N\sum_{k=0}^{2^{n_R}-1} \alpha_j|\widetilde{\lambda_k}\rangle_\text{R} \otimes |u_j\rangle_\text{M}.
\end{equation}

After encoding the estimated eigenvalues in the register qubits, the AQE maps them into the amplitude of the ancilla qubit. The resulting state of the ancilla is 
\begin{equation}
\sum_{j=1}^N\sum_{k=0}^{2^{n_R}-1}\left(\sqrt{1-\frac{C^2}{\widetilde{\lambda_k^2}}}|0\rangle_\text{A}+\frac{C}{\widetilde{\lambda_k}}|1\rangle_\text{A}\right)\alpha_j|\widetilde{\lambda_k}\rangle|u_j\rangle,
\end{equation}
 with normalization constant $C\leqslant 1$ chosen to be $\mathcal{O}\left(1/\kappa\right)$.

Finally, the inverse QPE must be performed to uncompute the estimated eigenvalues $|\lambda_j\rangle$ on the register qubits and reset them to their initial state of $|0\rangle^{\otimes n_R}$. After this step, the resulting state of the system is given by
\begin{equation}\label{EQ_final_state}
\sum_{j=1}^N\sum_{k=0}^{2^{n_R}-1}
\left(\sqrt{1-\frac{C^2}{\widetilde{\lambda_k^2}}}|0\rangle_\text{A}+\frac{C}{\widetilde{\lambda_k}}|1\rangle_\text{A}\right)\otimes|0\rangle_\text{R}\otimes \alpha_j |u_j\rangle_\text{M}
\end{equation}

The ancillary qubit must be measured in the $Z$-basis to obtain the normalized solution of the linear equation. A successful measurement outcome of $|1\rangle_\text{A}$ results in the state
\begin{equation}\label{EQ_measurement}
\frac{1}{C}\sum_{j=1}^N\frac{\alpha_j}{\widetilde{\lambda_k}}|u_j\rangle_\text{M},
\end{equation}
representing the solution of the linear equation, up to a normalization factor $C$. The solution is a pure state when the matrix $A$ is perfectly $n_R$-estimated. However, if there is an eigenvalue of $A$ that is not perfectly estimated, the total state in Eq.\eqref{EQ_final_state} is an entangled state, causing the solution state in Eq.~\eqref{EQ_measurement} to be a mixed state.

\subsection{Proposal for digital implementation: the ancilla quantum encoding}\label{sec42}

The digital implementation of the direct and inverse QPE in this algorithm is relatively simple, but, to our knowledge, there is no known explicit implementation for the AQE subroutine. In this section, we propose a systematic method for implementing the AQE that is independent of the problem matrix $A$ and does not require classical hybridization, as found in the HHL algorithm. Due to its relevance, we focus on computing $f(\lambda) = 1/\lambda$, but this subroutine can perform any function $f(A)$ that satisfies the conditions outlined in Ref.~\cite{HHL2009}.

In the literature, the AQE routine has been circumvented by employing different tricks. For instance, by using previous knowledge about the eigenvalues of the matrix $A$. Then, it is straightforward to tailor the best set of rotations to map those values into the amplitude of the ancilla qubit. Several experimental implementations in different experimental platforms~\cite{Photonic_HHL_2013, NMH_HHL_2014, Photonic_HHL_2014, JWP2017} have followed this approach to solve $2\times 2$ linear system of equations. There is an alternative implementation of the HHL for which no spectral information of the matrix $A$ is required. In Ref.~\cite{LJL2019} a quantum-classical hybridization of the algorithm is proposed. They repeatedly perform the QPE to obtain a $n_R$-bit description of the eigenvalues of $A$. Then, they determine the simplest circuit implementation to perform the AQE, tailored for those eigenvalues. Once the AQE is determined, it is possible to perform the complete HHL algorithm. However, this alternative fails if the vector $|b\rangle$ is not efficiently prepared. Despite the intrinsic interest of these approaches, both of them jeopardize the advantage of the algorithm, since they presume previous knowledge about the eigenvalues of the matrix $A$.

\begin{figure}[t]
    \centering
    \includegraphics[width=0.98\textwidth]{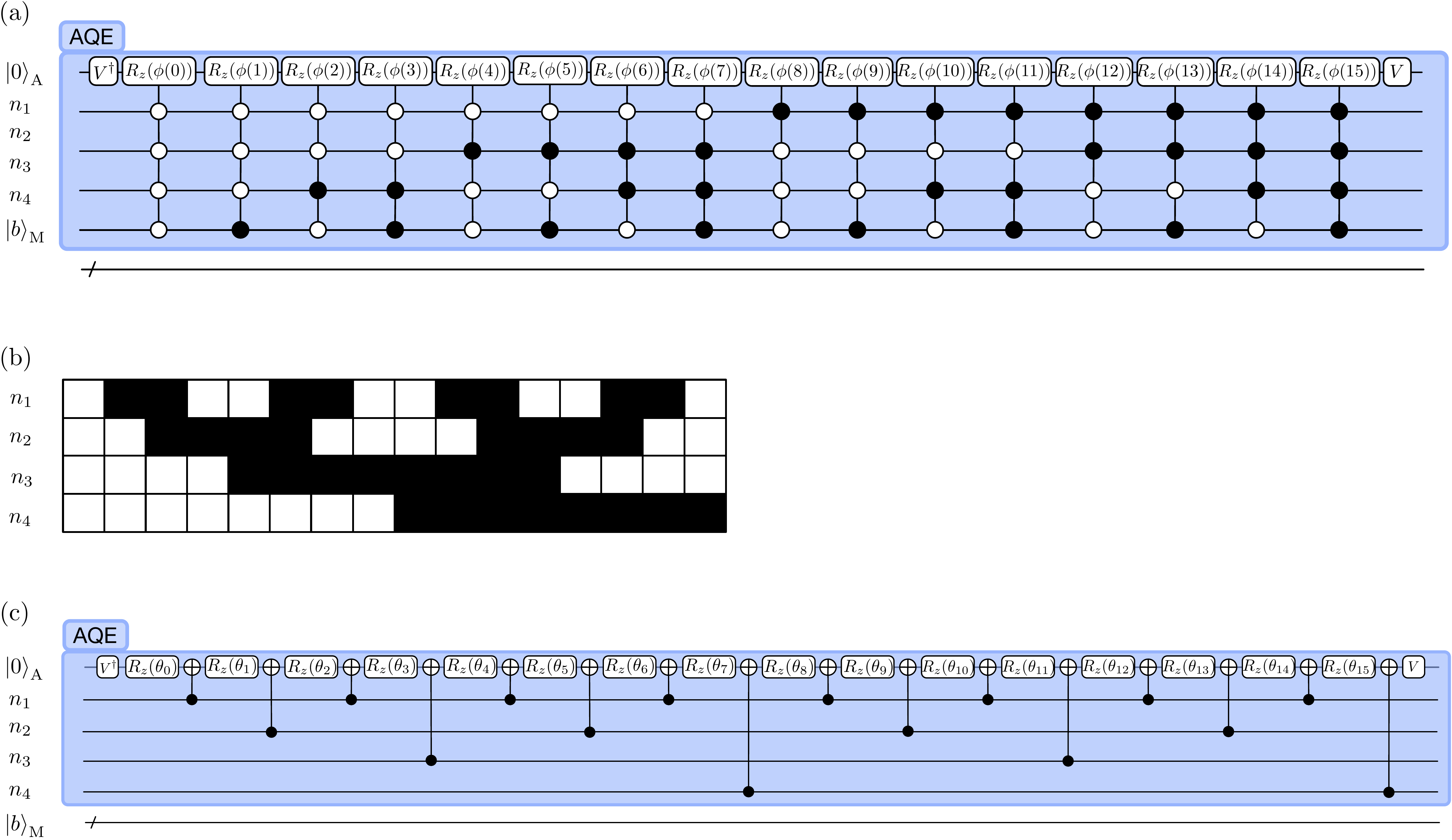}
    \caption{{\bf (a) Multi-qubit decomposition of the AQE step for a ${\bf n_R=4}$ qubits example.} The rotations $V$ and $R_z\left(\phi\left(p\right)\right)$ performed on the ancilla qubit are defined in Eqs.~\eqref{eqV} and~\eqref{eqRz}, respectively. {\bf (b) Binary reflected ${\bf 4}$-bit Gray code to define the positions of the control nodes.} The black and white rectangles refer to bit values one and zero, respectively. {\bf (c) AQE implementation using only single- and two-qubit gates.} Using the $4$-bit Gray code and following Ref.~\cite{MVBS2004}, it is possible to decompose the multi-qubit AQE implementation into single-qubit rotations around the $z$-axis and CNOTs. The value of each $\theta_i$ follows from Eq.~\eqref{eqThetaJ}.}
    \label{Fig2chap4}
\end{figure}

The AQE maps the superposition state of the registers into the amplitude of the ancilla qubit by means of the application of different controlled rotations. These rotation operations control each register qubit and act on the ancilla. To perform the AQE step, we will first rotate the system by applying the unitary gate $V$, which is defined as
\begin{equation}\label{eqV}
V = \frac{1}{\sqrt{2}}\begin{pmatrix}
-i & i\\
1 & 1
\end{pmatrix},
\end{equation}
and then, apply a series of rotations around the $Z$-axis in the ancilla qubit controlled by the state of the register qubits. This operation is defined as
\begin{equation}
U_{\text{AQE}} = \sum_{p=0}^{2^{n_R}-1} R_z\left(-\phi(p)\right)_\text{A} \otimes |\vec{b}_p\rangle\langle \vec{b}_p|_\text{R},
\end{equation}
where,
\begin{eqnarray}
R_z \left(\phi(p)\right) &=& \begin{pmatrix}
e^{-i\phi(p)/2} & 0\\
0 & e^{i\phi(p)/2}
\end{pmatrix}, \label{eqRz}\\
\phi(p) &=& \left\{ \begin{array}{lr}
0  & \text{if } p = 0\\
2 \arcsin\frac{1}{p} & \text{otherwise}
\end{array}\right. .
\end{eqnarray}
The bit-string $\vec{b}_p$ is the binary representation of the decimal number $p$, in other words, $p=\sum_{i=0}^{n_R} (\vec{b}_p)_i 2^{n_R-i}$. In Fig.~\ref{Fig2chap4}(a), we show an example of the AQE routine for a $n_R=4$ qubits case.

The AQE involves the application of several multi-controlled gates, leading to an exponential increase in the number of gates required for its implementation. According to Ref.~\cite{MVBS2004}, a general quantum gate acting on $n_R$ qubits, {\it i. e.} a unitary transformation with $4^{n_R}$ degrees of freedom, can be decomposed in $4^{n_R}$ single-qubit gates and $4^{n_R} - 2^{n_R+1}$ CNOT gates, which might be too demanding for NISQ devices. In our case, we employ $2^{n_R}$ multi-controlled gates with only $1$ degree of freedom for each of them to perform the AQE, which substantially reduces the number of single-qubit and CNOT gates required to decompose them. Following Ref. \cite{MVBS2004}, our $2^{n_R}$ multi-controlled gates can be decomposed into $2^{n_R}$ single qubit gates plus $2^{n_R}$ CNOT gates. Therefore, the depth of the AQE step depends  exponentially on the number of register qubits. This number is related to the precision and error of the HHL algorithm, as discussed in Ref. \cite{HHL2009}.

In order to implement the multi-controlled operations using only single- and two-qubit gates acting on the ancilla, we adopt the approach from Ref.\cite{MVBS2004}. They provide an equivalent circuit that performs subsequent rotations with modified angles, alternating with CNOT gates between control and target qubits. As a result, the operation can be implemented with the circuit depicted in Fig. \ref{Fig2chap4}(c). Here, the angles $\theta_i$ are the modified angles of the single-qubit rotations, related to the original angles by
\begin{equation}
\mqty(\theta_1 \\ \theta_2 \\ \vdots \\ \theta_{2^{n_R}})
=
M^{-1} 
\mqty(\phi (0) \\ \phi(1) \\ \vdots \\ \phi(2^{n_R}-1)),
\end{equation}
where $M$ and $M^{-1}$ are defined as:
\begin{equation}
M_{ij} = (-1)^{\text{bin}(i-1) \cdot g(j-1)}
\text{ and }
M^{-1} = \frac{1}{2^{n_R}} M^{T}.
\end{equation}
Here, $\text{bin}(i)$ is the $n_R$-bit binary representation of the integer $i$, $g(j)$ is the $j$-th string of the binary reflected Gray code (counting from 0), and the dot represents the bit-wise product. By substituting the expression for $\phi(b)$, then
\begin{equation}\label{eqThetaJ}
\theta_i = \frac{1}{2^{n_R}} \sum_{j=1}^{2^{n_R}-1} (-1)^{\text{bin}(j) \cdot g(i-1)} \arcsin \frac{1}{j}.
\end{equation}

The accuracy of the algorithm's output in comparison with the actual solution $|x\rangle$ of the equation $A|x\rangle = |b\rangle$ is expressed as $|| |x\rangle - |\tilde{x}\rangle || = \mathcal{O}(\kappa/2^{n_R})$, where $|\tilde{x}\rangle$ is the solution obtained by the algorithm and $\kappa$ is the condition number of the problem matrix $A$. To estimate the expected value of an observable $\Omega$ with a limited number of samples $N_s$, the error of the estimation scales as $\mathcal{O}(1/\sqrt{N_s})$. Therefore, the overall expected error, $\epsilon$, of estimating $\Omega$ in $|x\rangle$ by measuring $|\tilde{x}\rangle$ is
\begin{equation}\label{eq12}
\epsilon = \mathcal{O}\left(\frac{1}{\sqrt{N_s}} + \frac{\eta}{2^{n_R}}\right),
\end{equation}
where $\eta$ is a constant of $\mathcal{O}(\kappa)$. To determine the optimal number of register qubits $n_R$, we must ensure that no term in Eq.~\eqref{eq12} is dominant. This implies that the sensible amount of register qubits $n_R$ is of $\mathcal{O}\left(\log\left[\kappa \sqrt{N_s}\right]\right)$, which only depends on the number of samples, but not on the size of the problem matrix $A$.

The {\it generalized}-HHL (gHHL) algorithm aims to compute $\vec{x} = f(A) \vec{b}$, where $f(A)$ is a matrix function. The method can be easily adapted to perform this task. The AQE is used to perform the following operation on the ancilla qubit,
\begin{equation}
|0\rangle_\text{A} |\widetilde{\lambda_k}\rangle_\text{R} \xrightarrow{\text{AQE}} \left(\sqrt{1-f(\lambda^2)}|0\rangle_\text{A} + f(\lambda)|1\rangle_\text{A}\right)|\widetilde{\lambda_k}\rangle_\text{R}.
\end{equation}
To accomplish this, we set the rotation angles of the multi-controlled operations to $\phi(p) = 2 \arcsin f(p/2^{n_R})$. However, the error bounds are only met if the function $f(\lambda)$ satisfies the condition
\begin{equation}
\left| \left( \frac{\text{d}f}{\text{d}x}\right)^2 \left(1+\frac{1}{1-f(x)^2}\right)\right|<\eta^2 \quad \text{in the interval } x\in [0,1),
\end{equation}
which can be derived from Lemma 2 in the supplementary material of Ref.~\cite{HHL2009}. Here, $\eta$ is an arbitrary constant of order $\mathcal{O}(1)$. This implies that, in practice, one would need to construct an auxiliary function $F$ that satisfies the bound and behaves similarly to $f(\lambda)$ in a given interval, similar to normalizing a vector for its use as a quantum state.

\subsection{Digital-analog implementation}\label{sec43}

As previously mentioned, the HHL algorithm can be broken down into three steps (QPE, AQE, and inverse QPE). QPE itself can be further divided into two subroutines: controlled-Hamiltonian evolution and inverse quantum Fourier transform. Dividing each step into smaller subroutines makes it easier to describe the HHL algorithm within the DAQC framework, as some of these subroutines have already been thoroughly discussed in previous chapters.

To perform the controlled-Hamiltonian evolution, which is the first part of the QPE, we assume that the problem matrix $A$ has an $M$-body decomposition, where $M << \log_2 \dim (A)$. This allows the matrix to be efficiently uploaded using the DAQC protocol, as described in Ref. \cite{Parra2020}. The next step in the QPE subroutine is the implementation of the inverse quantum Fourier transform on the register qubits, which was discussed in detail in chapter \ref{chapter3}.

\begin{figure}[t]
    \centering
    \includegraphics[width=0.94\textwidth]{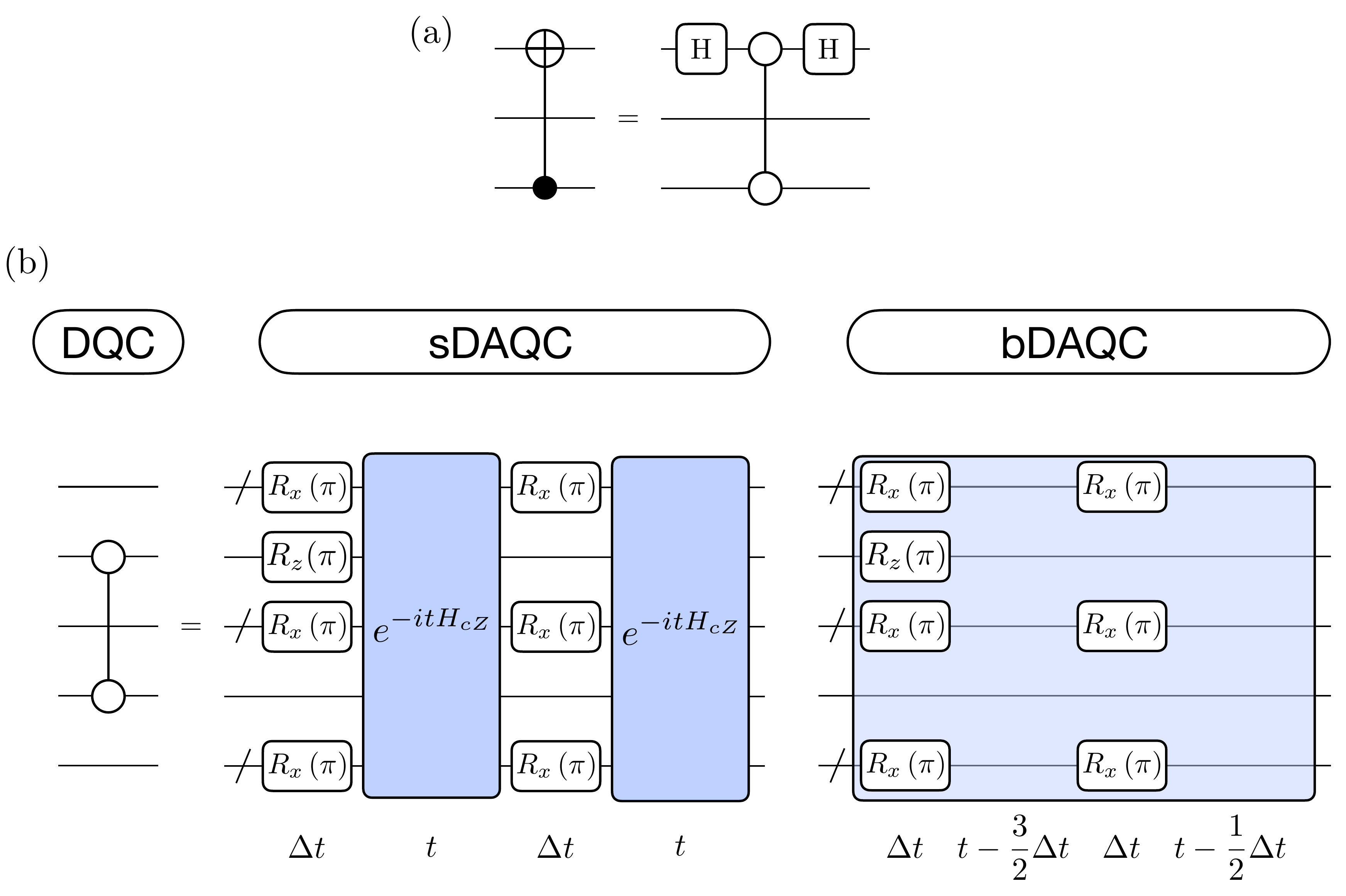}
    \caption{{\bf(a) Decomposition of a CNOT gate in terms of Hadamard and cZ gates.} {\bf (b) DAQC decomposition of a $cZ$ gate.} To perform a $cZ$ gate using sDAQC, two analog blocks with duration $t=\pi/8$ are necessary. For the digital blocks, all qubits except those in the $cZ$-gate must undergo a $\pi$-radian rotation around the $x$-axis before and after the first analog blocks. In the case of the $cZ$-gate qubits, a rotation around the $z$-axis should be applied either before or after the analog block, as the interaction Hamiltonian commutes with the $R_z(\pi)$ gate. In bDAQC, the single-qubit rotations are performed after the interaction.}
    \label{Fig3chap4}
\end{figure}

To implement the AQE step under the DAQC paradigm, we first decompose it into single-qubit rotations (Hadamard gates and rotations around the X-axis) and controlled-Z (cZ) gates, as shown in Fig. \ref{Fig3chap4}(a). To obtain a sDAQC decomposition, we construct the Hamiltonian of the cZ gates, which takes the form of a two-body Ising Hamiltonian:
\begin{equation}
H_\text{cZ} = -\frac{1}{2}\sigma_z\otimes\mathbb{I}+\frac{1}{4} \sigma_z\otimes\sigma_z.
\end{equation}
The cZ gate can be implemented using either the step-wise or the bang approach of the DAQC framework. Fig. \ref{Fig3chap4}(b) shows a scheme of both protocols for the cZ gate.

Without loss of generality, we assume that the interaction between the qubits of our quantum processor is a homogeneous two-body Ising Hamiltonian, $H_{\text{int}}$. Each $cZ$ gate performed between the ancilla and one of the register qubits can then be decomposed into two analog blocks as
\begin{equation}
U_\text{cZ} = e^{i\frac{\pi}{4} Z_\text{A}\otimes Z_\text{R}}=\left( \bigotimes_{\substack{k=1\\ k\neq i}}^{n_R}X_k\right) e^{i\frac{\pi}{8}H_I}  \left( \bigotimes_{\substack{k=1\\ k\neq i}}^{n_R}X_k\right) e^{i\frac{\pi}{8}H_I}.
\end{equation}
This decomposition is illustrated in Fig. \ref{Fig3chap4}(b) with the sDAQC implementation of a $cZ$ gate.

To implement the complete AQE in the DAQC paradigm, we decompose every $cZ$ into digital and analog blocks, as previously described. The analog blocks take the same amount of time $t$ to act, and the digital steps take a fixed time $\Delta t$ to be applied, as illustrated in Fig. \ref{Fig3chap4}(b). As it has been repeatedly mentioned in this thesis, in terms of scalability it is preferable to follow the bDAQC approach. The explicit DAQC description of the AQE step includes at most $2^{n_R}$ analog blocks and $(n_R + 1) 2^{n_R} + 1$ single-qubit gates, which can be improved with optimization techniques.

\subsection{Co-designed processor architecture for the HHL algorithm}\label{sec44}

\begin{figure}[t]
    \centering
    \includegraphics[width=0.70\textwidth]{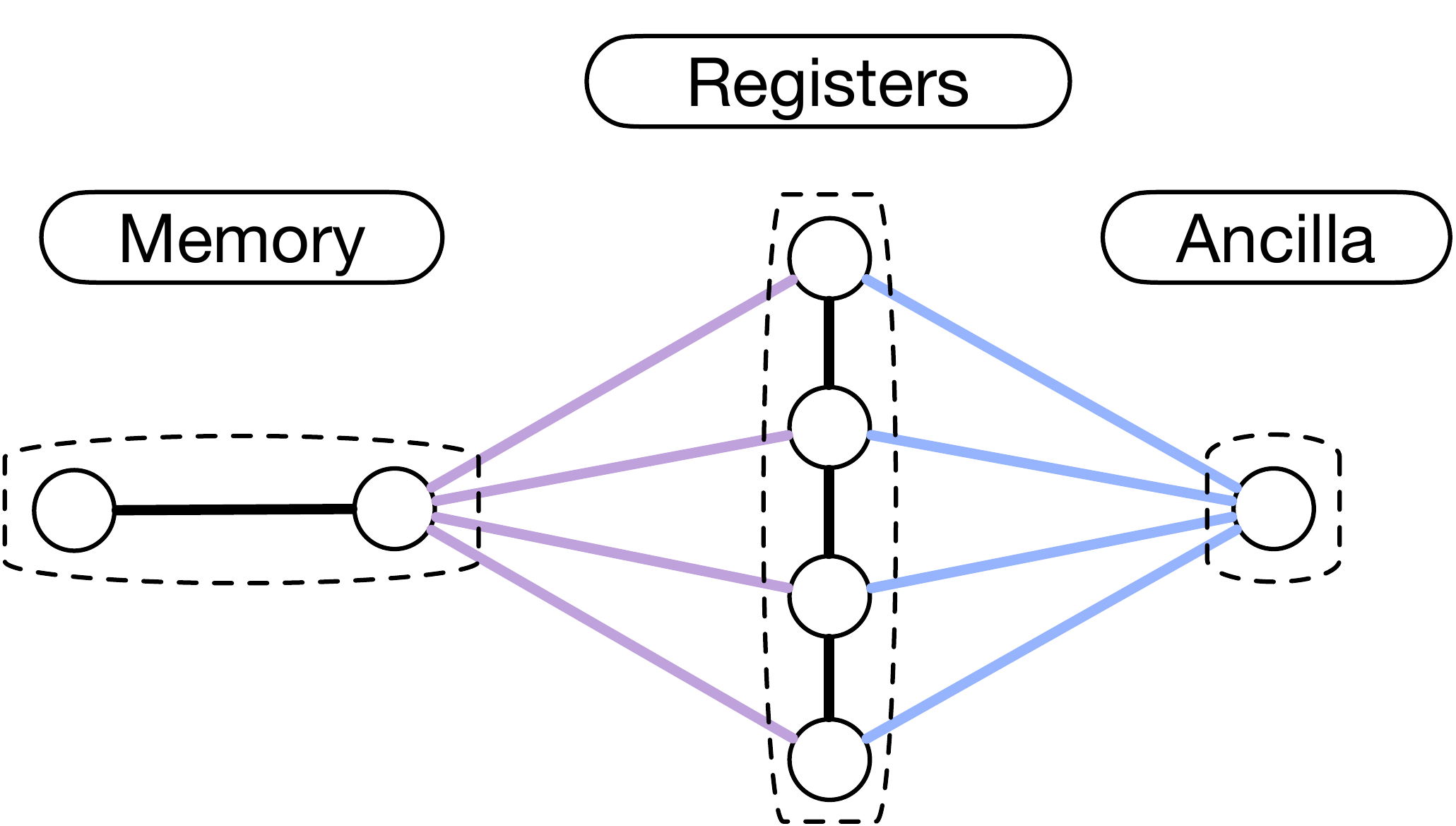}
    \caption{ {\bf Kite-like connectivity of the quantum processor.} The register qubits must be interconnected with each other, the ancilla qubit, and the memory qubits. However, the ancilla and memory qubits do not require a direct connection.}
    \label{Fig4chap4}
\end{figure}

The concept of co-design in quantum computing involves the simultaneous design of a quantum processor together with the implementation of a specific algorithm. This approach leverages the unique features and limitations of a quantum platform to optimize both the architecture and implementation of the algorithm, resulting in a more efficient and effective system. In particular, for the implementation of the HHL algorithm on NISQ devices, in Ref. \cite{Martin2022} we propose a co-designed ``kite-like" architecture that reduces the number of required SWAP gates.

The HHL algorithm requires specific connections between its three sets of qubits: the register, ancilla, and memory qubits. This co-designed architecture, shown in Fig.~\ref{Fig4chap4}, takes into account these connections and implements a kite-like structure that satisfies the requirements of the different steps in the HHL algorithm: the register qubits are connected among them and, simultaneously, to the ancilla and memory qubits, whereas the ancilla and the memory sets need not be connected directly. This optimization of the architecture leads to improved performance and reduced computational overhead.

It is important to note that the term ``qubits'' in this context refer to logical qubits, which may be constituted by a group of physical qubits that work together to perform the necessary operations and connectivity. By customizing both the architecture and implementation of the HHL algorithm for the specific features of the quantum platform, the co-design approach aims to maximize the performance and efficiency of the system.

\vspace{1cm}
With the conclusion of this chapter, we bring to an end the matryoshka construction of algorithms that we started with the QFT in chapter \ref{chapter3}. In the next chapter, we will examine the application of the DAQC framework to variational quantum algorithms.


\section[Digital-analog quantum approximate optimization algorithm]{Digital-analog quantum\\ approximate optimization algorithm}
\label{sec5}
\label{chapter5}

\fancyhead[RO]{5\quad DA QUANTUM APPROXIMATE OPTIMIZATION ALGORITHM}

%
%

\thiswatermark{\put(1,-302){\color{l-grey}\rule{70pt}{60pt}}
\put(70,-302){\color{grey}\rule{297pt}{60pt}}}

%
%

\vfill

\lettrine[lines=2, findent=3pt,nindent=0pt]{I}{n} this chapter, we will show how the DAQC paradigm can also be used with variational algorithms. These algorithms utilize classical optimization techniques and parametrized quantum circuits to increase the applicability of noisy quantum processors that have few qubits. One such algorithm that is most likely to prove useful in the near term is the quantum approximate optimization algorithm (QAOA) \cite{Farhi2014}. This chapter contains the work developed in Ref. \cite{Headley2020}, in which we show that the QAOA is a natural algorithm for the DAQC paradigm.

In Section \ref{sec51}, we introduce and thoroughly explain the QAOA. In Section \ref{sec52}, we expand the algorithm to the DAQC framework. Finally, in Section \ref{sec53}, we compare the performance of the QAOA algorithm under both the DQC and DAQC paradigms.

\subsection{The Quantum Approximate Optimization Algorithm} \label{sec51}

The QAOA algorithm is a hybrid quantum-classical algorithm that operates in discrete time and is utilized for computing solutions to combinatorial optimization problems. Originally, the algorithm was found to produce superior approximation ratios compared to the best classical algorithm for solving MAX-E3LIN2 problems \cite{Farhi2014_2}. However, this result was later surpassed by a quantum-inspired classical algorithm \cite{Barak2015}. Furthermore, research by Jiang \textit{et al.} \cite{Jiang2017} has demonstrated that QAOA can replicate the square root scaling of Grover's search algorithm, achieving the same speed-up without the need for the Grover mixing operator. Additionally, Hadfield \textit{et al.} have discovered that QAOA driving operators can be adjusted in such a way that a diverse range of problems can be solved without resorting to high-order penalty terms, which are typically utilized in annealing- or adiabatic-based approaches \cite{hadfield2019quantum}.

\begin{figure}[t]
    \centering
    \includegraphics[width=0.98\textwidth]{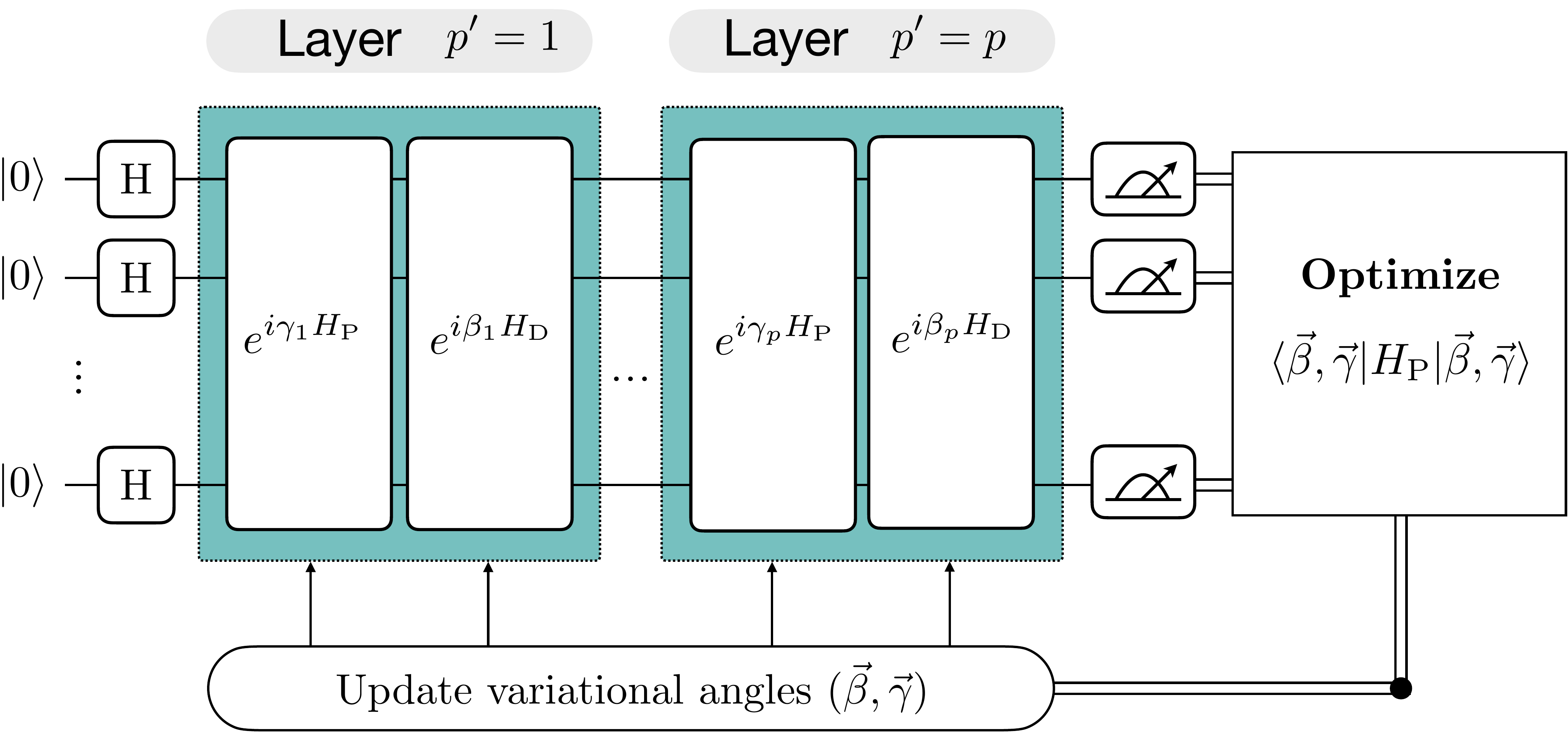}
    \caption{{\bf QAOA state preparation circuit.} The algorithm alternates the application of $H_\text{P}$ and $H_\text{D}$ Hamiltonians to an equal superposition of qubits. The resulting qubit states are then measured and a cost is calculated. A classical optimization algorithm is utilized to minimize this cost by varying the angles $\gamma$ and $\beta$.}
    \label{Fig1chap5}
\end{figure}

QAOA is a discrete-time hybrid quantum-classical algorithm for computing solutions to problems in combinatorial optimization. The classical optimizer tunes the $2p$ parameters $\vec \gamma, \vec \beta$ of a quantum circuit to maximize an objective function corresponding to high-quality solutions of a combinatorial optimization problem. In QAOA, an ansatz state 
\begin{equation}
\ket{\vec \beta,\vec\gamma}  =\prod_{p'=1}^p e^{i\beta_{p'}H_{\rm D}} e^{i\gamma_{p'} H_{\rm P}}\ket+^{\otimes n} \label{QAOA_state}
\end{equation}
is generated on a quantum processor using $p$ repetitions of two Hamiltonians: a problem Hamiltonian $H_{\text{P}}$ and a driver Hamiltonian $H_{\text{D}}$. A schematic quantum circuit of the algorithm is shown in Fig. \ref{Fig1chap5}. The problem and the driver Hamiltonians are defined as
\begin{equation}
    H_{\text{P}} = \sum_{z=0}^{2^n-1} C(z)|z\rangle\langle z|, \quad \text{and } H_{D} = \sum_{i=1}^n\sigma_x^{(i)}, 
\end{equation}
where $C$ is the objective function value for input strings $z$, and $\sigma_x^{(i)}$ refers to the Pauli matrix applied to the $i$-th qubit of the system. $H_\text{P}$ is chosen for its simplicity since it is a non-interacting Hamiltonian, while still enabling population transfer between any two given states.

To find a solution, the QAOA circuit is repeatedly executed, and the output string is measured to calculate the expectation value of $H_{\text{P}}$ under the QAOA ansatz state for the current parameters, as shown in Figure \ref{Fig1chap5}. This value is handed to a classical optimizer, with the goal of producing new parameters through a classical black-box optimization strategy. The expectation value of the problem Hamiltonian,
\begin{equation}
    \langle H_{\text{P}} \rangle_{\vec \beta,\vec\gamma} = \langle\vec \beta,\vec\gamma | H_{\text{P}} |\vec \beta,\vec\gamma \rangle,
\end{equation}
is then recomputed, and the process is repeated until a satisfactory solution to the problem is found. The success of QAOA is measured by the {\it mean approximation ratio}, defined as
\begin{equation}
\frac{\langle H_{\text{P}} \rangle_{\vec \beta,\vec\gamma}}{\max_{\psi}\langle\psi|H_{\text{P}}|\psi\rangle}.
\end{equation}

Combinatorial optimization problems with clauses encompassing at most two bits can be expressed in terms of two-qubit $ZZ$ interactions and single-qubit-$Z$ rotations. Problems in which the clauses are local to more bits require higher order terms and are therefore generally out of reach of NISQ quantum computers. In the literature, there are two well-studied optimization problems, namely MAX-CUT and MAX-2-SAT, that involve only terms of order up to $2$. MAX-CUT is defined on a problem graph whose vertices are binary variables, and the goal is to partition the graph in a way that maximizes the number of edges between different parts of the partition. The clauses of the problem or edges of the equivalent graph are of the type XOR between problem variables. XOR admits the truth table $00,01,10,11 \to 0,1,1,0$ which can be decomposed into a $Z$-based Hamiltonian following theorem 10 of Ref. \cite{hadfield2018quantum}. A MAX-CUT clause, therefore, manifests in the problem Hamiltonian as 
\begin{equation}
H_{\text{C,}jk} = \frac{1}{2}\left(\mathbb{I} - \sigma_z^{(j)}\sigma_z^{(k)}\right) = \begin{pmatrix}
0 & 0 & 0 & 0\\
0 & 1 & 0 & 0 \\
0 & 0 & 1 & 0 \\
0 & 0 & 0 & 0
\end{pmatrix}.
\end{equation}
The identity in this expression has no effect other than to keep the Hamiltonian non-negative such that the diagonal corresponds to the number of edges a given allocation cuts.

\subsection{Digital-Analog Quantum Approximate Optimization Algorithm} \label{sec52}

\begin{figure}
    \centering
    \includegraphics[width=0.98\textwidth]{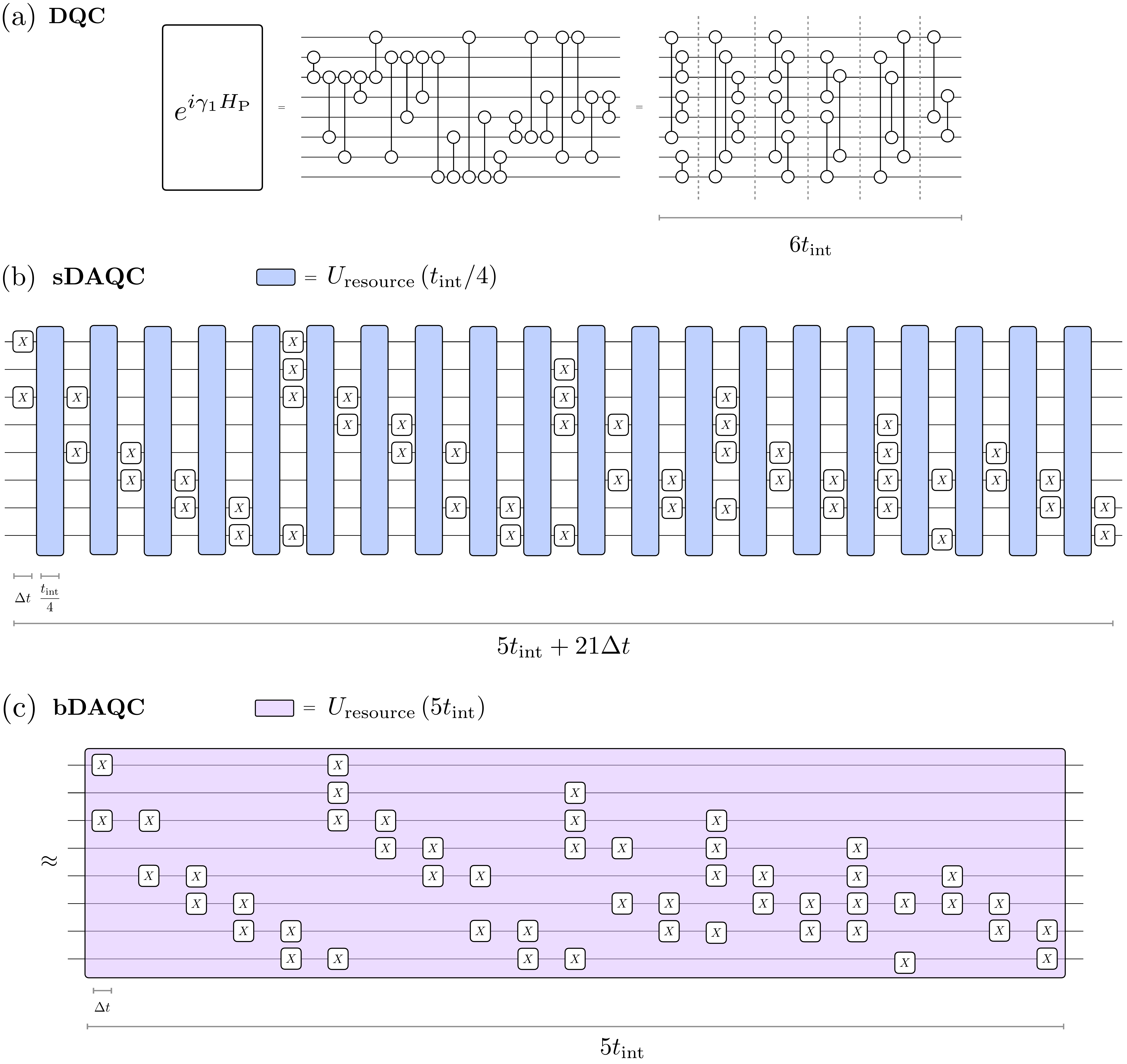}
    \caption{{\bf Circuit compilations for an 8-qubit MAX-CUT problem on a 5-regular graph. (a) DQC decomposition of the MAX-CUT problem Hamiltonian into $\sigma_z\sigma_z$ interactions.} The lines connected to two open circles represent $\sigma_z\sigma_z$ interactions applied for time $t_\text{int}$. On the right-hand side, we show how to parallelize the circuit into six time steps, each of which sees one qubit interact with only one other qubit at a time. Six time steps correspond to the maximum degree of a vertex plus one. {\bf (b) sDAQC decomposition of the same circuit.} A resource all-to-all homogeneous Ising Hamiltonian is turned on and off, punctuated by single-qubit gates. The $X$ gate corresponds to a rotation of $\pi$ radians around the $x$-axis and taxes $\Delta t$ time to be applied. This decomposition requires 20 uses of the resource Hamiltonian and 50 single-qubit operations. The time taken to apply the problem Hamiltonian is $5t_\text{int} + 21\Delta t$. {\bf (c) Decomposition of the same problem into the bDAQC scheme.} The resource Hamiltonian is turned on throughout the entire procedure. This circuit only approximates the time evolution invoked by the QAOA problem Hamiltonian but can be carried out in time $5t_\text{int}$ and also with 49 single-qubit $X$ operations. In the limit of infinitely fast $X$ gates, (c) is equivalent to (a) and (b).}
    \label{Fig2chap5}
\end{figure}

Let us assume that the device Hamiltoninan is as follows
\begin{equation}
    H_{\rm{Device}} (t) = f(t)H_{\rm{int}}+\alpha\sum_{i=1}^n \left( x_i(t)\sigma_x^{(i)}+z_i(t)\sigma_z^{(i)}\right),
\end{equation}
with
\begin{equation}
    H_{\rm{int}} = \sum_{j<k}^{n} r_{jk}\sigma_z^{(j)}\sigma_z^{(k)}.
\end{equation}
Meaning that the interaction Hamiltonian of the quantum processor of interest is $H_{\rm{int}}$ and that we can apply rotations around $x$- and $z$-axis.

In the sDAQC scheme, we assume control over the parameters $f$, $x_i$, and $z_i$, each of them taking values from $\{0,1\}$. In the bDAQC scheme, $f$ is always set to $1$ and only the single-qubit parameters may be altered. The single-qubit terms are stronger than the interaction Hamiltonian $H_{\rm{int}}$ by the factor $\alpha \geq 1$ and, in typical applications, $\alpha$ is expected to fall between $10-1000$ depending on architecture \cite{ballance2016high,linke2017experimental}. Though current devices tend to exhibit a ratio of single-qubit rotation speed to interaction strength at the lowest end of this range, they have little to gain from faster single-qubit operations, since they are typically limited by two-qubit interaction times and fidelity. We, therefore, expect that a device optimizing for DA applications could be engineered for greatly higher ratios $\alpha$. During the driving in bDAQC, all single-qubit $x_i$ operations are set to $1$, $z_i$ terms to $0$, and the interaction Hamiltonian is also active for the driver step, giving a driver Hamiltonian of
\begin{equation}
H_{\rm{bDAQC-Driver}} = \sum_{j<k} r_{jk}\sigma_z^{(j)}\sigma_z^{(k)} + \alpha \sum_{i=1}^n\sigma_x^{(i)}
\end{equation}
applied for device time given by the variational parameter $\beta$ divided by the driver strength $\alpha$ with $\beta \in [0,\pi]$. During the analog block steering operations, we use a similar Hamiltonian in which only a specific set of single-qubit $x_i$ terms are active. As described in chapter \ref{chapter2}, we wish to implement a full $X$-gate before and after each interaction block. The time to apply this gate will be $\Delta t = \frac\pi{\alpha}$. Applying the digital-analog QAOA (DA-QAOA) device Hamiltonian for a single QAOA layer thus affects the following unitary
\begin{equation}
    U_{\rm DA-QAOA} = \mathcal{T} \exp(-i\int_{t = 0}^{t_{\text{total}}}  H_{\rm{Device}} (t)\,\,dt), \label{DAQAOA_UNITARY}
\end{equation}
with $\vec x(t)$ defined by the matrix inversion procedure described in chapter \ref{chapter2}, $\mathcal{T}$ is the time-ordering meta-operator and $\vec z(t)$ used in the case that we are solving a SAT problem. $t_{\rm total}$ is the sum of all times in the non-negative DAQC time vector multiplied by the variational parameter $\gamma$ in addition to the driving time $\beta / \alpha$. A depiction of this problem Hamiltonian used to apply a MAX-CUT problem Hamiltonian is presented in Fig. \ref{Fig2chap5}. It represent the implementation of $1$ layer of the problem Hamiltonian $H_{\rm{P}}$. To complete the layer, SQG from the driver Hamiltonian are required.

\subsection{Performance of the QAOA under different computational paradigms}\label{sec53}

In this section, we present the main computational results of the work in Ref. \cite{Headley2020}. In section \ref{sec531}, we compare the performance of bDA-QAOA in comparison with the standard QAOA circuit for a set of chosen randomly generated problems. Afterward, in section \ref{sec532} we analyze this further and show the benefits of the DAQC framework in the application of variational quantum algorithms. Finally, in section \ref{sec533}, we examine the impact of experimental errors inherent to NISQ computing on the DAQC framework when utilizing variational algorithms such as the QAOA.

\subsubsection{Performance of the bDA-QAOA}\label{sec531}

\begin{figure}[t]
    \centering
    \includegraphics[width=0.98\textwidth]{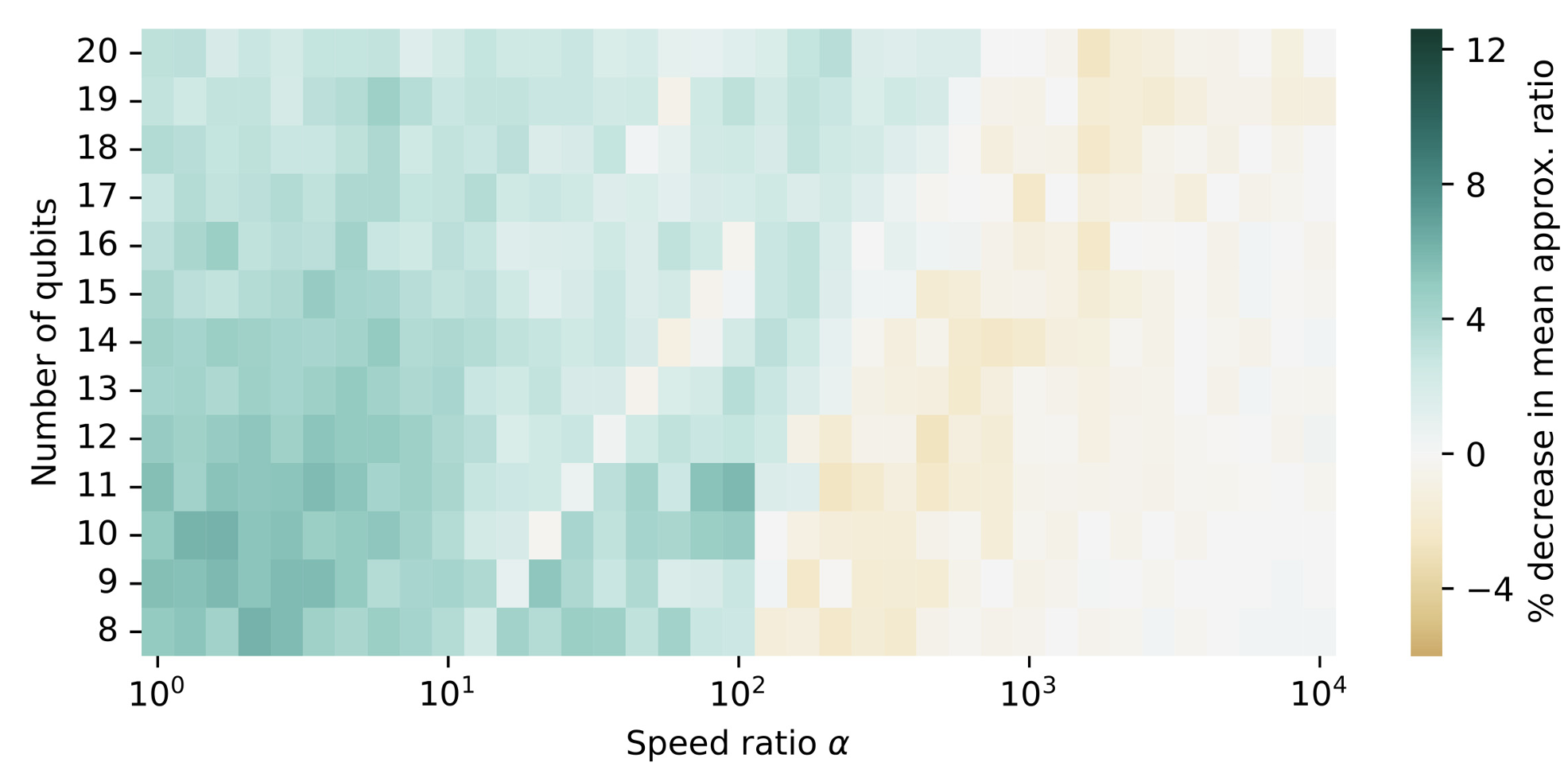}
    \caption{{\bf Average percentage difference between the mean approximation ratio of bDA-QAOA and error-free QAOA.} We have calculated it by averaging over $50$ randomly generated MAX-CUT problems with a constant filling factor of $p_{\rm clause} = 0.7$. Blue (dark region on the far left side of the figure) colors indicate that the bDA-QAOA ansatz state is worse than that provided by error-free QAOA, whereas shades of brown (the light gray region immediately left of the white region of zero error) indicate an improvement of the bDAQC over error-free QAOA. On the $x$-axis, the ratio $\alpha$ of single-qubit to problem Hamiltonian term strength is seen where error-free QAOA exists in the limit as this ratio becomes infinite.}
    \label{Fig3chap5}
\end{figure}

In bDA-QAOA, we perform QAOA using the ansatz state prepared by applying QAOA layers of the form described in equation \eqref{DAQAOA_UNITARY} as
\begin{equation}
    |\vec \beta,\vec\gamma\rangle^{\alpha,\rm{DA}} = U_{\alpha\textrm{-DA-QAOA}}|+\rangle^{\otimes n}.
\end{equation}

The bDA-QAOA algorithm introduces errors through the misspecification of the problem and driver Hamiltonians. Among these two, the misspecification of the problem Hamiltonian is likely to cause more significant error, as it requires a longer time to perform on the device and it is not tailored to a specific problem. This problem is not unique to quantum optimization and is known in quantum annealing literature as ``$J$ chaos'' where key aspects of the problem are not accurately represented in the dynamics of the annealing device. Without error mitigation strategies, these issues can severely impact the performance of adiabatic quantum computing \cite{pearson2019analog}.

bDA-QAOA is related to quantum random walk algorithms \cite{callison2019finding} and adiabatic quantum computing in that both the problem Hamiltonian and single-qubit driving operators are executed simultaneously. Therefore, one might expect that running the problem Hamiltonian alongside the driver in QAOA would not have a negative effect, as quantum random walks and diabatic quantum computing are active research areas \cite{morley2019quantum, Albash2015}. Simulations of bDA-QAOA, where the interaction Hamiltonian is identical to the problem Hamiltonian, showed no significant detrimental impact when compared to error-free QAOA, as no DA steering single-qubit operations were required.

However, coherent errors in the bDA-QAOA algorithm, which utilizes a non-problem-specific interaction Hamiltonian, are expected to be more detrimental than errors in QAOA with an always-on active problem Hamiltonian. These errors will result in a less problem-specific QAOA ansatz state, leading to a lower expected approximation ratio at a given depth. Figure \ref{Fig3chap5} displays the mean approximation ratio achieved by the bDA-QAOA ansatz state. For high values of $\alpha$, bDA-QAOA performs similarly to error-free QAOA, as expected. In the intermediate regime, minor increases in the mean approximation ratio are observed. Conversely, in the case of low $\alpha$, we observe consistently worse results for bDA-QAOA, due to problem misspecification induced by coherent DAQC errors.

\subsubsection{Variational resilience of DA-QAOA to digital-analog errors}\label{sec532}

\begin{figure}[t]
    \centering
    \includegraphics[width=0.98\textwidth]{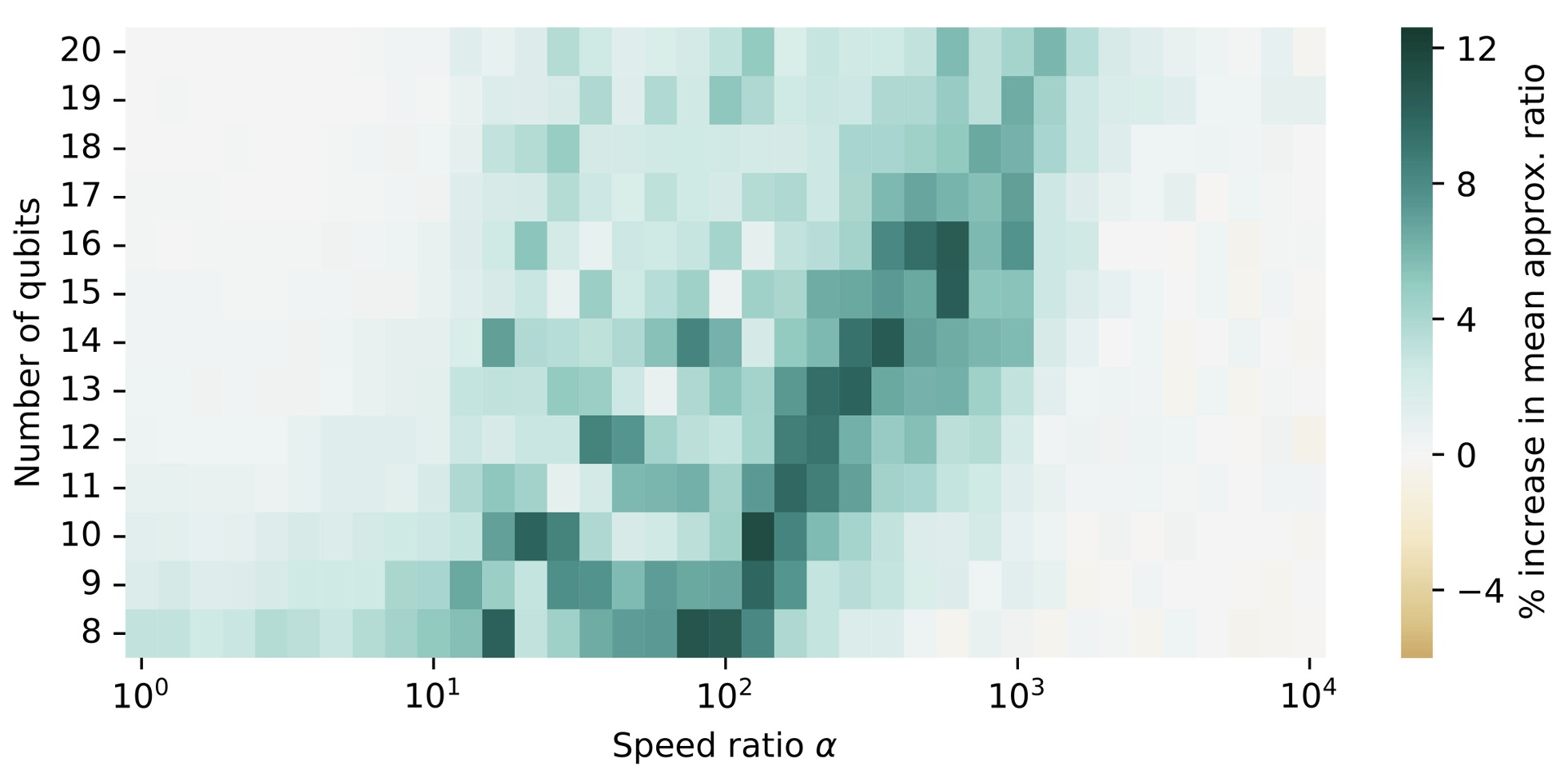}
    \caption{{\bf Percentage difference in mean approximation ratio attained by bDA-QAOA at parameters maximizing error-free QAOA and bDA-QAOA with optimized parameters.} We have computed it over 50 randomly generated MAX-CUT problems with a constant filling factor $p_{\rm clause} = 0.7$. On the $x$-axis, the ratio $\alpha$ of single-qubit to problem Hamiltonian term strength is displayed, with error-free QAOA existing in the limit as this ratio approaches infinity. Darker colors imply that for the concerned speed ratio and qubit number, the variational nature of QAOA can account for differences between bDA-QAOA and the ideal algorithm. The plot highlights the advantage of using a variational algorithm like QAOA compared to a non-variational algorithm in the presence of coherent errors in the DAQC framework.}
    \label{Fig4chap5}
\end{figure}

Variational quantum algorithms, like QAOA, are considered to have better error tolerance compared to non-variational algorithms. This is because the classical optimizer can compensate for systematic coherent over- or under-rotations and other errors \cite{mcclean2016theory}, making variational algorithms a suitable choice for NISQ quantum computing. QAOA finds the optimal parameters $\vec \beta^*$, $\vec\gamma^*$ by maximizing $\langle H_{\textrm{P}} \rangle_{\vec \beta,\vec\gamma}$. However, when the QAOA ansatz operators are changed to those of bDAQC, the parameters $\vec \beta^*$, $\vec\gamma^*$ that maximize $\langle H_{\textrm{P}} \rangle_{\vec \beta,\vec\gamma}$ may not necessarily maximize $\langle H_{\text{P}} \rangle_{\vec \beta,\vec\gamma}^{\alpha,\rm{DA}}$, where 
\begin{equation}
\langle H_{\textrm{P}} \rangle^{\alpha,\rm{DA}}_{\vec \beta,\vec\gamma} = \langle\vec \beta,\vec\gamma|^{\alpha,\rm{DA}} H_{\textrm{P}} |\vec \beta,\vec\gamma\rangle^{\alpha,\rm{DA}}.
\end{equation}

In fact, Fig. \ref{Fig4chap5} suggests that this is not the case and shows that a significant increase in the success probability of QAOA results from the variational freedom of the algorithm. Figure \ref{Fig4chap5} should be understood to demonstrate the parameter regimes for which it makes more sense to perform a variational algorithm such as QAOA rather than a fixed gate sequence algorithm such as the quantum Fourier transform. For high $\alpha$, the error introduced by the scheme is negligible, and both variational algorithms and fixed sequence algorithms will perform similarly. In the middle of the plot, a dark turquoise band can be seen implying that, while a non-variational algorithm will have low fidelity due to the presence of DA-induced coherent errors, the variational algorithm still functions. For sufficiently low $\alpha$, we enter a regime in which even the variational algorithm fails to recover any performance through altering parameters. We interpret that this lack of ability of DA-QAOA to absorb error in the low $\alpha$ regime is a manifestation of barren plateaus in the objective function \cite{mcclean2018barren}. Barren plateaus are a feature discovered to occur in the optimization landscapes of quantum neural networks. When parameterized random circuits are used as ansatze in variational algorithms, the gradient of the objective function with respect to the variational parameters is observed to become exponentially small in the number of qubits used. When $\alpha$ reduces to a certain value, we interpret that the DA-QAOA ansatz loses specificity to the problem Hamiltonian of interest. The variational form used for optimization no longer bears similarity to the objective function used and is, consequently, no better an ansatz than a random quantum circuit. At this point of low $\alpha$, we observe that the gradient of our objective function with respect to the variational parameters $\vec \gamma, \vec \beta$ tends to become prohibitively small and the approximation ratio attained therefore varies little with differing parameters.

\subsubsection{Sensitivity to other errors}\label{sec533}

Next to the errors discussed in the previous sections that are imminent to the hardware simplification provided by our digital-analog approach, the algorithm is exposed to other sources of errors common to NISQ computing. As detailed error budgets of concrete hardware are currently hard to determine, we would like to qualitatively evaluate their impact on our technique. 

On the one hand, single-qubit gate errors induced by decoherence measured by $T_{1/2}$ will have a full impact on this algorithm as these are repeatedly executed. Small errors in the rotation axis will also have a full impact as they can be mistaken for a modified problem Hamiltonian. Errors in the rotation angle can be expected to be less critical as some of them can be accommodated in the classical optimization process. So all in all, single-qubit errors have the same if somewhat smaller impact than in a compiled gate model QAOA. 

Two-qubit gates do not appear directly in our scheme thus avoiding two-qubit gate control errors as well as the additional entry points for noise through fast two-qubit control ports. However, the interaction mediated by the problem Hamiltonian can still create entangled states, which decay faster than non-entangled states. Notably, an $n$-qubit GHZ state dephases in a time $T_2/n$ \cite{dur2004stability}. The precise degree of entanglement needed for a specific problem instance is currently unknown for any quantum optimization algorithms. Yet, we can summarize that the sensitivity of digital-analog QAOA to two-qubit errors is lower than the compiled version. Given a single qubit error rate, alongside the total execution time of the algorithm relative to $T_2$, the depth at which this algorithm can be faithfully executed could be inferred.

In this estimate, we need to keep in mind whether coherent over-rotation errors have an effect different from incoherent errors. This case could occur if they interfered in a structured way. Given the randomization effect of the problem Hamiltonian to any state, this is unlikely and we expect that their impact is faithfully represented by the measured fidelity.

\newpage
In this chapter, we have highlighted the advantages of applying the DAQC paradigm for performing the QAOA algorithm. Our findings show that the DAQC setting allows for efficient implementation of the problem Hamiltonian, with minimal overhead for swapping resources. Additionally, we have shown that QAOA in the digital-analog setting is more resilient to errors compared to pre-programmed algorithms, making it a promising approach for near-term quantum devices designed for combinatorial optimization. This research offers a new avenue for the development of NISQ-era devices and brings us closer to realizing a quantum advantage for solving real-world problems.


\section[Digital-analog simulations using the cross--resonance effect]{Digital-analog simulations\\ using the cross--resonance effect}
\label{sec6}
\label{chapter6}

\fancyhead[RO]{6\quad DA SIMULATIONS USING THE CROSS-RESONANCE EFFECT}

%
%

\thiswatermark{\put(1,-302){\color{l-grey}\rule{70pt}{60pt}}
\put(70,-302){\color{grey}\rule{297pt}{60pt}}}

%
%

\lettrine[lines=2, findent=3pt,nindent=0pt]{A}{t} the moment, superconducting circuits have been established as a leading quantum platform in terms of controllability and scalability, mainly caused by the introduction of the transmon qubit~\cite{Koch2007}. Implementations controlled by microwave pulses have achieved very low errors on single-qubit gates~\cite{McKay2017}, and the most common two-qubit gate for fixed frequency transmons is based on the cross-resonance (CR) interaction \cite{Paraoanu2006, Rigetti2010, Chow2011}. In this chapter, we explore the potential use of the CR effect as an interaction Hamiltonian between transmon qubits, that would allow us to implement the DAQC paradigm in superconducting circuit quantum processors.

In section \ref{sec61}, we derive the effective interaction Hamiltonian of the CR effect. First, in section \ref{sec611}, we consider a CR gate interaction between two superconducting qubits in order to obtain a purely non-local, in a particular frame, effective interaction Hamiltonian. Further, in section \ref{sec612}, we consider a multi-qubit extension and derive the generalized effective multi-qubit two-local Hamiltonian. Next, in section \ref{sec62}, we consider how the multi-qubit Hamiltonian may be toggled into a variety of forms using digital single-qubit gates. Employing this set of Hamiltonians, we design DAQC protocols to simulate Ising (section \ref{sec631}), $\sigma_x\sigma_y$ (section \ref{sec632}), and Heisenberg (section \ref{sec633}) spin models. The resulting DAQC sequences are in some cases Trotter-error-free in 1D. We compute the Trotter error when it is present and find that it is reduced by a constant factor with respect to a DQC decomposition of the same model. In section \ref{sec64}, we discuss some practical implementations of our findings, and in section \ref{sec65}, we conclude the chapter with a discussion and a summary of the extensive work that we present. The content of this chapter is based on the work developed in \cite{GonzalezRaya2021}.

\subsection{Deriving the effective cross-resonance Hamiltonian}
\label{sec61}

\begin{figure}[t]
    \centering
    \includegraphics[width=0.90 \textwidth]{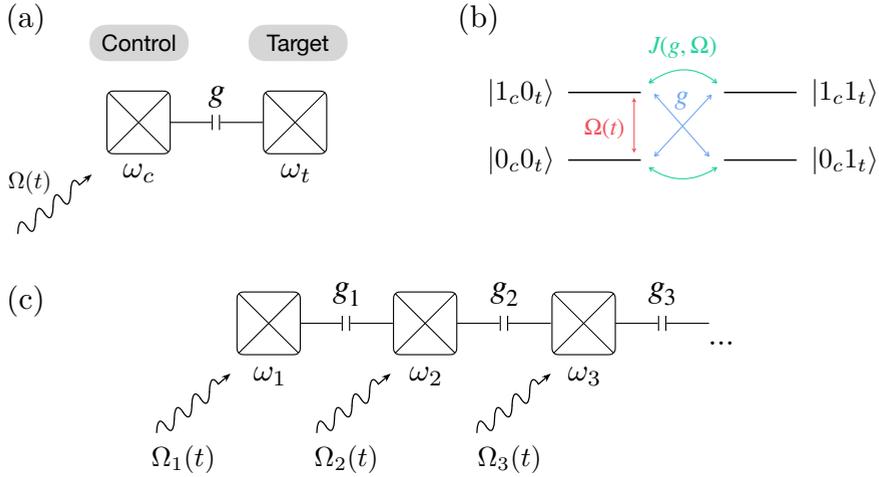}
    \caption{{\bf Graphical representation of the cross-resonance effect: (a) }Two qubits, the first one being the control qubit with resonance frequency $\omega_{c}$ and the second one the target qubit with resonance frequency $\omega_{t}$, are interacting with strength $g$. The control qubit is driven at the resonance frequency of the target qubit, with driving amplitude $\Omega(t)$. {\bf(b)} State space representation of the transitions between levels of the control and target qubits, in the presence of a driving of amplitude $\Omega(t)$ on the control qubit. The effective CR interaction is described by strength $J(g,\Omega)$. {\bf(c)} $N$ qubits with nearest neighbor interaction, all of them are driven at the resonance frequency of their neighbor to the right, illustrating the scenario we describe in section \ref{sec612}}
    \label{Fig1chap6}
\end{figure}

In this section, we present the effective CR Hamiltonian, following the derivation introduced in Ref. \cite{Rigetti2010}. We first introduce the two-qubit scenario, in order to develop an intuition for the effective coupling, and then generalize the results to the case of $N$ qubits. Further details of the calculations, supporting the main text, can be found in Appendix~\ref{AppAchap6}. Note that, in this thesis, we are working with $\hbar =1$.

\subsubsection{Two qubits}\label{sec611}

Our starting point is the laboratory frame Hamiltonian, written as
\begin{eqnarray}
\nonumber H_\text{LAB} &=& \frac{1}{2}(\omega_{1}^{q}\sigma_z^{(1)} +\omega_{2}^{q}\sigma_z^{(2)}) + \Omega_{1}\sigma_x^{(1)}\cos(\omega_{1} t+\phi_{1}) \\
&& + \Omega_{2}\sigma_x^{(2)}\cos(\omega_{2} t+\phi_{2}) + \frac{g}{2}\sigma_x^{(1)}\sigma_x^{(2)},
\label{eq:Ham2Q}
\end{eqnarray}
where $\sigma_x^{(i)}, \sigma_y^{(i)}, \sigma_z^{(i)}$ are the Pauli matrices supported on site $i$, and $\omega_{k}^{q}$ and $\omega_{k}$ are the resonance and the driving frequencies of qubit $k$, respectively. $\Omega_{k}$ represents the amplitude of the driving field, while $g$ denotes the strength of the interaction between the qubits. 

The effective Hamiltonian is derived by applying a series of unitary transformations --- described in detail in Appendix~\ref{AppA1chap6} --- to Eq.~\ref{eq:Ham2Q}. First, we apply a double rotation into the frame co-rotating at the driving frequency of the qubits ($\omega_1, \omega_2$). After this, we apply the rotating wave approximation (RWA), valid for  $\omega_1,\omega_2 \gg \delta_i = \omega_{1}^{q} - \omega_{1}, \Omega_i, g$, to drop fast terms rotating with frequency $\pm 2 \omega_1, \pm 2 \omega_2, \pm (\omega_1+\omega_2)$. We then proceed by applying two new rotations in order to express the Hamiltonian in a more convenient frame, named the quad frame (QF). In this frame, all local terms are eliminated and the result is a purely two-local Hamiltonian. The next step is to consider the case in which we drive the first qubit at the resonance frequency of the second qubit, $\omega_1 = \omega_{2}^{q}$, while the second one is not driven, as can be seen in Fig.~\ref{Fig1chap6}(a). After a final RWA, valid for $\Omega_1 \gg g$ or $\delta \gg g$, we end up with the effective Hamiltonian
\begin{equation}\label{eq:H_QF}
H_\text{QF} = \frac{g\Omega_{1}}{4\delta}(\cos\phi_{1}\sigma_x^{(1)}\sigma_x^{(2)} + \sin\phi_{1}\sigma_x^{(1)}\sigma_y^{(2)}).
\end{equation}
As $\phi_{1}$ is a controllable phase, we can set $\phi_{1}=0$, resulting in
\begin{equation}\label{H_eff_QF_2Q}
H_\text{QF} = \frac{g\Omega_{1}}{4\delta} \sigma_x^{(1)}\sigma_x^{(2)}.
\end{equation}

\subsubsection{$N$ qubits}\label{sec612}

The $N$-qubit Hamiltonian, in the laboratory frame, is given by
\begin{equation}
H_\text{LAB} = \sum_{k=1}^{N}\left[ \frac{\omega^{q}_{k}}{2}\sigma_z^{(k)} + \Omega_{k}\sigma_x^{(k)}\cos(\omega_{k}t+\phi_{k}) \right] +\sum_{k=1}^{N-1}\frac{g_{k}}{2}\sigma_x^{(k)}\sigma_x^{(k+1)}
\label{eq:HamilNq}
\end{equation}
We proceed by moving to the QF by means of appropriate rotations (see Appendix~\ref{AppA2chap6} for details). The driving field is then applied to all qubits at the resonance frequency of their neighbor to the right, as shown in Fig.~\ref{Fig1chap6}(c), except for case of open boundary conditions in which the last qubit is not driven. 
Similar to the two-qubit case, the frame transformations re-express the Hamiltonian in a purely two-local form. Keeping only terms linear in $\Omega_i/\delta_i$, and neglecting fast oscillating terms $\delta \gg g$ by RWA, we arrive at the effective Hamiltonian 
\begin{equation}\label{H_eff_QF_complete}
    H_\text{QF} = \sum_{k=1}^{N-1}\frac{g_{k}\Omega_{k}}{4\delta_{k}}\sigma_x^{(k)}\left(\sigma_y^{(k+1)}\sin(\phi_{k}-\phi_{k+1}) - \sigma_z^{(k+1)}\cos(\phi_{k}-\phi_{k+1})\right).
\end{equation}
Once again, we have the freedom to set $\phi_{k}=\phi$ for all $k$. The Hamiltonian then reduces to
\begin{equation}\label{H_QF_eff}
H_\text{QF} = \sum_{k=1}^{N-1}J_{k}\sigma_x^{(k)}\sigma_z^{(k+1)},
\end{equation}
where we have defined $J_{k}=-g_{k}\Omega_{k}/4\delta_{k}$. As seen in the two-qubit case, the Hamiltonian only contains two-qubit interaction terms. In the next sections, we will discuss the use of this Hamiltonian to generate the analog dynamics of a DA computation.

\subsection{Digital-analog dynamics from the CR Hamiltonian}\label{sec62}

We take Eq.~\ref{H_QF_eff} as a starting point, and consider $\Omega_{k} = \Omega$, $\delta_{k}=\delta$, $g_{k} = g$, $J_{k}=J$, for simplicity. Then, we write the effective Hamiltonian in the QF as
\begin{equation}\label{H_control}
H_\text{A} = J\sum_{k=1}^{N-1}\sigma_x^{(k)}\sigma_z^{(k+1)}.
\end{equation}

\subsubsection{Synthesis Error}\label{621}

Given that the effective Hamiltonian is the centerpiece of the simulation protocols, we need to estimate the synthesis error associated with the fact that it is an approximation of the original Hamiltonian. In the weak-driving regime $\Omega_{k} \ll \delta_{k}$, the original Hamiltonian without the QF RWA is
\begin{eqnarray}
H^\text{org} &=& \frac{g}{4} \sum_{k=1}^{N-1} \bigg\{ \left(\sigma_z^{(k)}\sigma_z^{(k+1)} + \sigma_y^{(k)}\sigma_y^{(k+1)}\right)\cos\delta t \nonumber\\
&& + \left(\sigma_y^{(k)}\sigma_z^{(k+1)} - \sigma_z^{(k)}\sigma_y^{(k+1)}\right)\sin\delta t \nonumber\\
&& -\frac{\Omega}{\delta} \left[ \sigma_x^{(k)}\sigma_z^{(k+1)} + \left(\sigma_z^{(k)}\cos2\delta t + \sigma_y^{(k)}\sin2\delta t\right)\sigma_x^{(k+1)}\right] \bigg\}.
\end{eqnarray}
In order to compute the synthesis error, we focus on the Frobenius norm, 
\begin{equation}
|| A ||_{F} = \sqrt{\tr(A^{\dagger}A)},    
\end{equation}
which provides an upper bound for the spectral norm. Let us compute the norm for the difference between the two Hamiltonians, $\Delta H = H^\text{org}-H_\text{A}$,
\begin{eqnarray}
\Delta H &=& \frac{g}{4} \sum_{k=1}^{N-1} \bigg\{ \left(\sigma_z^{(k)}\sigma_z^{(k+1)} + \sigma_y^{(k)}\sigma_y^{(k+1)}\right)\cos\delta t \nonumber\\
&& + \left(\sigma_y^{(k)}\sigma_z^{(k+1)} - \sigma_z^{(k)}\sigma_y^{(k+1)}\right)\sin\delta t  \nonumber \\
&& -\frac{\Omega}{\delta} \left(\sigma_z^{(k)}\cos2\delta t + \sigma_y^{(k)}\sin2\delta t\right)\sigma_x^{(k+1)} \bigg\}.
\end{eqnarray}
The latter part of this operator contributes with $\Omega^{2}/\delta^{2}$ to the Frobenius norm, so we will neglect that part in the approximation $\Omega/\delta \ll 1$. The rest can be written as
\begin{eqnarray}
\Delta H &=& \frac{g}{4} \sum_{k=1}^{N-1} \bigg\{ \left(\sigma_z^{(k)}\cos\delta t + \sigma_y^{(k)}\sin\delta t\right) \sigma_z^{(k+1)} \nonumber\\
&& + \left(\sigma_y^{(k)}\cos\delta t - \sigma_z^{(k)}\sin\delta t\right) \sigma_y^{(k+1)} \bigg\},
\end{eqnarray}
which corresponds to the result of a rotation given by $U_{k} = e^{-i\delta t \sigma_x^{(k)}/2}$. This norm can be computed analytically by rewriting the last expression as
\begin{equation}
\Delta H = \frac{g}{4}\sum_{k=1}^{N-1}U_{k}^{\dagger}\left(\sigma_z^{(k)}\sigma_z^{(k+1)}+\sigma_y^{(k)}\sigma_y^{(k+1)}\right)U_{k}.
\end{equation}
Then, we see that the only terms that survive the trace of $(\Delta H)^{\dagger}\Delta H$, which is computed as
\begin{equation}
    \frac{g^{2}}{16} \sum_{k,k'=1}^{N-1}U_{k}^{\dagger}\left(\sigma_z^{(k)}\sigma_z^{(k+1)}+\sigma_y^{(k)}\sigma_y^{(k+1)}\right)U_{k}U_{k'}^{\dagger}\left(\sigma_z^{(k')}\sigma_z^{(k'+1)}+\sigma_y^{(k')}\sigma_y^{(k'+1)}\right)U_{k'},
\end{equation}
are those which satisfy $k=k'$. Consequently, we obtain
\begin{equation}
\tr\left[(\Delta H)^{\dagger}\Delta H\right] = \frac{g^{2}}{16}\tr\left(2\sum_{k=1}^{N-1}\mathbb{I}\right) = \frac{g^{2}}{8}(N-1)\tr(\mathbb{I}),
\end{equation}
where $\mathbb{I}$ actually represents $\bigotimes_{k=1}^{N} \mathbb{I}_{k}$. We want to set the normalization to $\tr(\mathbb{I})=1$, which corresponds to a factor of $2^{-N/2}$ on the Frobenius norm, since
\begin{equation}
\left|\left| \bigotimes_{k=1}^{N} \mathbb{I}_{k} \right|\right|_{F} = 2^{N/2}.
\end{equation}
Then, we find the Frobenius norm for $N$ qubits ($N\geq 2$) to be
\begin{equation}
|| \Delta H ||_{F} = \frac{g}{2\sqrt{2}} \sqrt{N-1}.
\end{equation}
See that this norm diverges with the square root of the number of qubits. Notice however that the Frobenius norm per qubit decreases with $N$. Furthermore, we have computed the norm of the difference between the propagators, $\Delta \mathcal{U} = \mathcal{U}^\text{org}-\mathcal{U}_\text{A}$,
\begin{equation}
|| \Delta \mathcal{U} ||_{F} = \frac{g}{\delta\sqrt{2}} \left| \sin\frac{\delta t}{2} \right| \sqrt{N-1}.
\end{equation}
Here, the propagators are computed up to the first order in the Dyson series. Again, the norm of the difference of propagators per qubit decreases with $N$. Note that, for $\delta t \ll 1$, 
\begin{equation}
|| \Delta \mathcal{U} ||_{F} \approx t \cdot || \Delta H ||_{F}. 
\end{equation}

The synthesis errors corresponding to the Hamiltonians derived in further sections can be found in Appendix~\ref{AppBchap6}.

\subsubsection{Hamiltonian toggling}\label{622}

Let us now consider DA quantum simulations of the spin-1/2 Ising, $XY$, and Heisenberg models in 1 and 2 dimensions. We designate the effective Hamiltonian in the QF, given in Eq.~\ref{H_QF_eff}, as our fundamental DAQC Hamiltonian from which all others will be generated. By rotating to the reference frame in which the Hadamard transformation is applied to all {\em even} qubits, i.e. $V_\text{H}^\text{even}=\bigotimes_i \text{H}^{(2i)}$, the Hamiltonian transforms into 
\begin{equation}\label{H_even}
H^\text{even}= J\sum_{k=1}^{\frac{N}{2}}\sigma_x^{(2k-1)}\sigma_x^{(2k)} + J\sum_{k=1}^{\frac{N-1}{2}}\sigma_z^{(2k)} \sigma_z^{(2k+1)}. 
\end{equation}
From this reference frame, Hadamard transforming all qubits will toggle the Hamiltonian into its odd form, i.e. translating the Hamiltonian by one site,
\begin{equation}\label{H_odd}
H^\text{odd}= J\sum_{k=1}^{\frac{N}{2}}\sigma_z^{(2k-1)}\sigma_z^{(2k)} + J\sum_{k=1}^{\frac{N-1}{2}}\sigma_x^{(2k)} \sigma_x{(2k+1)}.  
\end{equation}

\subsubsection{Two-dimensional generalization}\label{623}

\begin{figure}[t]
    \centering
    \includegraphics[width=0.80\textwidth]{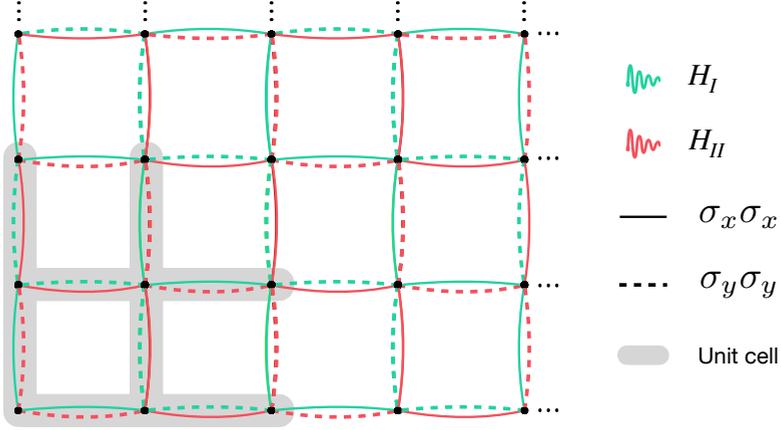}
    \caption{{\bf Illustration of analog Hamiltonian interactions on a 2-dimensional lattice.} The green (red) lattice represents the Hamiltonian given by Eq.~\ref{eq:H_{I}} (\ref{eq:H_{II}}). Vertices correspond to qubits in a 2D lattice and the solid and dashed edges correspond to the $\sigma_x\sigma_x$ and $\sigma_y\sigma_y$ interactions, respectively.}
    \label{Fig2chap6}
\end{figure}

Let us also consider the extension of the Hamiltonian to two dimensions. Consider a single target qubit in a two-dimensional lattice that is driven at the frequencies of its neighbors in the $+\hat{i}$ and $+\hat{j}$ directions. This realizes a $\sigma_x^{(c)} \sigma_z^{(t)}$-interaction between the control qubit located at $(i,j)$ and target qubits at sites $(i+1,j)$ and $(i,j+1)$. The extension of $H^\text{odd}$ in Eq.~\ref{H_odd} is
\begin{eqnarray}
\nonumber H_{2D}^\text{odd} &=& J \bigg[ \sum_{i,j=1}^{\frac{N}{2}} \sigma_z^{(2i-1,2j-1)}\left(\sigma_z^{(2i-1,2j)}+\sigma_z^{(2i,2j-1)}\right)\\
&&+ \sum_{i,j=1}^{\frac{N-1}{2}} \sigma_z^{(2i,2j)}\left(\sigma_z^{(2i,2j+1)}+\sigma_z^{(2i+1,2j)}\right) \nonumber\\
\nonumber &&+ \sum_{i=1}^{\frac{N}{2}}\sum_{j=1}^{\frac{N-1}{2}} \sigma_x^{(2i-1,2j)}\left(\sigma_x^{(2i-1,2j+1)}+\sigma_x^{(2i,2j)}\right) \\
&&+ \sum_{i=1}^{\frac{N-1}{2}}\sum_{j=1}^{\frac{N}{2}} \sigma_x^{(2i,2j-1)}\left(\sigma_x^{(2i,2j)}+\sigma_x^{(2i+1,2j-1)}\right) \bigg],
\end{eqnarray}
where summations run over repetitions of the unit cell illustrated in Fig.~\ref{Fig2chap6}. Likewise, the extension of $H^\text{even}$ in Eq.~\ref{H_even} is $H_{2D}^\text{even} = H_{2D}^{odd}(x\leftrightarrow z)$, which is easily realized by applying a Hadamard on every site of the lattice. Applying a global $R_x(\pi/2)=e^{-i\pi \sigma_x/4}$ transformation on Hamiltonian $H_{2D}^\text{even}$, we obtain
\begin{eqnarray}\label{eq:H_{I}}
\nonumber H_\text{I} &=& J\sum_{i,j=1}^{\frac{N}{2}} \left(\sigma_x^{(2i-1,2j-1)} \sigma_x^{(2i,2j-1)} + \sigma_y^{(2i,2j-1)} \sigma_y^{(2i+1,2j-1)}  \right.\\
\nonumber &&   + \sigma_y^{(2i-1,2j)} \sigma_y^{(2i,2j)} + \sigma_x^{(2i,2j)} \sigma_x^{(2i+1,2j)} \\
\nonumber&&  + \sigma_x^{(2i-1,2j-1)} \sigma_x^{(2i-1,2j)}+ \sigma_y^{(2i,2j-1)} \sigma_y^{(2i,2j)} \\
&& \left. + \sigma_y^{(2i-1,2j)} \sigma_y^{(2i-1,2j+1)} + \sigma_x^{(2i,2j)} \sigma_x^{(2i,2j+1)}\right),
\end{eqnarray}
where we have simplified the summation limits by considering that the Hamiltonian acts on a system with periodic boundary conditions. If we rotate $H_{2D}^{odd}$ by $R_x(\pi/2)$, we have
\begin{eqnarray}\label{eq:H_{II}}
\nonumber H_\text{II} &=& J\sum_{i,j=1}^{\frac{N}{2}} \left(\sigma_y^{(2i-1,2j-1)} \sigma_y^{(2i,2j-1)} + \sigma_x^{(2i,2j-1)} \sigma_x^{(2i+1,2j-1)} \right. \\
\nonumber &+& \sigma_x^{(2i-1,2j)} \sigma_x^{(2i,2j)} + \sigma_y^{(2i,2j)} \sigma_y^{(2i+1,2j)} \\
\nonumber &+& \sigma_y^{(2i-1,2j-1)} \sigma_y^{(2i-1,2j)} + \sigma_x^{(2i,2j-1)} \sigma_x^{(2i,2j)}  \\ 
&+& \left. \sigma_x^{(2i-1,2j)} \sigma_x^{(2i-1,2j+1)} + \sigma_y^{(2i,2j)} \sigma_y^{(2i,2j+1)} \right).
\end{eqnarray}
Note that $H_{II}$ is just a translation of $H_{I}$ by the vector $(1,1)$. The interactions described by these Hamiltonians are represented in Fig.~\ref{Fig2chap6}, where $H_\text{I}$'s and $H_\text{II}$'s interactions are illustrated by the green and red edges, respectively. In both cases, the solid (dashed) edges correspond to $\sigma_x\sigma_x \: (\sigma_y\sigma_y)$ interactions between adjacent qubits, and the summations in Eqs.~\ref{eq:H_{I}}, and \ref{eq:H_{II}} correspond to a tiling of the 2D lattice using the unit cell, highlighted in blue in Fig.~\ref{Fig2chap6}.

\subsection{Many-body compilation}\label{sec63}
We now discuss how to simulate a variety of paradigmatic spin models with the Hamiltonians discussed above.

\subsubsection{Ising model}\label{sec631}
So far we have considered a multi-qubit framework in which we drive all qubits at the resonance frequency of their neighbors to the right. For this particular case, let us now explore a scenario in which we drive only odd or even qubits, which can be achieved by tuning the system's parameters in the following way:
\begin{eqnarray}
\nonumber k \, \text{control} &\rightarrow& \{ \omega_{k}=\omega_{k+1}^{q}, \varphi_{k}(t) = \delta_{k+1}t + \phi_{k}-\phi_{k+1}, \\
\nonumber && \eta_{k}\approx\delta_{k}, \sin\xi_{k}\approx 1, \cos\xi_{k}\approx\frac{\Omega_{k}}{\delta_{k}}\}, \\
\nonumber k \, \text{target} &\rightarrow& \{ \varphi_{k}(t) = (\omega_{k}-\omega_{k+1})t -\phi_{k+1}, \Omega_{k}=0, \delta_{k}=0, \\
&& \eta_{k}=0, \phi_{k}=0, \sin\xi_{k}=0, \cos\xi_{k}=1\},
\end{eqnarray}
where the qubit we drive is the control qubit and its neighbor to the right is the corresponding target qubit. Assuming we drive only odd qubits, the choice of parameters leads to a particular QF transformation, represented by
\begin{equation}
V_\text{QF}^{\text{odd}} = \bigotimes_{k\,\text{odd}} V_\text{QF}^{(k)}V_{I}^{(k+1)},
\end{equation}
where $V_\text{QF}^{(k)}$ is the QF transformation applied on qubit $k$ (this transformation is discussed in Appendix~\ref{AppCchap6}), and $V_{I}^{(k+1)}=e^{-\frac{it}{2}\omega^{q}_{k+1}\sigma_z^{(k+1)}}$ is the transformation to the interaction picture of qubit $k+1$. After applying a RWA, by keeping the static terms, we write the Hamiltonian in the QF as
\begin{equation}
H_\text{QF}^{\text{odd}} = J \sum_{k=1}^{\frac{N}{2}} \sigma_x^{(2k-1)}\left(\sigma_x^{(2k)}\cos\phi + \sigma_y^{(2k)}\sin\phi\right),
\end{equation}
after setting $\delta_{2k-1}=\delta$, $\Omega_{2k-1}=\Omega$, $g_{2k-1}=g$, $\phi_{2k-1}=\phi$, and defining $J=g\Omega/4\delta$. See that this is a straightforward multi-qubit extension of the Hamiltonian in Eq.~\ref{eq:H_QF}. If we do the same, in the case in which we drive only even qubits, the transformation is
\begin{equation}
V_\text{QF}^{\text{even}} = \bigotimes_{k\,\text{even}} V_\text{QF}^{(k)}V_{I}^{(k+1)},
\end{equation}
and we obtain
\begin{equation}
H_\text{QF}^{\text{even}} = J \sum_{k=1}^{\frac{N-1}{2}} \sigma_x^{(2k)}\left(\sigma_x^{(2k+1)}\cos\phi + \sigma_y^{(2k+1)}\sin\phi\right).
\end{equation}
Considering $\phi=0$, these Hamiltonians transform into
\begin{equation}
 H_\text{QF}^{\text{odd}} = J \sum_{k=1}^{\frac{N}{2}} \sigma_x^{(2k-1)}\sigma_x^{(2k)}, \quad \text{and }\quad
H_\text{QF}^{\text{even}} = J \sum_{k=1}^{\frac{N-1}{2}} \sigma_x^{(2k)}\sigma_x^{(2k+1)},
\end{equation}
and we see that $[H_\text{QF}^{\text{odd}},H_\text{QF}^{\text{even}}]=0$. If we rotate all qubits by a Hadamard gate, we obtain
\begin{eqnarray}
\nonumber V^{\dagger}_{\text{H}}H_\text{QF}^{\text{odd}}V_{\text{H}} &=& J \sum_{k=1}^{\frac{N}{2}} \sigma_z^{(2k-1)}\sigma_z^{(2k)} \equiv H_{1}, \\
V^{\dagger}_{\text{H}}H_\text{QF}^{\text{even}}V_{\text{H}} &=& J \sum_{k=1}^{\frac{N-1}{2}} \sigma_z^{(2k)}\sigma_z^{(2k+1)} \equiv H_{2},
\end{eqnarray}
which leads to
\begin{equation}
H_{ZZ} = H_{1}+H_{2} = J\sum_{k=1}^{N-1} \sigma_z^{(k)}\sigma_z^{(k+1)}.
\end{equation}

\begin{figure}[t]
\centering
\includegraphics[width=0.8 \textwidth]{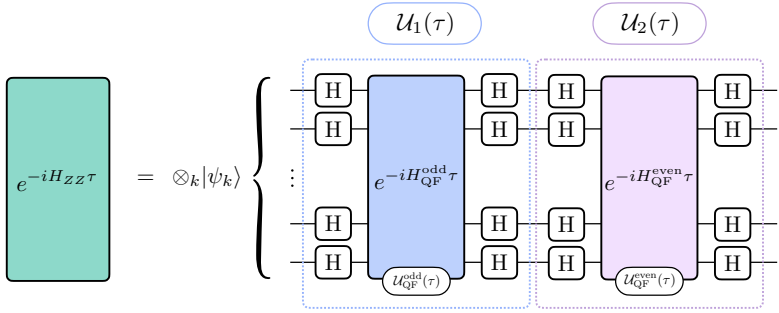}
\caption{{\bf Digital-analog quantum circuit to simulate the evolution under Hamiltonian $H_{ZZ}$ for a time $\tau$.} This simulation is carried out by transforming all qubits by $V_\text{QF}^{\text{even}}$, which entails transforming even qubits to the QF ($V_\text{QF}$) and odd qubits to the interaction picture ($V_{I}$), in a setup in which only even qubits are being driven. In this scenario, the analog propagator $\mathcal{U}_\text{QF}^{\text{even}}$ -- which describes the evolution under analog Hamiltonian $H_\text{QF}^{\text{even}}$ -- is then conjugated by Hadamard gates ($\text{H}$) on all qubits. This segment of the circuit simulates the evolution given by $\mathcal{U}_{2}(\tau)$. The circuit is repeated with the QF transformation being applied to odd qubits and the interaction picture transformation to even qubits ($V_\text{QF}^{\text{odd}}$), while only odd qubits are driven. This segment simulates the evolution given by $\mathcal{U}_{1}(\tau)$.}
\label{Fig3chap6}
\end{figure}

This sequence for simulating the evolution of $H_{ZZ}$ can be interpreted as the combination of two blocks: the first one represents the evolution given by $\mathcal{U}_{1}(t)=e^{-iH_{1}t}$, where we only drive odd qubits, and the second one represents the evolution given by $\mathcal{U}_{2}(t)=e^{-iH_{2}t}$, where we only drive even qubits, both in a frame rotated by Hadamard gates. The integrity of these simulation blocks relies on the fact that $[H_{1},H_{2}]=0$, meaning that the pairwise combination of propagators is exact. Then, the propagators corresponding to the two blocks can exactly describe the evolution of the whole, 
\begin{equation}
\mathcal{U}_{ZZ}(t) = e^{-i H_{ZZ}t} = e^{-i (H_{1}+H_{2}) t} = \mathcal{U}_{1}(t) \,\mathcal{U}_{2}(t).
\end{equation}
The propagator corresponding to $H_{ZZ}$ is computed as
\begin{equation}
\mathcal{U}_{ZZ}(t)|\psi\rangle = \mathcal{U}_{1}(t)\,\mathcal{U}_{2}(t)|\psi\rangle = V^{\dagger}_{\text{H}} \, \mathcal{U}_\text{QF}^{\text{odd}}(t) V_{\text{H}}\: V^{\dagger}_{\text{H}} \, \mathcal{U}_\text{QF}^{\text{even}}(t) V_{\text{H}}|\psi\rangle
\end{equation}
where $\mathcal{U}_\text{QF}^{\text{odd}}(t)$ and $\mathcal{U}_\text{QF}^{\text{even}}(t)$ are the propagators generated by $H_\text{QF}^{\text{odd}}$ and $H_\text{QF}^{\text{even}}$, respectively. The former is achieved by rotating all qubits by $V_{\text{QF}}^{\text{odd}}$, and the latter is achieved by rotating all qubits by $V_{\text{QF}}^{\text{even}}$. Furthermore, $V_{\text{H}} = \bigotimes_{k} \text{H}^{(k)}$, where $\text{H}^{(k)} = e^{i\pi/2} e^{-i\pi \sigma_y^{(k)}/4}e^{-i\pi \sigma_z^{(k)}/2}$ represents the application of a Hadamard gate on qubit $k$. The simulation protocol comprises the following steps:
\begin{enumerate}
\item Prepare an initial product state $\otimes_{k}|\psi_{k}\rangle$.
\item Apply Hadamard gates, $\text{H}$, on all qubits.
\item Let the states evolve according to the underlying analog Hamiltonian with analog propagator $\mathcal{U}_\text{QF}^{\text{even}}(\tau)$ for time $\tau$.
\item Let the states evolve according to the underlying analog Hamiltonian with analog propagator $\mathcal{U}_\text{QF}^{\text{odd}}(\tau)$ for time $\tau$.
\item Apply Hadamard gates, $\text{H}$, on all qubits.
\end{enumerate}
Note that operations that consist of the application of a unitary $U$, followed by their inverse $U^{\dagger}$, render the identity as the result, and thus are not mentioned in the simulation protocol steps. However, these operations are included in the figures for illustrative purposes. Due to the idiosyncrasies of the Hamiltonians derived in this protocol, we benefit from the absence of Trotter error, which implies no limits on application time, $\tau$, of the block. This block is represented in Fig.~\ref{Fig3chap6}. To evolve a state $|\psi\rangle$ with Hamiltonian $H_{\sigma_z\sigma_z}$, one must re-apply the block $M$ times where the total evolution time is $T=M\tau$. Gate-based quantum circuits describe the application of quantum gates following the usual flow of time. That is, from left to right, following the order in which the operators are applied on a quantum state represented by a ket.

\subsubsection{$XY$ model}\label{sec632}

Let us now describe a protocol to simulate a $XY$ model in which all adjacent spins interact by $\sigma_x\sigma_x + \sigma_y\sigma_y$ terms. 

\subsubsection*{1D Simulation}

\begin{figure}[t]
\centering
\includegraphics[width=0.98 \textwidth]{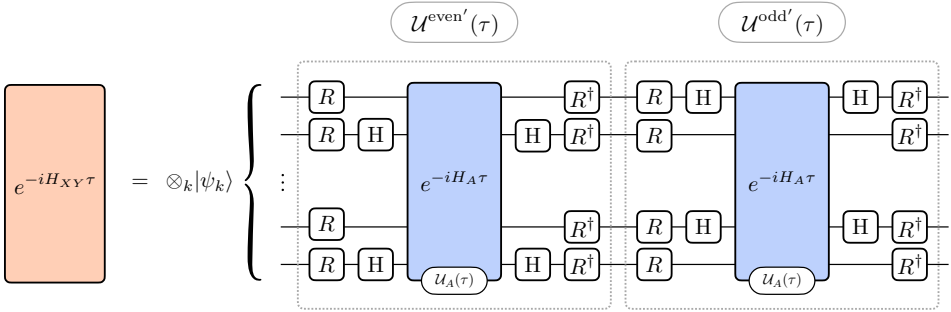}
\caption{{\bf Digital-analog quantum circuit to simulate the evolution of an initial quantum state under Hamiltonian $H_{XY}$ for a time $\tau$.} This simulation is carried out by conjugating the analog propagator $\mathcal{U}_{A}$ -- which describes the evolution under analog Hamiltonian $H_{A}$ -- by Hadamard gates ($\text{H}$) on even qubits, combined with rotations $R$ on all qubits, which are $x$ rotations by $R_{x}(\pi/2)$. This segment of the circuitsimulates the evolution given by $\mathcal{U}^{\text{even}'}(\tau)$. The circuit is then repeated with Hadamard gates applied to odd qubits, and this segment simulates the evolution given by $\mathcal{U}^{\text{odd}'}(\tau)$.}
\label{Fig4chap6}
\end{figure}

In the 1D case, we start from the Hamiltonians in Eqs. \ref{H_even}, and \ref{H_odd}. By performing the same $R_{x}(\pi/2) \equiv R$ about each qubit, we find
\begin{eqnarray}
\nonumber H^{\text{odd}'} &=& R^{\dagger} H^{\text{odd}} R =  J\sum_{k=1}^{\frac{N}{2}}\sigma_y^{(2k-1)}\sigma_y^{(2k)} + J\sum_{k=1}^{\frac{N-1}{2}}\sigma_x^{(2k)} \sigma_x^{(2k+1)}, \\ 
H^{\text{even}'} &=& R^{\dagger} H^{\text{even}} R =  J\sum_{k=1}^{\frac{N}{2}}\sigma_x^{(2k-1)}\sigma_x^{(2k)} + J\sum_{k=1}^{\frac{N-1}{2}}\sigma_y^{(2k)} \sigma_y^{(2k+1)} ,
\end{eqnarray}
which, upon summing, realize the 1D $XY$ chain Hamiltonian
\begin{equation}
H_{XY} = H^{\text{even}'}+ H^{\text{odd}'}  = J\sum_{k=1}^{N-1}\left(\sigma_x^{(k)}\sigma_x^{(k+1)} + \sigma_y^{(k)} \sigma_y^{(k+1)}\right).
\end{equation}
The key to this protocol is that $[H^{\text{even}'}, H^{\text{odd}'}]=R^{\dagger}[H^{\text{even}}, H^{\text{odd}}]R=0$, which implies
\begin{equation}
\mathcal{U}_{XY}(t) = e^{-i H_{XY}t} = e^{-i(H^{\text{odd}'} + H^{\text{even}'})t} = \mathcal{U}^{\text{odd}'}(t)\, \mathcal{U}^{\text{even}'}(t).
\end{equation}
This allows us to decompose the total $XY$ propagator into the product of two toggled Hamiltonians which results in a Trotter-error-free dynamics simulation protocol. The propagator $\mathcal{U}_{XY}(t)$ is further decomposed as
\begin{equation}
\mathcal{U}_{XY}(t)|\psi\rangle = \mathcal{U}^{\text{odd}'}\, \mathcal{U}^{\text{even}'}|\psi\rangle = W^{\text{odd}\dagger}\, \mathcal{U}_{A}(t) W^{\text{odd}}\, W^{\text{even}\dagger}\mathcal{U}_{A}(t) W^{\text{even}}|\psi\rangle,
\end{equation}
where $\mathcal{U}_{A}(t)$ is the original analog propagator generated by $H_{A}$, of Eq.~\ref{H_control}, and
\begin{eqnarray}
W^{\text{odd}} &=& \bigotimes_{k \,\text{odd}} \text{H}^{(k)}R^{(k)}R^{(k+1)},\\
W^{\text{even}} &=& \bigotimes_{k \,\text{even}} \text{H}^{(k)} R^{(k-1)}R^{(k)},
\end{eqnarray}
where $\text{H}^{(k)}$ and $R^{(k)}$ represent the application of a Hadamard gate and a $\pi/2$ $x$-rotation, respectively, on qubit $k$. The simulation protocol consists of the following steps:
\begin{enumerate}
\item Prepare an initial product state $\otimes_{k}|\psi_{k}\rangle$.
\item Apply a $x$-$\pi/2$ rotation on all qubits with $R_{x}(\pi/2)$. 
\item Apply Hadamard gates, $\text{H}$, on even qubits.
\item Let the states evolve according to the underlying analog Hamiltonian with analog propagator $\mathcal{U}_{A}(\tau)$ for time $\tau$.
\item Apply Hadamard gates, $\text{H}$, on all qubits.
\item Let the states evolve according to the underlying analog Hamiltonian with analog propagator $\mathcal{U}_{A}(\tau)$ for time $\tau$.
\item Apply Hadamard gates, $\text{H}$, on odd qubits.
\item Undo the $x$-$\pi/2$ rotation on all qubits by $R_{x}^{\dagger}(\pi/2)$.
\end{enumerate}
The entire sequence of operations needed to evolve by the $XY$ Hamiltonian is depicted in Fig.~\ref{Fig4chap6}. To evolve for a total time $T$ with Hamiltonian $H_{XY}$, one must re-apply the block $M=T/\tau$ times. Note that the three layers of single qubit rotations in between evolution by the analog propagators simplify into the product of single-qubit gates, which in this case simplifies to $R^{\dagger}\text{H} R = (x+y)/\sqrt{2}$.

\subsubsection*{2D Simulation and Digital vs. Digital-Analog Trotter Errors}
\label{sec:2D_XY}

\begin{figure}[t]
    \centering
    \includegraphics[width=0.98 \textwidth]{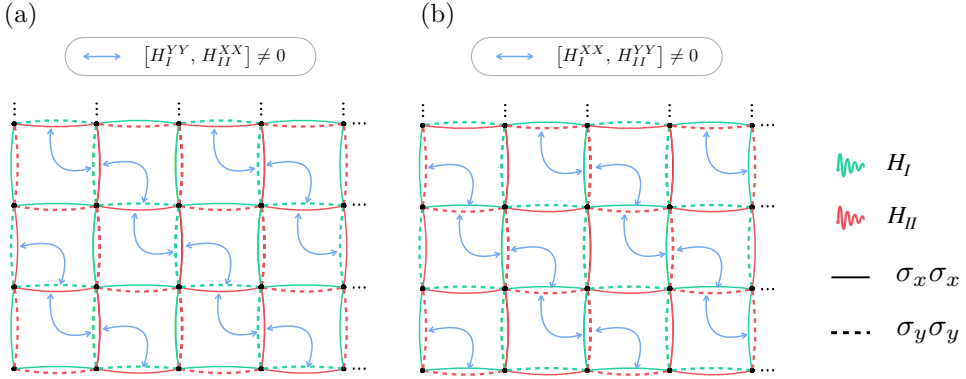}
    \caption{{\bf Lattice representation of the interactions featured on $H_{I}$ (green) and $H_{II}$ (red).} The blue arrows indicate the non-commuting terms between $H_{I}$ and $H_{II}$. These Hamiltonians are split into $xx$ and $yy$ - supported operators, $H_{i} = H_{i}^{XX} + H_{i}^{YY}$, in order to identify the two non-commuting operators: {\bf (a)} Non-zero terms of $[H_{I}^{YY},H_{II}^{XX}]$, {\bf (b)} Non-zero terms of $[H_{I}^{XX},H_{II}^{YY}]$. Jointly, these terms estimate the total Trotter error of the DAQC decomposition.}
    \label{Fig5chap6}
\end{figure}

The same two-Hamiltonian decomposition may be performed in two dimensions, taking the Hamiltonians in Eqs.~\ref{eq:H_{I}}, and \ref{eq:H_{II}}, such that $H^{2D}_{XY}=H_{I}+H_{II}$. However, since $[H_{I},H_{II}] \ne 0$ in two dimensions, we must resort to an approximate Trotter decomposition of the 2D $XY$ propagator. If we compute $[H_{I},H_{II}]$ we find 16 non-commuting terms.

Let us now compare the errors arising from a first order Trotter decomposition of our target evolution unitary. 
Overall, our goal is to determine the gate complexity of an approximate product decomposition $\widetilde{\mathcal{U}}_{XY}^{2D}$ such that $||\mathcal{U}_{XY}^{2D}(\tau) - \widetilde{\mathcal{U}}_{XY}^{2D}(\tau)|| \le \epsilon$ for an $\epsilon$ of our choosing. Here the target propagator is generated by exponentiating the target Hamiltonian $H^{2D}_{XY}$, while $\widetilde{\mathcal{U}}_{XY}^{2D} $ is generated by a first order Trotter decomposition which may be implemented through our DAQC Hamiltonians or through a digitized decomposition.

A first order Trotterization approximates an operator exponential of two generally non-commuting operators, $\alpha$ and $\beta$, as  $e^{\Delta t \alpha} e^{\Delta t \beta} = e^{\Delta t (\alpha+\beta)} + \mathcal{O}(\Delta t^2 [\alpha,\beta])$ by discarding the $\Delta t^2$ terms in the small $\Delta t$ regime. This quantity can be made arbitrarily small by breaking up the total evolution time into sufficiently small pieces $\Delta t=\tau/N$. Bounding the error in the DAQC case reduces to computing $||[H_{I},H_{II}]||$. Breaking down each Hamiltonian into its $X$ and $Y$ components, such that $H_{i} = H^{XX}_{i} + H^{YY}_{i}$, simplifies the commutator norm to $||[H^{YY}_{I},H^{XX}_{II}] + [H^{XX}_{I},H^{YY}_{II}]||=||A+B||$. These operators are 
\begin{eqnarray}
\label{eq:A}
A &=& J^{2} \sum_{i,j,i',j'} \left[\sigma_y^{(2i,2j-1)} \sigma_y^{(2i+1,2j-1)} + \sigma_y^{(2i-1,2j)} \sigma_y^{(2i,2j)} \right.  \\
&& + \sigma_y^{(2i,2j-1)} \sigma_y^{(2i,2j)} + \sigma_y^{(2i-1,2j)} \sigma_y^{(2i-1,2j+1)}, \nonumber \\
  &&  \sigma_x^{(2i',2j'-1)} \sigma_x^{(2i'+1,2j'-1)}+\sigma_x^{(2i'-1,2j')} \sigma_x^{(2i',2j')} \nonumber \\
  && \left.+ \sigma_x^{(2i',2j'-1)} \sigma_x^{(2i',2j')} + \sigma_x^{(2i'-1,2j')} \sigma_x^{(2i'-1,2j'+1)}\right] \nonumber \\ 
  &=& - 2i J^{2} \sum_{i,j} \left[\sigma_x^{(2i,2j+1)}\sigma_y^{(2i+1,2j)}\sigma_z^{(2i+1,2j+1)} + \sigma_x^{(2i+1,2j-1}\sigma_y^{(2i,2j)}\sigma_z^{(2i,2j-1)} \right. \nonumber \\
 && \left. + \sigma_x^{(2i,2j}\sigma_y^{(2i-1,2j+1)}\sigma_z^{(2i-1,2j)} + \sigma_x^{(2i,2j-1)}\sigma_y^{(2i-1,2j)}\sigma_z^{(2i,2j)} + (y \leftrightarrow x)\right], \nonumber
\end{eqnarray}

\begin{eqnarray}
\label{eq:B}
B &=& J^{2} \sum_{i,j,i',j'} \left[\sigma_x^{(2i-1,2j-1)} \sigma_x^{(2i-1,2j)} + \sigma_x^{(2i-1,2j-1)} \sigma_x^{(2i,2j-1)} \right. \\
&&  + \sigma_x^{(2i,2j)} \sigma_x^{(2i,2j+1)} + \sigma_x^{(2i,2j)} \sigma_x^{(2i+1,2j)}, \nonumber \\
&& \sigma_y^{(2i'-1,2j'-1)} \sigma_y^{(2i'-1,2j')} + \sigma_y^{(2i'-1,2j'-1)} \sigma_y^{(2i',2j'-1)} \nonumber \\ 
&& \left. + \sigma_y^{(2i',2j')} \sigma_y^{(2i',2j'+1)} + \sigma_y^{(2i',2j')} \sigma_y^{(2i'+1,2j')}\right] \nonumber \\
\nonumber &=& 2i J^{2} \sum_{i,j} \left[\sigma_x^{(2i,2j)}\sigma_y^{(2i-1,2j+1)}\sigma_z^{(2i,2j+1)} + \sigma_x^{(2i+1,2j)}\sigma_y^{(2i,2j+1)}\sigma_z^{(2i,2j)}\right.  \\
&& \left. + \sigma_x^{(2i,2j)}\sigma_y^{(2i+1,2j-1)}\sigma_z^{(2i+1,2j)} +\sigma_x^{(2i,2j-1)}\sigma_y^{(2i-1,2j)}\sigma_z^{(2i-1,2j-1)} + (y \leftrightarrow x)\right].\nonumber 
\end{eqnarray}

Alternatively, from visually inspecting supports and Pauli character of the Hamiltonians $H_I$ and $H_{II}$ denoted in red and green in Fig.~\ref{Fig5chap6}, we can see that there are 8 terms per unit cell in A and that there are likewise 8 similar, but differently supported terms in B. Summing over the two sets of terms in the bulk, we obtain 

{\footnotesize
\begin{eqnarray}
    \nonumber ||[H_{I},H_{II}]|| &=& ||2i J^{2} \sum_{i,j} (-1)^{i+j}  \sigma_z^{(i,j)} [(\sigma_x^{(i-1,j)} \sigma_y^{(i,j-1)} - \sigma_x^{(i,j+1)} \sigma_y^{(i+1,j)}) + (x\leftrightarrow y)] || \\ 
    \nonumber &\le& 2 J^{2} \sum_{i,j} ||(-1)^{i+j} \sigma_z^{(i,j)} [(\sigma_x^{(i-1,j)} \sigma_y^{(i,j-1)} - \sigma_x^{(i,j+1)} \sigma_y^{(i+1,j)}) + (x\leftrightarrow y)] || \\
    &=& 2 J^{2} N^2 ||\sigma_z^{(i,j)} [(\sigma_x^{(i-1,j)} \sigma_y^{(i,j-1)} - \sigma_x^{(i,j+1)} \sigma_y^{(i+1,j)}) + (x\leftrightarrow y)]|| \leq 8 J^{2} N^2,\nonumber
    \label{eq:DATrottercommutator}
\end{eqnarray}
}
where we have used the triangle inequality on the spectral norms of the operators. 

In order to get a better insight into the performance of the DA computation of the two-dimensional $XY$ model, we need to compare the Trotter error of both DQC and DAQC approaches. This error is proportional to the commutator of $\left[H _I,H_{II} \right]$ given in Eq.~\ref{eq:DATrottercommutator} in the DAQC case. In the purely digital case, the commutator that we need to compute is $\left[H_{xx},H_{yy} \right]$, where $H_{xx}$ contains all $\sigma_x \sigma_x$ qubit interactions and $H_{yy}$ all the $\sigma_y\sigma_y$ interactions. Independent of the order in which the gates are implemented, the digital error is bounded by 
\begin{eqnarray}
\label{eq:DD_comm}
\nonumber ||[H_{xx},H_{yy}]|| &=& || J^{2} \sum_{i,j} [(\sigma_x^{(i,j)} \sigma_x^{(i+1,j)} + \sigma_x^{(i,j)} \sigma_x^{(i,j+1)}) ,  (x\leftrightarrow y)] || \\ 
&\leq& 24 J^{2} N^2
\end{eqnarray}
where the final factor arises from a product of the factor of two for the $N^2$ vertical and horizontal edges, a factor of $6$ counting all the non-commuting $\sigma_y\sigma_y$ neighbors of each $\sigma_x\sigma_x$ interaction, and a final factor of two arising from the Pauli commutation relations. Alternatively, by analyzing the form of Eqs.~\ref{eq:A} and \ref{eq:B} we note that the $A$ and $B$ components of the commutator can be identified with free fermions hopping along the diagonal loops of the two-dimensional lattice as defined by the blue arrows in Fig.~\ref{Fig5chap6}. Next, we Jordan-Wigner transform to a Majorana representation, take periodic boundary conditions, and Fourier transform along the loops. As a result, the spectral norm of $A$ and $B$ is tightened from $\mathcal{O}((NJ)^2) \rightarrow \mathcal{O}((J)^2)$ which removes the extensive factor. This tighter bound is proved in Appendix~\ref{AppDchap6}. Likewise, we may use similar techniques to decompose the Digital commutator of Eq.~\ref{eq:DD_comm} into a sum of three times as many free fermion Hamiltonians. The resulting ratio of purely DQC to DAQC commutator norms is still a factor of three. In either case, the DAQC protocol improves the Trotter error bound by a constant factor of three. This constant factor speedup can be used to extend the simulation time by the same factor.

\subsubsection{Heisenberg model}\label{sec633}

\begin{figure}[t]
\centering
\includegraphics[width=0.98 \textwidth]{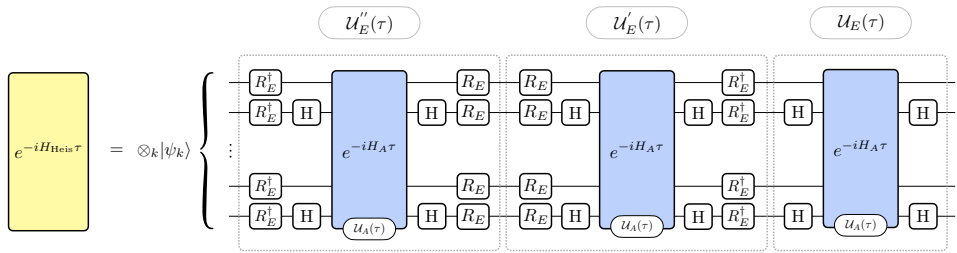}
\caption{{\bf Digital-analog quantum circuit to simulate the evolution of an initial quantum state under Hamiltonian $H_{\text{Heis}}$ for a time $\tau$.} This simulation is carried out by conjugating the analog propagator $\mathcal{U}_{A}$ -- which describes the evolution under analog Hamiltonian $H_{A}$ -- by Hadamard gates ($\text{H}$) on even qubits. It is additionally conjugated by $R_{E}^{2} = -R_{E}^{\dagger}$, where $R_{E}$ is a cyclic transformation that allows us to obtain all $S_{i}S_{i+1}$ interactions. This segment of the circuit simulates the evolution given by $\mathcal{U}_{E}^{''}(\tau)$. The first repetition of the circuit simulates the evolution given by $\mathcal{U}_{E}^{'}(\tau)$. This is done by conjugating $\mathcal{U}_{A}(\tau)$ by Hadamard gates on even qubits, followed by a permutation by $R_{E}$. Then, the circuit is repeated a final time, solely conjugating the analog propagator by Hadamard gates on even qubits, to simulate the evolution given by $\mathcal{U}_{E}(\tau)$.}
\label{Fig6chap6}
\end{figure}

We now consider the task of simulating the more complex Heisenberg spin model. The Hamiltonian describing the Heisenberg chain in 1 dimension is $H_{\text{Heis}}= \sum_i \bm{S}_i \cdot \bm{S}_{i+1}$, with $\bm{S}=(\sigma_x,\sigma_y,\sigma_z)$. Consider the general Bloch sphere rotation $R_{\hat{n}}(2\theta) = e^{-i\theta(\hat{n}\cdot\vec{\sigma})}$. We can set the angle $\theta$ such that this rotation is cyclic; that is, $\theta=\pi/(3\sqrt{3})$ leads to a cyclic permutation $\sigma_x\rightarrow \sigma_z$, $\sigma_y\rightarrow \sigma_x$ and $\sigma_z\rightarrow \sigma_y$. This transformation is realized by 
\begin{equation}
R_E \equiv R_{\hat{n}}\left(2\frac{\pi}{3\sqrt{3}}\right)=e^{-i\frac{\pi}{3\sqrt{3}}(\sigma_x+\sigma_y+\sigma_z)} = \frac{1}{2}\left[\mathbb{I}-i(\sigma_x+\sigma_y+\sigma_z)\right]
\end{equation}
which can easily be implemented on individual qubits by the Euler decomposition $R_E=e^{-i \sigma_y \frac{ \pi }{4}} e^{-i \sigma_z \frac{ \pi }{4}}$. The cyclic nature of this transformation is manifested through the property $R_{E}^{3}=-\mathbb{I}$. Applying this transformation on all qubits of the Hamiltonian in Eq.~\ref{H_even} zero, one, and two times, leads to the following Hamiltonians,
\begin{eqnarray}
H_E &=& H^{\text{even}} = J\sum_{k=1}^{\frac{N}{2}}\sigma_x^{(2k-1)}\sigma_x^{(2k)} + J\sum_{k=1}^{\frac{N-1}{2}}\sigma_z^{(2k)} \sigma_z^{(2k+1)}, \\
\nonumber H_E' &=& R_{E}^{\dagger}H^{\text{even}}R_{E} = J\sum_{k=1}^{\frac{N}{2}}\sigma_z^{(2k-1)}\sigma_z^{(2k)} + J\sum_{k=1}^{\frac{N-1}{2}}\sigma_y^{(2k)} \sigma_y^{(2k+1)},\\
\nonumber H_E''&=& R_{E}^{2 \dagger}H^{\text{even}}R_{E}^{2} = J\sum_{k=1}^{\frac{N}{2}}\sigma_y^{(2k-1)}\sigma_y^{(2k)} + J\sum_{k=1}^{\frac{N-1}{2}}\sigma_x^{(2k)} \sigma_x^{(2k+1)}.
\end{eqnarray}
Summing them together, we obtain the Heisenberg Hamiltonian,
\begin{equation}
H_{\text{Heis}} = H_{E} + H_{E}'+H_{E}'' = J\sum_{k=1}^{N-1}(\sigma_x^{(k)}\sigma_x^{(k+1)}+\sigma_y^{(k)}\sigma_y^{(k+1)}+\sigma_z^{(k)}\sigma_z^{(k+1)}).  
\end{equation}
In this case, the Hamiltonians do not commute with each other, which means that the construction of the propagator will include Trotter error (analyzed below),
\begin{eqnarray}
\nonumber \mathcal{U}_{\text{Heis}}(t) &=& e^{-i H_{\text{Heis}}t} = e^{-i(H_{E} + H'_{E} + H''_{E})t} \\
&=& \mathcal{U}_{E}(t)\: \mathcal{U}'_{E}(t)\: \mathcal{U}''_{E}(t) + \mathcal{O}(J^2 t^{2}).
\end{eqnarray}
The propagator $\mathcal{U}_{\text{Heis}}(t)$ is constructed as
\begin{eqnarray}
\mathcal{U}_{\text{Heis}}(t)|\psi\rangle &\approx& \mathcal{U}_{E}(t)\: \mathcal{U}'_{E}(t)\: \mathcal{U}''_{E}(t)|\psi\rangle \\
\nonumber &=& V_\text{H}^{\text{even}\, \dagger}\;\mathcal{U}_{A}(t) \: V_\text{H}^{\text{even}}\\
\nonumber && R_{E}^{\dagger}\, V_\text{H}^{\text{even}\dagger}\: \mathcal{U}_{A}(t)\, V_\text{H}^{\text{even}} \, R_{E}\\
&& R_{E}^{2\; \dagger}\, V_\text{H}^{\text{even}\, \dagger}\: \mathcal{U}_{A}(t)\,  V_\text{H}^{\text{even}}\, R_{E}^{2} |\psi\rangle,
\end{eqnarray}
where $\mathcal{U}_{A}(t)$ is the analog propagator generated by $H_{A}$, and $V_\text{H}^{\text{even}} = \bigotimes_{i} \text{H}^{(2i)}$. $\text{H}^{(k)}$ represents the application of a Hadamard gate on qubit $k$. This protocol comprises the following steps:
\begin{enumerate}
\item Prepare an initial product state $\otimes_{k}|\psi_{k}\rangle$.
\item Apply the cyclic transformation twice with $R_{E}^{2}$, which is equivalent to $R_{E}^{\dagger}$, on all qubits.
\item Apply Hadamard gates, $\text{H}$, on even qubits.
\item Let the states evolve according to the underlying analog Hamiltonian with analog propagator $\mathcal{U}_{A}(\tau)$ for time $\tau$.
\item Apply Hadamard gates, $\text{H}$, on even qubits.
\item Undo the double cyclic transformation by applying $R_{E}$ on all qubits.
\item Apply the cyclic transformation with $R_{E}$ on all qubits.
\item Apply Hadamard gates, $\text{H}$, on even qubits.
\item Let the states evolve according to the underlying analog Hamiltonian with analog propagator $\mathcal{U}_{A}(\tau)$ for time $\tau$.
\item Apply Hadamard gates, $\text{H}$, on even qubits.
\item Undo the cyclic transformation with $R_{E}^{\dagger}$ on all qubits.
\item Apply Hadamard gates, $\text{H}$, on even qubits.
\item Let the states evolve according to the underlying analog Hamiltonian with analog propagator $\mathcal{U}_{A}(\tau)$ for time $\tau$.
\item Apply Hadamard gates, $\text{H}$, on even qubits.
\end{enumerate}
This sequence of quantum gates constitutes a block, which can be seen in Fig.~\ref{Fig6chap6}. To evolve with Hamiltonian $H_{\text{Heis}}$ for a total time $T$, one must re-apply the block $M=T/\tau$ times. 

\subsubsection*{Digital vs. Digital-Analog Synthesis Errors}

In order to quantify the computational benefit of this method, let us compute and compare the above Trotterized error against that of a digitized two-local decomposition. A digitized decomposition we will employ alternating layers of $\sigma_x \sigma_x$, $\sigma_y \sigma_y$, and $\sigma_z \sigma_z$ interactions applied to all even bonds, followed by the same operator action on odd bonds. Such a decomposition is based on the fact that all interactions, on a single bond, commute, but the interactions on adjacent bonds, which share a single spin, do not commute. To first order, the Trotter error is given by 
\begin{eqnarray}
J^2\norm{[\bm{S}_{i-1}\cdot \bm{S}_i,\bm{S}_i \cdot \bm{S}_{i+1}]} &=& J^2\norm{\sum_{\mu,\nu}\sigma^\mu_{i-1} [\sigma^\mu_i,\sigma^\nu_i] \sigma^\nu_{i+1}} \nonumber \\
&=& 2 J^2\norm{\bm{S}_{i-1}\cdot\bm{S}_i \cross \bm{S}_{i+1}} \nonumber \\
 & \le & 12J^2, 
\end{eqnarray}
where we have used the fact that $\bm{S}_{i-1}\cdot\bm{S}_i \cross \bm{S}_{i+1}$ contains 6 Pauli terms. For a 1D Heisenberg chain, the total commutator is bounded by $12 J^2 N$. Meanwhile, on the DAQC side we need to bound 
\begin{eqnarray}
e^{-it (H_E+H_E'+H_E'')} &=& e^{-it H_E} e^{-it (H_E'+H_E'')} \\ &+& \mathcal{O}(t^2 [H_E,H_E'+H_E''])\nonumber \\ 
&=& e^{-it H_E} e^{-it H_E'} e^{-it H_E''}\nonumber  \\ 
&+& \mathcal{O}(t^2 ([H_E,H_E'+H_E''] + [H_E',H_E''])) \nonumber 
\end{eqnarray}
These commutators are
\begin{eqnarray}
[H_{E},H'_{E}] &=& 2 i J^2\sum_{k} \sigma_x^{(k-1)}\sigma_z^{(k)}\sigma_y^{(k+1)}, \\
\nonumber [H_{E},H''_{E}] &=& -2 i J^2\sum_{k} \sigma_z^{(k-1)}\sigma_x^{(k)}\sigma_y^{(k+1)}, \\
\nonumber [H'_{E},H''_{E}] &=& 2 i J^2\sum_{k} \sigma_z^{(k-1)}\sigma_y^{(k)}\sigma_x^{(k+1)},
\end{eqnarray}
and their sum can be bounded by
\begin{eqnarray}
\nonumber && ||[H_{E},H'_{E}] + [H_{E},H''_{E}] + [H'_{E},H''_{E}]|| \\
\nonumber &=& 2 J^{2} ||\sum_{k} \sigma_x^{(k-1)}\sigma_z^{(k)}\sigma_y^{(k+1)} - \sigma_z^{(k-1)}\sigma_x^{(k)}\sigma_y^{(k+1)} + \sigma_z^{(k-1)}\sigma_y^{(k)}\sigma_x^{(k+1)}|| \\
\nonumber & \leq & 2 J^{2} \sum_{k} || \sigma_x^{(k-1)}\sigma_z^{(k)}\sigma_y^{(k+1)} - \sigma_z^{(k-1)}\sigma_x^{(k)}\sigma_y^{(k+1)} + \sigma_z^{(k-1)}\sigma_y^{(k)}\sigma_x^{(k+1)}|| \\
&\leq& 6 J^{2}N.
\end{eqnarray}
We again find that the bound on the error in the DAQC protocol is smaller by a constant factor than in the DQC approach.

\subsection{Practical Implementation}\label{sec64}

In order to experimentally realize our DA simulation protocols in an accurate manner further practical experimental steps are required. The critical steps for doing so, whose details depend on the user's specific goals, are broadly partitioned as either i) characterization or ii) Hamiltonian optimization. Since each of these steps bring its own theoretical and experimental challenges, we now describe promising paths forward for each step.

A critical step towards validating the accuracy of DA simulations, thereby quantifying their error, is to accurately characterize the analog many-body Hamiltonian at the center of our protocols. While conceptually simple, the characterization of a many-body Hamiltonian is not scalable (with exponentially growing complexity) by naive process tomography~\cite{Nielsen2000}. To aid in the scalable characterization of our Hamiltonians, we note that all of the expected interactions are geometrically local and, using this information, one should take advantage of Hamiltonian tomography schemes with polynomial growing model spaces as constricted by locality
~\cite{Shabani2011, daSilva2011, Qi2017, Bairey2019}. Hamiltonian estimation is further complicated by interactions coupling the principle system to unwanted environmental degrees of freedom and, to address this complication, we point the interested reader to recently developed open quantum systems characterization techniques
~\cite{Bairey2020, Dumitrescu2020}. Additionally, Bayesian Hamiltonian learning
~\cite{Granade2012} techniques may also be considered, although efficient importance sampling is required to adequately update models in this case. 

After experimentally identifying the dominant interactions, a natural next step is to eliminate unwanted couplings. Our analog Hamiltonian arises from a model relying on a two-level approximation and perturbation theory in $\Omega/\delta$. However, it is known that the CR-operation comes with a variety of additional terms~\cite{Sheldon2016,Magesan2020,Malekakhlagh2020}, such as $\sigma_z^{(2)}, \sigma_y^{(2)}, \sigma_z^{(1)} \sigma_z^{(2)}$ as well as spectator phase errors, in practice. One may consider a few routes in order to combat these additional terms. For example, tailoring echo sequences can eliminate certain unwanted interactions~\cite{Ku2020} and, in addition, it has been shown that residual single qubit interactions can be removed by applying active cancellation tones
~\cite{Sheldon2016}. Another promising avenue for removing residual interactions comes from judiciously arranging, or actively controlling, qubit frequencies or their relative anharmonicities. For example, Ref.~\cite{Malekakhlagh2020} provides a detailed analysis of the role qubit frequencies play and have shown that certain bands in the space of frequency detunings (see regions I and IV in Fig.~4 (f)) maximize the signal-to-noise ratio $\abs{\frac{\sigma_z\sigma_x}{\sigma_z\sigma_z}}$. Even more recent work~\cite{Kandala2020} has highlighted how additional fixed frequency coupling elements, which dress the qubit level spacings, may also remove unwanted $\sigma_z\sigma_z$ interactions. 

Lastly, instead of removing the residual couplings, one may leverage the additional interactions to define new classes of analog Hamiltonians. These analog Hamiltonians would be useful in simulating the dynamics of different spin models. In the limit that these additional terms are sufficiently small, one would expect them to contribute as disorder or small fluctuations in the system parameters. In this case, the (low energy theory and effective) model is expected to still lie in the parent model's universality class. Alternatively, outside this limit, the presence of the additional terms may potentially enrich the computational capability of the analog Hamiltonian as applied to more complex spin models.

\subsection{Discussions}\label{sec65}

In this chapter, we start with a Hamiltonian based on the Rabi model describing two superconducting qubits interacting through the CR effect and propose an extension to a multi-qubit scenario. The resulting Hamiltonian is transformed into a reference frame where only two-body interactions remain, resulting in our analog Hamiltonian. With it, we have assembled a Hamiltonian toolbox through toggling by different single-qubit gates. 

The variety of Hamiltonians we have obtained were efficiently combined to simulate Ising, $XY$, and Heisenberg spin models on a 1-dimensional chain, as well as the $XY$ model on a 2-dimensional lattice. For the 1D Ising and $XY$ models, our simulation protocols are Trotter-error free up to first order in $\Omega/\delta$, meaning that the full time evolution is given by a single DA block. For the 2D $XY$ and 1D Heisenberg chain, we were able to reduce the error in a first order Trotter approximation by a constant factor of $3$ for 2D $XY$ and of $2$ for the Heisenberg chain. Our techniques, therefore, extend the duration of possible time evolutions by a constant factor. While the constant factor improvement does not provide a polynomial speedup in the asymptotic limit, it does provide a meaningful and practical advantage for near-term, noisy, simulations. A natural avenue of future research could be to explore the possible reach of quantum computation by offering a larger collection of analog Hamiltonians which naturally arise in superconducting platforms. Going beyond our simple Trotter analysis, it would also be interesting to investigate the scaling improvements resulting from the use of the DA Hamiltonians within more advanced product formulas \cite{Tran2020} or alternative Hamiltonian simulation techniques \cite{Berry2015}. 

Finally, we have provided a succinct discussion regarding the steps which are necessary to implement our DA protocols in practice. In doing so we outlined promising routes toward scalable characterization and tailoring the precise nature of the analog interactions. Another issue that must be tackled is the problem of geometrically designing the qubit detunings such that all qubits are kept within a particular range. Then, given these detunings, one should increase or decrease the individual driving to maintain a constant ratio $\Omega/\delta$ for all neighboring pairs. In reality, one must go beyond this simple approximation and will need to calibrate each of the individual drivings as the CR interaction may be highly sensitive to resonances that depend not only on the detuning but also on the qubit's anharmonicities~\cite{Malekakhlagh2020}.


\section[Conclusions and perspectives]{Conclusions and perspectives}\label{chapter7}
\fancyhead[LE]{\rightmark}
\fancyhead[RO]{\leftmark}


\thiswatermark{\put(1,-280){\color{l-grey}\rule{70pt}{42pt}}
\put(70,-280){\color{grey}\rule{297pt}{42pt}}}

%
%

\vfill

\lettrine[lines=2, findent=3pt,nindent=0pt]{I}{n} this thesis, we present a comprehensive overview of the work carried out in the field of DAQC over the course of four years. Our aim is to provide a comprehensive reference guide for the implementation of quantum algorithms within the DAQC framework. Through our research, we demonstrate that this computational paradigm presents a promising solution for the constraints of NISQ devices. We have shown that the DAQC framework enables the implementation of quantum algorithms with lower resource consumption, resulting in high-performing solutions. Additionally, we have equipped the reader with a toolbox for simulating various important Hamiltonians within the DAQC framework, which serve as building blocks for the quantum algorithms we have presented in this thesis. These algorithms demonstrate the full potential of the DAQC framework and its ability to effectively tackle complex quantum computing problems.

The thesis begins with an exploration of the construction of DA quantum algorithms based on a combination of single-qubit rotations and multiqubit entangling dynamics, referred to as digital and analog blocks respectively. We provide systematic protocols for constructing generic Ising, $2$-body, and $M$-body Hamiltonians. Through our analysis, we compare the bDAQC and sDAQC approaches of the paradigm and conclude that the bDAQC protocol enhances the fidelity of the global quantum dynamics by reducing the impact of switching on/off the entangling evolution. We also delve into the most relevant noise sources in current superconducting quantum processors and evaluate the resilience of the DAQC paradigm under realistic implementation circumstances. These results, in conjunction with the systematic protocols outlined in the thesis, enable us to successfully adapt several quantum algorithms.

Our first algorithm of interest is the QFT. This algorithm serves as a critical component in many other quantum algorithms, such as the QPE. To evaluate its performance under realistic conditions, we apply the noise model presented earlier and test the performance of both the QFT and the QPE using both the DQC and DAQC paradigms. Our results demonstrate that the fidelity achieved by the DAQC paradigm improves with an increasing number of qubits.

Next, we delve into the study of the HHL algorithm for solving linear systems of equations. Despite its broad range of applications, this algorithm poses a significant challenge for NISQ devices due to its high number of two-qubit gates and specific qubit connectivity requirements. Additionally, an efficient implementation of the AQE subroutine remains unclear. To tackle these challenges, we provide a systematic approach to implement the AQE subroutine using both the DQC and DAQC paradigms. To mitigate the impact of noise sources when running a quantum algorithm, we also propose a technique to reduce them by minimizing algorithm's depth. This is achieved by co-designing the quantum processor to be tailored to the specific implementation of the desired algorithm, thus reducing the number of SWAP gates needed to artificially connect physically unconnected qubits. We obtain a kite-like quantum processor for the implementation of a generalized version of the HHL algorithm, which is at the heart of many quantum subroutines across various fields. Our findings make a significant contribution to the field and highlight the importance of optimizing quantum algorithms for NISQ devices.

In our quest to extend the application of the DAQC paradigm, we explored its potential in the field of variational quantum algorithms. Our focus was on QAOA and we proposed its implementation using DAQC methods. Our results indicate that the banged DAQC approach provides a more efficient representation of the problem Hamiltonian, with reduced swapping requirements. Additionally, our findings demonstrate that, under the DAQC paradigm, QAOA is more robust to errors compared to traditional approaches, making it a promising candidate for NISQ devices aimed at solving combinatorial optimization problems. This study, together with Ref. \cite{Michel2023}, marks a significant step towards integrating variational algorithms into the DAQC framework and holds great potential for future research and exploration.

Finally, focusing on superconducting circuits quantum processors, we use the cross-resonance effect as the analog component of the DAQC paradigm. We proposed an extension of the cross-resonance interaction between two superconducting qubits to a multiqubit system. By transforming the resulting Hamiltonian into a reference frame that only includes two-body interactions, we were able to create a toolbox of analog Hamiltonians by toggling different single-qubit gates. Our results, combined with the proposed architecture in \cite{Albarran2022}, provide a strong foundation for continued exploration of quantum algorithms implementation using current technology.

In conclusion, this thesis has presented a comprehensive study on the application of the DAQC paradigm to a wide range of quantum algorithms. Through the combination of single-qubit rotations and multiqubit entangling dynamics, it has been demonstrated a systematic approach to construct generic Ising, $2$-body, and $M$-body Hamiltonian. Our findings showed that the bDAQC protocol offers improved fidelity compared to the stepwise approach, and our noise analysis revealed that it is possible to reduce the impact of noise sources by co-designing the quantum processor. We have applied the DAQC paradigm to the QFT, QPE, HHL, and QAOA quantum algorithms and have shown that it is possible to improve their performance under realistic noise circumstances. Additionally, the study of CR dynamics within the DAQC paradigm provides a promising direction for future research in the field of superconducting circuits quantum processors. Overall, this thesis provides a significant contribution to the field of quantum computing and demonstrates the potential of the DAQC paradigm as a robust and flexible framework for implementing quantum algorithms on NISQ devices. We believe that our work will serve as a foundation for further research and development in this area.

The results of this thesis demonstrate the potential of the DAQC paradigm in solving various problems in quantum computing. With the development of new and more advanced quantum processors, the potential applications of DAQC algorithms will only continue to grow. Furthermore, the study of these algorithms sheds light on the possible techniques to optimize the implementation of quantum algorithms in NISQ devices and reduce their susceptibility to errors. As we look to the future, it is clear that there is much room for further exploration in this field, as the DAQC paradigm has the potential to play a critical role in the development of practical quantum computing solutions. The future direction of this work could include studying the scalability of DAQC algorithms, investigating the use of DAQC in hybrid quantum-classical algorithms, and exploring new techniques to mitigate errors and improve performance. In any case, this thesis represents an important step toward unlocking the full potential of DAQC for quantum computing.


\titleformat{\section}[display]
{\vspace*{160pt}
\bf\sffamily \Huge}
{{\textcolor{black}{\thesection}}. #1}
{0pt}
{#1}
[]
\titlespacing*{\section}{75pt}{10pt}{40pt}[40pt]

\section*{Appendices}
\phantomsection
\addcontentsline{toc}{section}{Appendices}

\fancyhead[RO]{APPENDICES}

\cleardoublepage

\appendix

\fancyhead[RO]{\leftmark}

\vfill
\titleformat{\section}[display]
{\vspace*{150pt}
\bfseries\sffamily \LARGE}
{\begin{picture}(0,0)\put(-64,-31){\textcolor{black}{\thesection}}\end{picture}}
{0pt}
{\textcolor{black}{#1}}
[]
\titlespacing*{\section}{80pt}{10pt}{50pt}[0pt]
\vfill
\section{Additional details and proofs regarding the fundamentals of DAQC}

\fancyhead[RO]{APP. A\quad ADDITIONAL PROOFS REGARDING DAQC}

\subsection{Universality}
\label{Appsec:Universality}
We briefly recall that a machine able to implement single qubit rotations and an entangling gate, e.g.,  a controlled-phase gate, has the ability to perform universal quantum computation efficiently. It can be easily proven that a two-qubit ZZ gate, $ZZ_{ij}(\Phi)=e^{-i \Phi \sigma_z^{(i)} \sigma_z^{(j)}}$, combined with single qubit rotations can be used to implement a controlled-phase gate $CZ_{ij}(\phi)=(Z_i(-\phi)\otimes Z_j(-\phi))ZZ_{ij}(\phi/2)$ (up to a general phase) where
\begin{equation}
Z(\phi)=\begin{pmatrix} 1 &0\\ 0 & e^{i\phi}\end{pmatrix},\,\, 
CZ(\phi)=\begin{pmatrix} 1 &0 &0 &0\\
0 &1 &0&0\\
0 & 0 &1&0\\
0&0&0& e^{-i2\phi}\end{pmatrix}.
\end{equation}

\subsection{Negative times in the Ising model}
\label{AppSec:Negative_times}

Some of the times $t_\alpha= \mathsf{M}_{\alpha\beta}^{-1}g_\beta(t_F/g)$ of the analog blocks to implement the Ising model can be negative. In an sDAQC protocol, one solution would be to do such analog evolutions with inverted coupling signs. However, in the bDAQC paradigm, we must keep the base entangling Hamiltonian $\bar{H}_{ZZ}$ untouched. There is a simple way to mimic such behavior by using an equivalent set of times that produces the same evolution. 

Given that the vector of times $t_{\alpha}=(t_1,t_2,...,t_{\mathrm{min}},...)$ with $t_{\mathrm{min}}<0$ and $t_{\mathrm{min}}< t_\alpha\,\forall \alpha$, there is an equivalent evolution with $N(N-1)/2-1$ blocks of times $\tilde{t}_{\alpha}=t_{\alpha}+|t_{\mathrm{min}}|e_1$ and an extra analog block with time $|t_{\mathrm{min}}|\lambda_1$, where $\lambda_1$ is the eigenvalue of a matrix $\mathsf{M}_{\alpha\beta}$ with corresponding constant eigenvector $e_1=\gamma(1,1,1...)^T$.

Let us first focus on the properties of the matrix $\mathsf{M}_{\alpha\beta}$ defined in Eq. (\ref{eq2.9}), created from the signs of applying sequentially gates $\sigma_x^{(n)}\sigma_x^{(m)}$ before and after the analog block evolutions. This matrix has an eigenvector $e_1=\gamma(1,1,1...)^T$ which corresponds to the eigenvalue $\lambda_1=N(N-9)/2+8$, that is negative for $N\in\{3,5,6\}$, and positive thereafter $N\in \mathbb{Z}\ge7$. Notice that we omit the case for $N=4$ as the matrix is non-invertible and different set of gates must be performed.

For $N<7$ we observe the following identity
\begin{eqnarray}
\mathsf{M}_{\alpha\beta}\left(t_{\alpha}+|t_{\mathrm{min}}|e_1-|t_{\mathrm{min}}|e_1\right)&=&\mathsf{M}_{\alpha\beta}\left(t_{\alpha}+|t_{\mathrm{min}}| e_1\right)-\lambda_1|t_{\mathrm{min}}|e_1\nonumber\\
&=&\mathsf{M}_{\alpha\beta}\tilde{t}_\alpha+|\lambda_1t_{\mathrm{min}}|e_1
\end{eqnarray}
which corresponds to an evolution with $N(N-1)/2-1$ blocks of times $\tilde{t}_\alpha$.  One block has zero time ($t_{\mathrm{min}}+|t_{\mathrm{min}}|=0$), and there is an extra analog block of time $|\lambda_1 t_{\mathrm{min}}|$ not sandwiched by any SQR.  The evolution decomposes into
\begin{equation}
	U_{ZZ}(t_F)=e^{i\sum_{\beta}g\mathsf{M}_{\alpha\beta}\tilde{t}_\alpha\sigma_z^{(j)} \sigma_z^{(k)}}e^{i|\lambda_1 t_{\mathrm{min}}| \sum_{\beta}g \sigma_z^{(j)} \sigma_z^{(k)}},
\end{equation}
where the dependence of $j,k$ on $\beta$ has not been explicitly written and is the same as in Eq. (\ref{eq:H_beta}).

\subsection{Independent coupling parameters in XZ model}
\label{Appsec:XZmodel_couplings}
The implementation of the XZ model with the protocol of Subsec. \ref{sec21} requires the solution of the following linear system
\begin{equation}
	\left[\alpha_j^{(\mu,s)}\alpha_k^{(\nu,s)}\right]\begin{pmatrix}
	g_{12}^{(1)}\\
	g_{12}^{(2)}\\
	g_{12}^{(3)}\\
	g_{12}^{(4)}\\
	g_{13}^{(1)}\\
	\vdots\\
	g_{N-1,N}^{(4)}
	\end{pmatrix}=\begin{pmatrix}
	g_{12}^{XX}\\
	g_{12}^{XZ}\\
	g_{12}^{ZX}\\
	g_{12}^{ZZ}\\
	g_{13}^{XX}\\
	\vdots\\
	g_{N-1,N}^{ZZ}
	\end{pmatrix},
\end{equation}
with the matrix  
\begin{eqnarray}
	\left[\alpha_j^{(\mu,s)}\alpha_k^{(\nu,s)}\right]=\bigoplus_{j<k}^N
	\begin{pmatrix} 
	S_{\theta_j}^1 S_{\theta_k}^1 & S_{\theta_j}^2S_{\theta_k}^2&S_{\theta_j}^3 S_{\theta_k}^3 & S_{\theta_j}^4 S_{\theta_2}^4\\
	S_{\theta_j}^1C_{\theta_k}^1 & S_{\theta_j}^2C_{\theta_k}^2&S_{\theta_j}^3 C_{\theta_k}^3 & S_{\theta_j}^4 C_{\theta_k}^4\\
	C_{\theta_j}^1S_{\theta_k}^1 & C_{\theta_j}^2S_{\theta_k}^2&C_{\theta_j}^3 S_{\theta_k}^3 & C_{\theta_j}^4 S_{\theta_k}^4\\
	C_{\theta_j}^1C_{\theta_k}^1 & C_{\theta_j}^2C_{\theta_k}^2&C_{\theta_j}^3 C_{\theta_k}^3 & C_{\theta_j}^4 C_{\theta_k}^4\\
	\end{pmatrix},\nonumber\\
\end{eqnarray}
where we have defined the parameters
\begin{equation}
S_{\theta_j}^s=\sin(\theta_j^{(s)})=\alpha_j^{(x,s)}\qquad \text{and } \qquad C_{\theta_j}^s=\cos(\theta_j^{(s)})=\alpha_j^{(z,s)}. \nonumber
\end{equation}
This matrix is invertible for a dense set of phase values. That is, the sets of phases that make it singular has measure zero. From a practical perspective, we do not want eigenvalues close to zero, because after inversion we would have long  simulating times. One useful and well-behaved array of phases is $\theta_w^{(s)}=\frac{s\pi(w)}{2(w+1)}$, with distance between two nearest-neighbour qubit phases scaling polynomially as
\begin{equation}
d(s,w)=|\theta_w^{(s)}-\theta_{w+1}^{(s)}|=\frac{s\pi}{2(w^2+3w+2)}.\nonumber
\end{equation} 

\subsection{$\mathbf{M}$-body Hamiltonians}
\label{Appsec:Mbody}
We extend here the steps explained in subsection \ref{sec22} to simulate a NN Hamiltonian evolution with up to $4$-body interactions like 
\begin{eqnarray}
H_{4b}&=&\sum_{j,\chi\eta} g_{(2,j)}^{\chi\eta}\sigma_{\chi}^{(j)}\sigma_{\eta}^{(j+1)}+\sum_{j,\chi\eta\gamma} g_{(3,j)}^{\chi\eta\gamma}\sigma_{\chi}^{(j)}\sigma_{\eta}^{(j+1)}\sigma_{\gamma}^{(j+2)}\nonumber\\
&&+\sum_{j,\chi\eta\gamma\rho}g_{(4,j)}^{\chi\eta\gamma\rho}\sigma_{\chi}^{(j)}\sigma_{\eta}^{(j+1)}\sigma_{\gamma}^{(j+2)}\sigma_{\rho}^{(j+3)},
\end{eqnarray}
where $\{\chi,\eta,\gamma,\rho\}=\{x,y,z\}$ and $j=\{1,...,N\}$, starting with NN fixed coupling ZZ Ising models. To create terms with support in all interactions by a generalized M{\o}lmer-S{\o}rensen type of gate, we need to interleave inhomogeneous Ising Hamiltonians with two different and rotated XX-Ising evolutions as
\begin{eqnarray}
H_1=e^{-iO_{XX}^{1}}H_{ZZ}^1e^{iO_{XX}^{1}},\\
H_2=e^{-iO_{XX}^{2}}H_{ZZ}^2e^{iO_{XX}^{2}},
\end{eqnarray}
where 
\begin{equation*}
    O_{XX}^{1}=\Phi_1\sigma_x^{(1)}\sigma_x^{(2)}+\Phi_3\sigma_x^{(3)}\sigma_x^{(4)}+\Phi_5\sigma_x^{(5)}\sigma_x^{(6)}+...
\end{equation*}
and its translationally shifted 
\begin{equation*}
    O_{XX}^{2}=\Phi_2\sigma_x^{(2)}\sigma_x^{(3)}+\Phi_4\sigma_x^{(4)}\sigma_x^{(5)}+\Phi_6\sigma_x^{(6)}\sigma_x^{(7)}+...
\end{equation*}
are built from evolutions of ZZ models rotated with SQRs in all qubits as
\begin{equation*}
    R=\otimes_j^N\left(\cos(\pi/4)\sigma_z^{(j)}+\sin(\pi/4)\sigma_x^{(j)}\right).
\end{equation*}
For $M=4$, $O_{XX}^{1}$ and $O_{XX}^{2}$ contain interacting operators separated by the interaction length $L=M/2=2$, e.g.,  $O_{XX}^{1}$ has a term $\sigma_x^{(1)}\sigma_x^{(2)}$ but not $\sigma_x^{(2)}\sigma_x^{(3)}$. Had we wanted to simulate a five-/six- (seven-/eight-) body Hamiltonian, we would need a different decomposition with 3 (4) translationally invariant sets of blocks. The ZZ-Ising NN Hamiltonians $H_{ZZ}^{s}=\sum_j g_j^{s} \sigma_z^{(j)}\sigma_z^{(j+1)}$, with $s=\{1,2\}$. 

$H_1$ contains all two-body and three-body terms with different supports but not in four-body terms, i.e.,  for a chain of 8 qubits it looks like
\begin{eqnarray}
	H_1&=&g_1^1\cos(2\theta_2)\sigma_z^{(1)}\sigma_z^{(2)}+g_1^1\sin(2\theta_2)\sigma_z^{(1)}\sigma_y^{(2)}\sigma_x^{(3)} +g_2^1\sigma_z^{(2)}\sigma_z^{(3 )}\nonumber\\
	&+&g_3^1\cos(2\theta_2)\cos(2\theta_4)\sigma_z^{(3)}\sigma_z^{(4 )}+g_3^1\sin(2\theta_2)\cos(2\theta_4)\sigma_x^{(2)}\sigma_y^{(3)}\sigma_z^{(4)}\nonumber\\
	&+&g_3^1\cos(2\theta_2)\sin(2\theta_4)\sigma_z^{(2)}\sigma_y^{(3)}\sigma_x^{(4)}+g_3^1\sin(2\theta_2)\sin(2\theta_4)\sigma_x^{(2)}\sigma_y^{(3)}\sigma_y^{(4)}\sigma_x^{(5)}\nonumber\\
	&+&g_4^1\sigma_z^{(4)}\sigma_z^{(5)} +g_5^1\cos(2\theta_4)\cos(2\theta_6)\sigma_z^{(5)}\sigma_z^{(6)}\nonumber\\
        &+&g_5^1\sin(2\theta_4)\cos(2\theta_6)\sigma_x^{(4)}\sigma_y^{(5)}\sigma_z^{(6)}\nonumber\\
	&+&g_5^1\cos(2\theta_4)\sin(2\theta_6)\sigma_z^{(4)}\sigma_y^{(5)}\sigma_x^{(6)}+g_5^1\sin(2\theta_4)\sin(2\theta_6)\sigma_x^{(4)}\sigma_y^{(5)}\sigma_y^{(6)}\sigma_x^{(7)}\nonumber\\
	&+&g_6^1\sigma_z^{(6)}\sigma_z^{(7)} +g_7^1\cos(2\theta_6)\sigma_z^{(7)}\sigma_z^{(8)}+g_7^1\sin(2\theta_6)\sigma_x^{(6)}\sigma_y^{(7)}\sigma_z^{(8)}.\label{eq:M_body_H_1}	
\end{eqnarray}

On the other hand, $H_2$ contains (again) terms in all supports for two-body and three-body interactions and the complementary four-body terms 
\begin{eqnarray}
H_2&=&g_1^2\sigma_z^{(1)}\sigma_z^{(2)}+g_2^2\cos(2\theta_1)\cos(2\theta_3)\sigma_z^{(2)}\sigma_z^{(3)}\nonumber\\
&+&g_2^2\sin(2\theta_1)\cos(2\theta_3)\sigma_x^{(1)}\sigma_y^{(2)}\sigma_z^{(3)}\nonumber\\
&+&g_2^2\cos(2\theta_1)\sin(2\theta_3)\sigma_z^{(2)}\sigma_y^{(3)}\sigma_x^{(4)} +g_2^2\cos(2\theta_1)\cos(2\theta_3)\sigma_x^{(1)}\sigma_y^{(2)}\sigma_y^{(3)}\sigma_x^{(4)}\nonumber\\
&+&g_3^2\sigma_z^{(3)}\sigma_z^{(4)}+g_4^2\cos(2\theta_3)\cos(2\theta_5)\sigma_z^{(4)}\sigma_z^{(5)}\nonumber\\
&+&g_4^2\sin(2\theta_3)\cos(2\theta_5)\sigma_x^{(3)}\sigma_y^{(4)}\sigma_z^{(5)}+g_4^2\cos(2\theta_3)\sin(2\theta_5)\sigma_z^{(4)}\sigma_y^{(5)}\sigma_x^{(6)}\nonumber\\
&+&g_4^2\cos(2\theta_3)\cos(2\theta_5)\sigma_x^{(3)}\sigma_y^{(4)}\sigma_y^{(5)}\sigma_x^{(6)} + g_5^2\sigma_z^{(5)}\sigma_z^{(6)}\nonumber\\
&+&g_6^2\cos(2\theta_5)\cos(2\theta_7)\sigma_z^{(6)}\sigma_z^{(7)}\nonumber\\
&+&g_6^2\sin(2\theta_5)\cos(2\theta_7)\sigma_x^{(5)}\sigma_y^{(6)}\sigma_z^{(7)}+g_6^2\cos(2\theta_5)\sin(2\theta_7)\sigma_z^{(6)}\sigma_y^{(7)}\sigma_x^{(8)}\nonumber\\
&+&g_6^2\sin(2\theta_5)\sin(2\theta_7)\sigma_x^{(5)}\sigma_y^{(6)}\sigma_y^{(7)}\sigma_x^{(8)}+g_7^2\sigma_y^{(7)}\sigma_z^{(8)}.\label{eq:M_body_H_2}
\end{eqnarray}

The constant coefficients of operators in (\ref{eq:M_body_H_1}) and (\ref{eq:M_body_H_2}) are entangled in groups of maximum size 4. For the simulation of the four-body generalized Hamiltonian, it is thus enough to sum 4 of each of the blocks $H_0=\sum_{k=1}^4 H_1^{(k)}+H_2^{(k)}$ to decouple the parameters of at least one term operating in each support. We have again a dense set of phases such that randomly chosen ones would make the system of equations invertible. A particular choice of sets that would work are $\theta_1^{(k)}=\theta_{1+4n}^{(k)}=\theta_2^{(k)}=\theta_{2+4n}^{(k)}=(2\pi k/3)$ and $\theta_3^{(k)}=\theta_{3+4n}^{(k)}=\theta_4^{(k)}=\theta_{4+4n}^{(k)}=(2\pi k/5)$ with $n=\{1,\dots,[N/4]\}$ and $k=\{1,\dots4\}$.

Finally, we use more local rotations to generate the arbitrary $M\le4$ Hamiltonian. In particular, we need to concatenate $3^{M}=81$ $H_0$ blocks, maximum number of independent parameters in an $M$-body interaction,interleaved by generalized SQRs $R^{(l)}=\otimes_j^N r_j^{(l)}\sigma_x^{(j)}+s_j^{(l)}\sigma_y^{(j)}+t_j^{(l)}\sigma_z^{(j)}$
\begin{equation}
H_{4b}=\sum_l R^{(l)} H_0^{(l)} R^{(l)},
\end{equation}
where $(r_j^{(l)},s_j^{(l)},t_j^{(l)})$ are unit-sphere cartesian decompositions that fulfill the constraint $|r_j^{(l)}|^2+|s_j^{(l)}|^2+|t_j^{(l)}|^2=1$. As it is common when simulating non-commuting Hamiltonians, we need to repeat the whole process for each Trotter step $e^{-iH_{4b}t}\approx (e^{-iH_{4b}t/n_T})^{n_T}$ to approximate the evolution.

\titleformat{\section}[display]
{\vspace*{150pt}
\bfseries\sffamily \LARGE}
{\begin{picture}(0,0)\put(-64,-29){\textcolor{black}{\thesection}}\end{picture}}
{0pt}
{\textcolor{black}{#1}}
[]
\titlespacing*{\section}{80pt}{10pt}{50pt}[0pt]
\vfill
\section{Detailed calculations regarding the cross-resonance effect}

\fancyhead[RO]{APP. B\quad DETAILS ON THE CROSS-RESONANCE EFFECT}

\subsection{CR Hamiltonian}\label{AppAchap6}

Here we provide further details of the derivation of the effective Hamiltonians described in section \ref{sec61} of chapter \ref{chapter6}.

\subsubsection{Two-qubit case} \label{AppA1chap6} 
The transformation that takes the Hamiltonian in Eq.~\ref{eq:Ham2Q} into a doubly rotating frame is given by
\begin{equation}
U_{12} = \exp\left[ -\frac{i}{2}((\omega_{1}t + \phi_{1})z_{1} + (\omega_{2}t + \phi_{2})z_{2})\right].
\end{equation}
This operation results in
\begin{eqnarray}
H_{2} &=& \frac{1}{2}(\delta_{1}z_{1} + \delta_{2}z_{2}) + \Omega_{1}\cos(\omega_{1}t+\phi_{1})[x_{1}\cos(\omega_{1}t+\phi_{1})-y_{1}\sin(\omega_{1}t+\phi_{1})] \nonumber \\ 
&+& \Omega_{2}\cos(\omega_{2}t+\phi_{2})[x_{2}\cos(\omega_{2}t+\phi_{2})-y_{2}\sin(\omega_{2}t+\phi_{2})] \nonumber \\
&+&\frac{g}{2}[x_{1}\cos(\omega_{1}t+\phi_{1})-y_{1}\sin(\omega_{1}t+\phi_{1})]\nonumber\\
&&[x_{2}\cos(\omega_{2}t+\phi_{2})-y_{2}\sin(\omega_{2}t+\phi_{2})],
\end{eqnarray}
with $\delta_{k} = \omega_{k}^{q} - \omega_{k}$. Next a rotating wave approximation (RWA) is performed by dropping terms proportional to $e^{\pm 2i\omega_{1}}$, $e^{\pm 2i\omega_{2}}$, and $e^{\pm i(\omega_{1}+\omega_{2})}$. The validity of this approximation relies on a time-average of the Hamiltonian and noting that $\Omega/(\omega_i+\omega_j)\ll 1$ and $g/(\omega_i+\omega_j)\ll 1$, $\forall i,j$. The remaining terms are either static, or rotating at $\pm(\omega_{1}-\omega_{2})$:
\begin{eqnarray}
H_{2} &=& \frac{1}{2}(\delta_{1}z_{1} + \delta_{2}z_{2}) + \frac{1}{2}(\Omega_{1}x_{1} + \Omega_{2}x_{2}) \nonumber\\
&+& \frac{g}{4}\left[ \cos\varphi_{12}(t) (x_{1}x_{2} + y_{1}y_{2}) + \sin\varphi_{12}(t) (x_{1}y_{2} - y_{1}x_{2}) \right],
\end{eqnarray}
where we defined $\varphi_{12}(t) = (\omega_{1}-\omega_{2})t + \phi_{1} - \phi_{2}$. Next we apply the rotation
\begin{equation}
U_{3} = \exp\left[ \frac{i}{2}(\xi_{1} y_{1} + \xi_{2} y_{2}) \right],
\end{equation}
with $\tan\xi_{k} = \delta_{k}/\Omega_{k}$. The resulting Hamiltonian is
\begin{eqnarray}
H_{3} &=& \frac{1}{2}\left(\frac{\Omega_{1}}{\cos\xi_{1}}x_{1} + \frac{\Omega_{2}}{\cos\xi_{2}}x_{2}\right) \nonumber \\
&+& \frac{g}{4} \{ \cos\varphi_{12}(t)[(x_{1}\cos\xi_{1}-z_{1}\sin\xi_{1})(x_{2}\cos\xi_{2}-z_{2}\sin\xi_{2}) + y_{1}y_{2}]  \nonumber \\
&&+ \sin\varphi_{12}(t)[(x_{1}\cos\xi_{1}-z_{1}\sin\xi_{1})y_{2} - y_{1}(x_{2}\cos\xi_{2}-z_{2}\sin\xi_{2}) ] \},
\end{eqnarray}
where we have used $\delta_{k}\cos\xi_{k}-\Omega_{k}\sin\xi_{k} = 0$ and $\delta_{k}\sin\xi_{k}+\Omega_{k}\cos\xi_{k} = \Omega_{k}/\cos\xi_{k}$. 
The last transformation is given by
\begin{equation}
U_{4} = \exp\left[ -\frac{it}{2}(\eta_{1}x_{1} + \eta_{2}x_{2})\right],
\end{equation}
where $\eta_{k} = \sqrt{\delta_{k}^{2} + \Omega_{k}^{2}}$, such that $\eta_{k} = \Omega_{k}/\cos\xi_{k} = \delta_{k}/\sin\xi_{k}$. This takes our Hamiltonian into the quad frame (QF),
\begin{eqnarray}
H_{4} &=& \frac{g}{4} \{ \cos\varphi_{12}(t) [ x_{1}x_{2}\cos\xi_{1}\cos\xi_{2} - x_{1}\cos\xi_{1}\sin\xi_{2}(z_{2}\cos\eta_{2}t + y_{2}\sin\eta_{2}t) \nonumber \\
&&- (z_{1}\cos\eta_{1}t + y_{1}\sin\eta_{1}t)x_{2}\sin\xi_{1}\cos\xi_{2} \nonumber\\
&&+ (z_{1}\cos\eta_{1}t + y_{1}\sin\eta_{1}t)(z_{2}\cos\eta_{2}t + y_{2}\sin\eta_{2}t)\sin\xi_{1}\sin\xi_{2} \nonumber \\
&&+ (y_{1}\cos\eta_{1}t - z_{1}\sin\eta_{1}t)(y_{2}\cos\eta_{2}t - z_{2}\sin\eta_{2}t) ] \nonumber\\
&&+ \sin\varphi_{12}(t) [ x_{1}\cos\xi_{1}(y_{2}\cos\eta_{2}t - z_{2}\sin\eta_{2}t) \nonumber \\
&&- (z_{1}\cos\eta_{1}t + y_{1}\sin\eta_{1}t)(y_{2}\cos\eta_{2}t - z_{2}\sin\eta_{2}t)\sin\xi_{1} \nonumber\\
&&- (y_{1}\cos\eta_{1}t - z_{1}\sin\eta_{1}t)x_{2}\cos\xi_{2} \nonumber \\
&&+ \sin\xi_{2}(y_{1}\cos\eta_{1}t - z_{1}\sin\eta_{1}t)(z_{2}\cos\eta_{2}t + y_{2}\sin\eta_{2}t)] \}.
\end{eqnarray}

Now, we consider the scenario in which we drive the first qubit at the resonance frequency of the second qubit by imposing that $\omega_1 = \omega_{2}^{q}$, while the second qubit is not driven, i.e. $\Omega_{2}=0$, $\eta_{2}=0$, $\delta_{2}=0$, $\omega_{2} = \omega_{2}^{q}$, $\xi_{2}=0$, $\phi_{2}=0$ which implies $\varphi_{12}(t) = \phi_{1}$. The resulting Hamiltonian is
\begin{eqnarray}
\nonumber H_{4} &=& \frac{g}{4} \{ \cos\phi_{1} [ x_{1}x_{2}\cos\xi_{1} - (z_{1}\cos\eta_{1}t + y_{1}\sin\eta_{1}t)x_{2}\sin\xi_{1} \nonumber\\
&&+ (y_{1}\cos\eta_{1}t - z_{1}\sin\eta_{1}t)y_{2} ] \\
&+& \sin\phi_{1} [ x_{1}y_{2}\cos\xi_{1} - (z_{1}\cos\eta_{1}t + y_{1}\sin\eta_{1}t)y_{2}\sin\xi_{1} \nonumber\\
&&- (y_{1}\cos\eta_{1}t - z_{1}\sin\eta_{1}t)x_{2} \},
\end{eqnarray}
where we see that static terms have developed from the slowly rotating terms we kept in the RWA, since with the cross-resonant driving $\omega_{1}-\omega_{2}=\omega_{2}^{q}-\omega_{2}=0$. Finally, we perform a second RWA by dropping any term proportional to $e^{\pm i\eta_{1}t}$, and keep only the static terms. Additionally, we consider the weak-driving regime ($\Omega_{1}/\delta_{1}\ll 1$), which simplifies $\cos\xi_{1}\approx \Omega_{1}/\delta_{1}$. Then, we arrive at the Hamiltonian
\begin{equation}
H_{4} = \frac{g\Omega_{1}}{4\delta_{1}} (x_{1}x_{2}\cos\phi_{1}+x_{1}y_{2}\sin\phi_{1}),
\end{equation}
presented in Eq.~\ref{eq:H_QF}. The validity of this approximation relies on $g/\eta_{1}\approx g/\delta_{1}\ll 1$, which is enforced in the weak-coupling regime. See that the remaining terms after this second RWA are those we kept as slow-rotating after the first RWA, and the terms neglected in this case oscillate with $\delta_{1}=\omega_{1}^{q}-\omega_{1}=\omega_{1}^{q}-\omega_{2}^{q}$. 

\subsubsection{N-qubit case}\label{AppA2chap6}
We start with the N-qubit Hamiltonian in the laboratory frame, given by Eq.~\ref{eq:HamilNq} in the main text. We can move to the QF by applying the following transformations
\begin{eqnarray}
U_{12} &=& \exp\left[ -\frac{i}{2}\sum_{k=1}^{N}(\omega_{k}t + \phi_{k})z_{k}\right], \\
U_{3} &=& \exp\left[ \frac{i}{2}\sum_{k=1}^{N}\xi_{k} y_{k} \right], \qquad \text{and } \\
U_{4} &=& \exp\left[ -\frac{it}{2}\sum_{k=1}^{N}\eta_{k} x_{k} \right].
\end{eqnarray}
Now, as stated in the main text, we drive all qubits at the resonance frequency of their neighbour to the right (except for the last one when applicable). This implies that $\omega_{k}=\omega_{k+1}^{q}$, $\varphi_{k}(t) = \delta_{k+1}t + \phi_{k} - \phi_{k+1}$ and, in the weak-driving regime $\Omega_{k} \ll \delta_{k}$, $\eta_{k}\approx \delta_{k}$. This results in
\begin{eqnarray}\label{H_QF_o}
H_{4} &=& \frac{1}{4} \sum_{k=1}^{N-1} g_{k}  \bigg\{ \cos(\delta_{k}t + \phi_{k}-\phi_{k+1}) (y_{k}y_{k+1}+z_{k}z_{k+1}) \nonumber\\
&+& \sin(\delta_{k}t + \phi_{k}-\phi_{k+1}) (y_{k}z_{k+1}-z_{k}y_{k+1}) \nonumber\\
&+& \frac{\Omega_{k}}{\delta_{k}} \Big[ \sin(\phi_{k}-\phi_{k+1}) x_{k}y_{k+1} - \cos(\phi_{k}-\phi_{k+1}) x_{k}z_{k+1} \Big] \nonumber\\
&-& \frac{\Omega_{k+1}}{\delta_{k+1}} \Big[ \sin[(\delta_{k}+\delta_{k+1})t + \phi_{k}-\phi_{k+1}] y_{k}x_{k+1}\nonumber\\
&+& \cos[(\delta_{k}+\delta_{k+1})t + \phi_{k}-\phi_{k+1}] z_{k}x_{k+1} \Big] \bigg\}.
\end{eqnarray}
The next step is to perform the RWA by neglecting all fast oscillating terms, with frequencies $\delta_k$ and $\delta_k+ \delta_{k+1}$, while keeping the static ones. The resulting Hamiltonian, in the QF, is given by
\begin{equation}
H_{4} = \sum_{k=1}^{N-1}\frac{g_{k}\Omega_{k}}{4\delta_{k}}x_{k}(y_{k+1}\sin(\phi_{k}-\phi_{k+1}) - z_{k+1}\cos(\phi_{k}-\phi_{k+1})),
\end{equation}
as appears in Eq.~\ref{H_eff_QF_complete}.

which, when written back in the DF results in 
\begin{equation}
H_{DF} = \frac{1}{2}\sum_{k=1}^{N}( \Omega_{k}x_{k} + \delta_{k}z_{k}) + \sum_{k=1}^{N-1}\frac{g_{k}\Omega_{k}}{4\delta_{k}}z_{k}(x_{k+1}\cos\varphi_{k}(t) + y_{k+1}\sin\varphi_{k}(t)).
\label{eq:effHamilNqDF}
\end{equation}
Again, for completeness, we can write this effective Hamiltonian 
\begin{eqnarray}
H_{\text{IBM}}^{eff} &=& \sum_{k=1}^{N-1}\frac{\Omega_{k}}{2}  \Big[ x_{k}\cos(\delta_{k}t-\phi_{k}) - y_{k}\sin(\delta_{k}t-\phi_{k}) \nonumber\\
&+& \frac{g_{k}}{2\delta_{k}}z_{k}(x_{k+1}\cos\phi_{k} + y_{k+1}\sin\phi_{k})\Big],
\end{eqnarray}
as well as the original one
\begin{eqnarray}
H_{\text{IBM}} &=& \frac{1}{2}\sum_{k=1}^{N-1} \Big[ \Omega_{k} (x_{k}\cos(\delta_{k}t-\phi_{k}) - y_{k}\sin(\delta_{k}t-\phi_{k})) \nonumber\\
&+& \frac{g_{k}}{2}(\cos\delta_{k}t (x_{k}x_{k+1} + y_{k}y_{k+1}) + \sin\delta_{k}t (x_{k}y_{k+1} - y_{k}x_{k+1})) \Big]
\label{eq:IBMHamilNq}
\end{eqnarray}
in IBM's frame. 

To move to the IBM's frame we need to apply the following transformation
\begin{equation}
V = \exp\left[ -\frac{i}{2}\sum_{k=1}^{N}(\delta_{k}t-\phi_{k})z_{k}\right],
\end{equation}
which leads directly to the equations shown in the main text.

\subsection{Synthesis errors}
\label{AppBchap6}
In this appendix, we want to show the synthesis errors corresponding to the toggled Hamiltonians. For the $XY$ model, the original Hamiltonian is
\begin{eqnarray}
H^{org}_{XY} &=& \frac{g}{4} \sum_{k=1}^{N-1} \bigg\{ - \frac{\Omega}{\delta}(y_{k}y_{k+1} + x_{k}x_{k+1}) + (x_{k}y_{k+1}+y_{k}x_{k+1}-2z_{k}z_{k+1})\cos\delta t  \nonumber\\
 &+& [z_{k}(y_{k+1}-x_{k+1}) + (x_{k}-y_{k})z_{k+1}]\sin\delta t \nonumber\\
&+& \frac{\Omega}{\delta} [z_{k}(y_{k+1}-x_{k+1})\sin 2\delta t - (y_{k}y_{k+1} + x_{k}x_{k+1})\cos 2\delta t] \bigg\}.
\end{eqnarray}
Then, the difference between original and effective Hamiltonians,
\begin{eqnarray}
\Delta H_{XY} &=& \frac{g}{4} \sum_{k=1}^{N-1} \bigg\{ (x_{k}y_{k+1}+y_{k}x_{k+1}-2z_{k}z_{k+1})\cos\delta t \nonumber\\
&+& [z_{k}(y_{k+1}-x_{k+1}) + (x_{k}-y_{k})z_{k+1}]\sin\delta t \nonumber\\
&+& \frac{\Omega}{\delta} [z_{k}(y_{k+1}-x_{k+1})\sin 2\delta t - (y_{k}y_{k+1} + x_{k}x_{k+1})\cos2\delta t] \bigg\},
\end{eqnarray}
constitutes the error we want to estimate. We find the Frobenius norm is given by
\begin{equation}
|| \Delta H_{XY} ||_{F} = \frac{g}{2}\sqrt{N-1} .
\end{equation}
On the other hand, the original $ZZ$ toggled Hamiltonian is
\begin{eqnarray}
H^{org}_{ZZ} &=& \frac{g}{4} \sum_{k=1}^{N-1} \bigg\{ \left[ z_{k}\frac{\Omega}{\delta} - x_{k}\cos\delta t + y_{k}\sin\delta t \right] z_{k+1}\nonumber\\
&+& (y_{k}\cos\delta t + x_{k}\sin\delta t)y_{k+1} \nonumber \\
&+& \cos\varphi_{k}(t) \Big[  z_{k}(z_{k+1}\frac{\Omega}{\delta} -x_{k+1}\cos\delta t + y_{k+1}\sin\delta t) \nonumber  \\
&+& y_{k}(y_{k+1}\cos\delta t + x_{k+1}\sin\delta t) \Big]\nonumber \\
&+& \sin\varphi_{k}(t) \Big[ -z_{k}(y_{k+1}\cos\delta t + x_{k+1}\sin\delta t)\nonumber\\
&+& y_{k}(z_{k+1}\frac{\Omega}{\delta} - x_{k+1}\cos\delta t + y_{k+1}\sin\delta t) \Big] \bigg\},
\end{eqnarray}
so that the difference is
\begin{eqnarray}
\nonumber \Delta H_{ZZ} &=& \frac{g}{4} \sum_{k=1}^{N-1} \bigg\{ (- x_{k}\cos\delta t + y_{k}\sin\delta t ) z_{k+1} + (y_{k}\cos\delta t + x_{k}\sin\delta t)y_{k+1} \\
\nonumber &+& \cos\varphi_{k}(t) \Big[  z_{k}(z_{k+1}\frac{\Omega}{\delta} -x_{k+1}\cos\delta t + y_{k+1}\sin\delta t) \nonumber\\
&+& y_{k}(y_{k+1}\cos\delta t + x_{k+1}\sin\delta t) \Big] \nonumber \\
&+& \sin\varphi_{k}(t) \Big[ -z_{k}(y_{k+1}\cos\delta t + x_{k+1}\sin\delta t) \nonumber\\
&+& y_{k}(z_{k+1}\frac{\Omega}{\delta} - x_{k+1}\cos\delta t + y_{k+1}\sin\delta t) \Big] \bigg\}.
\end{eqnarray}
The Frobenius norm is then given by
\begin{equation}
|| \Delta H_{ZZ} ||_{F} = \frac{g}{2\sqrt{2}}\sqrt{N-1}\sqrt{2+\cos\delta t\cos(\varphi_{k}(t)-\delta t)+\frac{\Omega}{\delta}\sin\delta t\sin\varphi_{k}(t)}.
\end{equation}

\subsection{Unitary transformation to the Quad Frame}
\label{AppCchap6}
In order to perform a quantum simulation on the QF, we need to translate the state of our circuit to this frame. Then, considering a simulation scenario in any IBM superconducting chip, we want to find a simple expression for the combination of rotations we need to apply in order to move from IBM's frame into the QF. For that, we will expand the product 
\begin{equation}
U^{\dagger}_{\text{IBM}}U_{12}U_{3}U_{4} = e^{\frac{it}{2}\sum_{k=1}^{N-1}\omega^{q}_{k}z_{k}} e^{-\frac{it}{2}\sum_{k=1}^{N-1}\omega_{k}z_{k}} e^{\frac{i}{2}\sum_{k=1}^{N-1}\xi_{k} y_{k}} e^{-\frac{it}{2}\sum_{k=1}^{N-1}\eta_{k} x_{k}},
\end{equation}
having set $\phi_{k}=\phi=0$. See that the first two exponentials can be combined, such that
\begin{equation}
U^{\dagger}_{\text{IBM}}U_{12}U_{3}U_{4} =  e^{\frac{it}{2}\sum_{k=1}^{N-1}\delta_{k}z_{k}} e^{\frac{i}{2}\sum_{k=1}^{N-1}\xi_{k} y_{k}} e^{-\frac{it}{2}\sum_{k=1}^{N-1}\eta_{k} x_{k}},
\end{equation}
where $\delta_{k} = \omega_{k}^{q}-\omega_{k}$. Now, it is satisfied that
\begin{equation}
 \exp\left[ i\sum_{k=1}^{N}\theta_{k} \sigma_{k} \right] = \prod_{k=1}^{N} e^{i\theta_{k}\sigma_{k}}
\end{equation}
for $\sigma = x$, $y$, or $z$. This means that we can write
\begin{equation}
U^{\dagger}_{\text{IBM}}U_{12}U_{3}U_{4} =  \prod_{k=1}^{N-1} e^{\frac{it}{2}\delta_{k}z_{k}} e^{\frac{i}{2}\xi_{k}y_{k}} e^{-\frac{it}{2}\eta_{k}x_{k}},
\end{equation}
and we can use the Euler form for Pauli matrices,
\begin{equation}
e^{ i \theta \sigma} = \cos\theta \, \mathbb{I} + i\sin\theta \, \sigma,
\end{equation}
to express these rotations as 
\begin{eqnarray}
U^{\dagger}_{\text{IBM}}U_{12}U_{3}U_{4} &=&  \prod_{k=1}^{N-1} (\cos\frac{\delta_{k}t}{2} \mathbb{I}_{k} + i\sin\frac{\delta_{k}t}{2} z_{k}) (\cos\frac{\xi_{k}}{2} \mathbb{I}_{k} + i\sin\frac{\xi_{k}}{2} y_{k}) \nonumber\\
&& (\cos\frac{\eta_{k}t}{2} \mathbb{I}_{k} - i\sin\frac{\eta_{k}t}{2} x_{k}).
\end{eqnarray}
Recall that, working in the regime $\Omega \ll \delta$, we had approximated $\eta \approx \delta$, $\sin\xi \approx 1$, and $\cos\xi \approx \Omega/\delta$. Knowing that $\sin\theta/2 = \sqrt{(1-\cos\theta)/2}$ and $\cos\theta/2 = \sqrt{(1+\cos\theta)/2}$, we can simplify
\begin{equation}
\cos\frac{\xi_{k}t}{2} \mathbb{I}_{k} + i\sin\frac{\xi_{k}t}{2} y_{k} \approx \frac{1}{\sqrt{2}} \left( \mathbb{I}_{k} + i y_{k} + \frac{\Omega_{k}}{2\delta_{k}}(\mathbb{I}_{k} - i y_{k})\right),
\end{equation}
where we have used $\sqrt{1\pm x}\approx 1\pm x/2$ for small $x$. In this expansion, we eventually find
\begin{equation}
U^{\dagger}_{\text{IBM}}U_{12}U_{3}U_{4} =  \prod_{k=1}^{N-1} \frac{1}{\sqrt{2}}\bigg[ \mathbb{I}_{k} + i y_{k}+ \frac{\Omega_{k}}{2\delta_{k}} ((\mathbb{I}_{k} - i y_{k}) \cos\delta_{k}t  + i(z_{k}-x_{k}) \sin\delta_{k}t ) \bigg],
\end{equation}
which we will denote by $V_{\text{QF}}$. Let us check the unitarity of this operator by computing
\begin{equation}
V_{\text{QF}}V_{\text{QF}}^{\dagger} = \mathbb{I} + \mathcal{O}\left(\frac{\Omega^{2}}{\delta^{2}}\right).
\end{equation}
The previous calculations were set in the weak-driving regime ($\Omega/\delta \ll 1$), considering terms up to first order in $\Omega/\delta$ and neglecting higher orders. This is consistent with the approximations we have made here, and thus the unitarity of $V_{\text{QF}}$ relies on these approximations.

\subsection{2D $XY$ model}
\label{AppDchap6}

\begin{figure}[t]
\centering
\includegraphics[width=0.85 \textwidth]{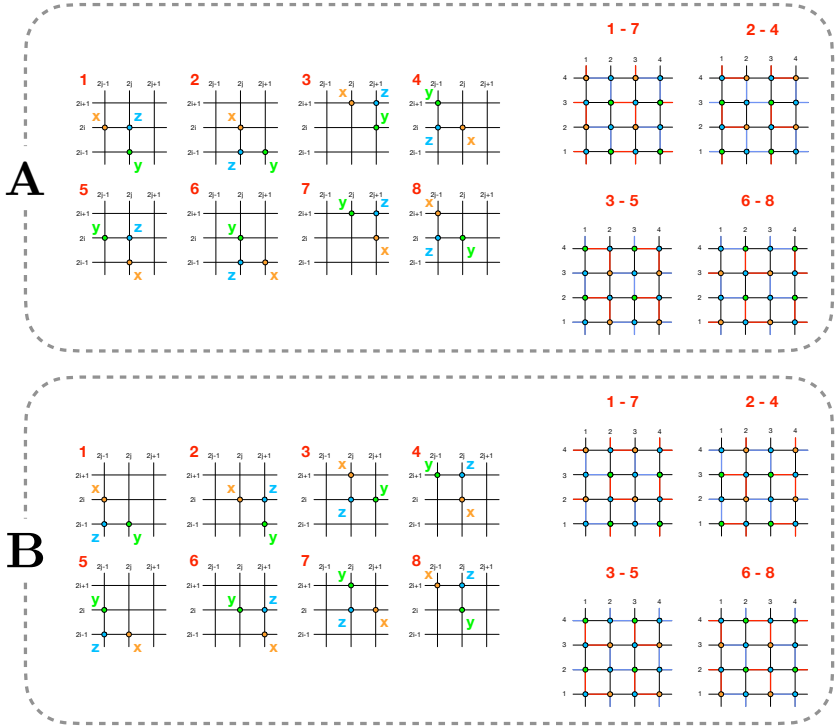}
\caption{{\bf Representation of the spin triplets, grouped in $A$ and $B$}. These triplets, inside either $A$ or $B$, can be grouped in pairs, which can be used to tile the entire lattice with staircase patterns. These tilings form strings with periodic boundary conditions along the diagonals of the lattice, which we transform from spins to fermions in order to estimate $||A||+||B||$, the Trotter error associated with the DA simulation of the $XY$ model in 2D.}
\label{Fig7chap6}
\end{figure}

In this appendix, we describe the transformation from a spin lattice to a string of fermions with 2-site hopping, which allows us to estimate more accurately the Trotter error associated to the simulation of the $XY$ model in 2 dimensions. This error will be given by the commutator $[H_{I},H_{II}]$, split into $A =  [H_{I}^{yy},H_{II}^{xx}]$ and $B = [H_{I}^{xx},H_{II}^{yy}]$. Both $A$ and $B$ include 8 terms, each one having the form $x$-$z$-$y$ at three different vertices, as we can see in Fig.~\ref{Fig7chap6}. Assembling pairs of these terms, joining $x_{i}z_{i+1}y_{i+2}$ with $y_{i+2}z_{i+3}x_{i+4}$ or $y_{i}z_{i+1}x_{i+2}$ with $x_{i+2}z_{i+3}y_{i+4}$, either in $A$ or in $B$, we can construct diagonal strings in the lattice, which we also represent in Fig.
~\ref{Fig7chap6}. These diagonals can then be thought of as 1-d strings, and given this outline, we can apply a Jordan-Wigner transformation and introduce Majorana fermion operators. Let us take a couple of terms (1 \& 7) in $B$ to illustrate this issue:
 \begin{equation}
 \sum_{i,j} (x_{2i,2j-1}z_{2i-1,2j-1}y_{2i-1,2j} + x_{2i,2j+1}z_{2i,2j}y_{2i+1,2j}),
 \end{equation}
can be turned into a string as
 \begin{equation}
 \sum_{j=1}^{N^{2}/4} (x_{4j-3}z_{4j-2}y_{4j-1} + y_{4j-1}z_{4j}x_{4j+1}).
 \end{equation}

The Jordan-Wigner transformation, followed by defining Majorana operators $\gamma^{(1)}_{i} = c_{i}+c^{\dagger}_{i}$ and $\gamma^{(2)}_{i} = -i(c^{\dagger}_{i}-c_{i})$, transforms spins as
\begin{equation}
x_{j}z_{j+1}y_{j+2} = -i \gamma_{j}^{(2)}\gamma_{j+2}^{(2)}
\end{equation}
and
\begin{equation}
y_{j}z_{j+1}x_{j+2} = i \gamma_{j}^{(1)}\gamma_{j+2}^{(1)}.
\end{equation}
These transformations lead to the Hamiltonian
\begin{equation}
-i \sum_{j=1}^{N^{2}/4} \left(\gamma^{(2)}_{4j-3}\gamma^{(2)}_{4j-1}-\gamma^{(1)}_{4j-1}\gamma^{(1)}_{4j+1}\right).
 \end{equation}
 The Fourier transform of Majorana operators is given by
\begin{eqnarray}
\nonumber \gamma^{(1)}_{j} &=& \frac{1}{\sqrt{N}} \sum_{k}\left(\gamma^{(1)}_{k}\cos kj + \gamma^{(2)}_{k}\sin kj \right), \\
\gamma^{(2)}_{j} &=& \frac{1}{\sqrt{N}} \sum_{k}\left(\gamma^{(2)}_{k}\cos kj - \gamma^{(1)}_{k}\sin kj \right),
\end{eqnarray}
together with
\begin{eqnarray}
\nonumber \sum_{j}\gamma^{(1)}_{j}\gamma^{(1)}_{j+2} &=& \frac{1}{2}\sum_{k}\Big[ \cos2k \left(\gamma^{(1)}_{k}\gamma^{(1)}_{k}+\gamma^{(2)}_{k}\gamma^{(2)}_{k}+\gamma^{(1)}_{k}\gamma^{(1)}_{-k}-\gamma^{(2)}_{k}\gamma^{(2)}_{-k} \right) \\
\nonumber &+& \sin2k \left(\gamma^{(1)}_{k}\gamma^{(2)}_{k}-\gamma^{(2)}_{k}\gamma^{(1)}_{k}-\gamma^{(1)}_{k}\gamma^{(2)}_{-k}-\gamma^{(2)}_{k}\gamma^{(1)}_{-k} \right)  \Big] \\
\nonumber \sum_{j}\gamma^{(2)}_{j}\gamma^{(2)}_{j+2} &=& \frac{1}{2}\sum_{k}\Big[ \cos2k \left(\gamma^{(1)}_{k}\gamma^{(1)}_{k}+\gamma^{(2)}_{k}\gamma^{(2)}_{k}-\gamma^{(1)}_{k}\gamma^{(1)}_{-k}+\gamma^{(2)}_{k}\gamma^{(2)}_{-k} \right) \\
&+& \sin2k \left(\gamma^{(1)}_{k}\gamma^{(2)}_{k}-\gamma^{(2)}_{k}\gamma^{(1)}_{k}+\gamma^{(1)}_{k}\gamma^{(2)}_{-k}+\gamma^{(2)}_{k}\gamma^{(1)}_{-k} \right)  \Big],
\end{eqnarray}
which leads to the expression of a string in Fourier space,
\begin{equation}
i \sum_{k} \left[ \cos 2k \left(\gamma^{(1)}_{k}\gamma^{(1)}_{-k}-\gamma^{(2)}_{k}\gamma^{(2)}_{-k} \right) - \sin 2k \left(\gamma^{(1)}_{k}\gamma^{(2)}_{-k}+\gamma^{(2)}_{k}\gamma^{(1)}_{-k} \right)\right].
\end{equation}
 The resulting matrix is block-diagonal, and it has two types of blocks. The first block, $\Gamma_{1}$, contains the elements $\cos 2k$ for $k\in[-\pi,\pi]$, and the second block, $\Gamma_{2}$, contains the elements $\sin 2k$. Both blocks have elements only along the anti-diagonal. Then, the matrix in Fourier space looks like
 \begin{equation}
 \begin{pmatrix} \Gamma_{1} & -\Gamma_{2} \\ -\Gamma_{2} & -\Gamma_{1} \end{pmatrix}.
 \end{equation}
We can write the eigenvalue problem as 
\begin{equation}
\det \left[  \begin{pmatrix} \Gamma_{1} & -\Gamma_{2} \\ -\Gamma_{2} & -\Gamma_{1} \end{pmatrix} -  \begin{pmatrix} \mathbb{I}\lambda_{+} & 0 \\ 0 & \mathbb{I}\lambda_{-} \end{pmatrix}\right] = 0,
\end{equation}
for $\lambda_{+}=\lambda$ and $\lambda_{-}=-\lambda$. Given that the determinant of this matrix is
\begin{equation}
\det(\Gamma_{1}+\Gamma_{2}\Gamma_{1}^{-1}\Gamma_{2})\det\Gamma_{1} = 1,\nonumber
\end{equation}
we have a hint that $\lambda = \pm 1$, and in fact 
\begin{equation}
\det \left[  \begin{pmatrix} \Gamma_{1} & -\Gamma_{2} \\ -\Gamma_{2} & -\Gamma_{1} \end{pmatrix} -  \begin{pmatrix} \mathbb{I}\lambda & 0 \\ 0 & -\mathbb{I}\lambda \end{pmatrix}\right] = (1-\lambda^{2})^{N^{2}/2} = 0
\end{equation}
As we can see, the eigenvalues of this matrix are $\lambda_{j}=\pm 1$. Then, the spectral norm of this pair of elements in the commutator is equal to one, for any lattice size. Since there are 4 total pairs in $B$, we will get that the norm can be estimated as $||B|| \leq 8 J^{2}$, taking into account the $2i$ factor from Pauli commutation relations, and the $J^{2}$ from the analog Hamiltonian. Given that the norm of $A$ can be estimated in the same way, we have
\begin{equation}
||[H_{I},H_{II}]|| = || - A+B || \leq ||A|| + ||B|| \leq 16 J^{2},
\end{equation}
where we have eliminated the dependence on the system size.


\gdef\thesubsection{}


%

\renewcommand{\refname}{Bibliography}

\bibliographystyle{X}

\let\oldbibliography\bibliography

\renewcommand{\bibliography}[1]{{%
\let\section\subsection
\oldbibliography{#1}}}

\end{document}